\documentclass[longauth]{aa}

\usepackage{graphicx}
\usepackage{natbib}
\usepackage{scalerel}

\usepackage[table]{xcolor}

\usepackage{txfonts}
\usepackage[pdfencoding=auto,psdextra]{hyperref}
\hypersetup{
    colorlinks=true,
    linkcolor=blue,
    filecolor=magenta,      
    urlcolor=blue,
    citecolor=blue
}
\makeatletter
\renewcommand*\aa@pageof{, page \thepage{} of \pageref*{LastPage}}
\makeatother

\usepackage[utf8]{inputenc}

\usepackage[switch, modulo]{lineno}

\usepackage{euclid}
\usepackage{siunitx}

\usepackage{glossaries}
\newacronym{edff}{EDF-F}{Euclid Deep Field Fornax}
\newacronym{edfall}{EDFs}{Euclid Deep Fields}

\newcommand{\AMICO}{\texttt{AMICO}\xspace}
\newcommand{\PZWav}{\texttt{PZWav}\xspace}
\newcommand{\tfive}{\theta_{\rm 500}}
\newcommand{\Mfive}{M_{500}}
\newcommand{\Rfive}{R_{500}}
\newcommand{\ntfive}{{\rm n}\theta_{\rm 500}}
\newcommand{\eROSITA}{eROSITA\xspace}
\newcommand{\sigmav}{\sigma_\mathrm{v}}
\newcommand{\mtwoc}{M2C\xspace}
\newcommand{\XMM}{XMM-\emph{Newton}\xspace}
\newcommand{\tway}{two-way\xspace}
\newcommand{\catT}{catalogue\,T\xspace}
\newcommand{\catM}{catalogue\,M\xspace}

\usepackage{glossaries}

\newacronym{esa}{ESA}{European Space Agency}
\newacronym{cmb}{CMB}{cosmic microwave background}
\newacronym{ecgc}{ECGC}{Euclid Catalogue of Galaxy Clusters}
\newacronym{ec}{EC}{Euclid Consortium}
\newacronym{des}{DES}{Dark Energy Survey}
\newacronym{rm}{RM}{\texttt{RedMaPPer}}
\newacronym{icm}{ICM}{intracluster medium}
\newacronym{sz}{SZ}{Sunyaev--Zeldovich}
\newacronym{mcxc}{MCXC-II}{Meta-Catalogue of X-ray Clusters II}
\newacronym{rosat}{ROSAT}{R\"ontgen Satellite}
\newacronym{rass}{RASS}{ROSAT All-Sky Survey}
\newacronym{mcsz}{MCSZ}{Meta-Catalogue of SZ Clusters}
\newacronym{spt}{SPT}{South Pole Telescope}
\newacronym{act}{ACT}{Atacama Cosmology Telescope}
\newacronym{lc2}{LC$^2$}{Literature catalogues of weak lensing clusters of galaxies}
\newacronym{mccd}{MCCD}{Meta-Catalogue of Cluster Dispersions}
\newacronym{sdss}{SDSS}{Sloan Digital Sky Survey}
\newacronym{desi}{DESI}{Dark Energy Spectroscopic Instrument}
\newacronym{snr}{S/N}{signal-to-noise-ratio}

\usepackage{cleveref}
\crefname{section}{Sect.}{Sects.}
\Crefname{section}{Section}{Sections}
\crefname{figure}{Fig.}{Figs.}
\Crefname{figure}{Figure}{Figures}
\crefname{table}{Table}{Tables}
\Crefname{table}{Table}{Tables}
\crefname{appendix}{Appendix}{Appendices}
\Crefname{appendix}{Appendix}{Appendices}

\begin{document}
%
% Put the title of your paper here:
%

\title{\Euclid preparation}
\subtitle{XCI. Validation method for the detections \\ in the Euclid Catalogue of Galaxy Clusters using external data}

%%%% Version Monday 8th of September 2025 09:26:34 AM UT												
%%%% Please do not edit the author list -- contact ECEB Bureau for changes
\newcommand{\orcid}[1]{} %% if already defined in aa.cls: comment, or use renewcommand			   
\author{Euclid Collaboration: J.-B.~Melin\thanks{\email{jean-baptiste.melin@cea.fr}}\inst{\ref{aff1}}
\and S.~A.~Stanford\orcid{0000-0003-0122-0841}\inst{\ref{aff2}}
\and A.~Widmer\orcid{0009-0005-4111-2716}\inst{\ref{aff3}}
\and P.~Tarr\'io\orcid{0000-0002-0915-0131}\inst{\ref{aff4}}
\and J.~G.~Bartlett\orcid{0000-0002-0685-8310}\inst{\ref{aff3},\ref{aff5}}
\and T.~Sadibekova\orcid{0000-0002-5162-4222}\inst{\ref{aff6}}
\and G.~W.~Pratt\inst{\ref{aff6}}
\and M.~Arnaud\inst{\ref{aff7},\ref{aff8}}
\and F.~Pacaud\orcid{0000-0002-6622-4555}\inst{\ref{aff9}}
\and T.~H.~Reiprich\orcid{0000-0003-2047-2884}\inst{\ref{aff9}}
\and A.~Biviano\orcid{0000-0002-0857-0732}\inst{\ref{aff10},\ref{aff11}}
\and S.~Bardelli\orcid{0000-0002-8900-0298}\inst{\ref{aff12}}
\and S.~Borgani\orcid{0000-0001-6151-6439}\inst{\ref{aff13},\ref{aff11},\ref{aff10},\ref{aff14},\ref{aff15}}
\and P.-S.~Corasaniti\orcid{0000-0002-6386-7846}\inst{\ref{aff16},\ref{aff17}}
\and S.~Ettori\orcid{0000-0003-4117-8617}\inst{\ref{aff12},\ref{aff18}}
\and A.~Finoguenov\orcid{0000-0002-4606-5403}\inst{\ref{aff19}}
\and Z.~Ghaffari\orcid{0000-0002-6467-8078}\inst{\ref{aff10},\ref{aff11}}
\and P.~A.~Giles\orcid{0000-0003-4937-8453}\inst{\ref{aff20}}
\and M.~Girardi\orcid{0000-0003-1861-1865}\inst{\ref{aff13},\ref{aff10}}
\and J.~B.~Golden-Marx\orcid{0000-0002-6394-045X}\inst{\ref{aff21}}
\and A.~H.~Gonzalez\orcid{0000-0002-0933-8601}\inst{\ref{aff22}}
\and M.~Klein\orcid{0000-0002-8248-4488}\inst{\ref{aff23}}
\and G.~F.~Lesci\orcid{0000-0002-4607-2830}\inst{\ref{aff24},\ref{aff12}}
\and M.~Maturi\orcid{0000-0002-3517-2422}\inst{\ref{aff25},\ref{aff26}}
\and B.~J.~Maughan\orcid{0000-0003-0791-9098}\inst{\ref{aff27}}
\and L.~Moscardini\orcid{0000-0002-3473-6716}\inst{\ref{aff24},\ref{aff12},\ref{aff28}}
\and M.~Pierre\orcid{0000-0003-2648-2469}\inst{\ref{aff6}}
\and M.~Radovich\orcid{0000-0002-3585-866X}\inst{\ref{aff29}}
\and P.~Rosati\orcid{0000-0002-6813-0632}\inst{\ref{aff30},\ref{aff12}}
\and J.~G.~Sorce\orcid{0000-0002-2307-2432}\inst{\ref{aff31},\ref{aff32}}
\and E.~Tsaprazi\inst{\ref{aff33}}
\and B.~Altieri\orcid{0000-0003-3936-0284}\inst{\ref{aff34}}
\and A.~Amara\inst{\ref{aff35}}
\and S.~Andreon\orcid{0000-0002-2041-8784}\inst{\ref{aff36}}
\and N.~Auricchio\orcid{0000-0003-4444-8651}\inst{\ref{aff12}}
\and C.~Baccigalupi\orcid{0000-0002-8211-1630}\inst{\ref{aff11},\ref{aff10},\ref{aff14},\ref{aff37}}
\and M.~Baldi\orcid{0000-0003-4145-1943}\inst{\ref{aff38},\ref{aff12},\ref{aff28}}
\and E.~Branchini\orcid{0000-0002-0808-6908}\inst{\ref{aff39},\ref{aff40},\ref{aff36}}
\and M.~Brescia\orcid{0000-0001-9506-5680}\inst{\ref{aff41},\ref{aff42}}
\and S.~Camera\orcid{0000-0003-3399-3574}\inst{\ref{aff43},\ref{aff44},\ref{aff45}}
\and G.~Ca\~nas-Herrera\orcid{0000-0003-2796-2149}\inst{\ref{aff46},\ref{aff47},\ref{aff48}}
\and V.~Capobianco\orcid{0000-0002-3309-7692}\inst{\ref{aff45}}
\and C.~Carbone\orcid{0000-0003-0125-3563}\inst{\ref{aff49}}
\and J.~Carretero\orcid{0000-0002-3130-0204}\inst{\ref{aff50},\ref{aff51}}
\and M.~Castellano\orcid{0000-0001-9875-8263}\inst{\ref{aff52}}
\and G.~Castignani\orcid{0000-0001-6831-0687}\inst{\ref{aff12}}
\and S.~Cavuoti\orcid{0000-0002-3787-4196}\inst{\ref{aff42},\ref{aff53}}
\and K.~C.~Chambers\orcid{0000-0001-6965-7789}\inst{\ref{aff54}}
\and A.~Cimatti\inst{\ref{aff55}}
\and C.~Colodro-Conde\inst{\ref{aff56}}
\and G.~Congedo\orcid{0000-0003-2508-0046}\inst{\ref{aff57}}
\and C.~J.~Conselice\orcid{0000-0003-1949-7638}\inst{\ref{aff58}}
\and L.~Conversi\orcid{0000-0002-6710-8476}\inst{\ref{aff59},\ref{aff34}}
\and Y.~Copin\orcid{0000-0002-5317-7518}\inst{\ref{aff60}}
\and F.~Courbin\orcid{0000-0003-0758-6510}\inst{\ref{aff61},\ref{aff62}}
\and H.~M.~Courtois\orcid{0000-0003-0509-1776}\inst{\ref{aff63}}
\and A.~Da~Silva\orcid{0000-0002-6385-1609}\inst{\ref{aff64},\ref{aff65}}
\and H.~Degaudenzi\orcid{0000-0002-5887-6799}\inst{\ref{aff66}}
\and G.~De~Lucia\orcid{0000-0002-6220-9104}\inst{\ref{aff10}}
\and H.~Dole\orcid{0000-0002-9767-3839}\inst{\ref{aff32}}
\and M.~Douspis\orcid{0000-0003-4203-3954}\inst{\ref{aff32}}
\and F.~Dubath\orcid{0000-0002-6533-2810}\inst{\ref{aff66}}
\and C.~A.~J.~Duncan\orcid{0009-0003-3573-0791}\inst{\ref{aff57},\ref{aff58}}
\and X.~Dupac\inst{\ref{aff34}}
\and S.~Dusini\orcid{0000-0002-1128-0664}\inst{\ref{aff67}}
\and S.~Escoffier\orcid{0000-0002-2847-7498}\inst{\ref{aff68}}
\and M.~Fabricius\orcid{0000-0002-7025-6058}\inst{\ref{aff69},\ref{aff70}}
\and M.~Farina\orcid{0000-0002-3089-7846}\inst{\ref{aff71}}
\and S.~Farrens\orcid{0000-0002-9594-9387}\inst{\ref{aff6}}
\and F.~Faustini\orcid{0000-0001-6274-5145}\inst{\ref{aff52},\ref{aff72}}
\and S.~Ferriol\inst{\ref{aff60}}
\and F.~Finelli\orcid{0000-0002-6694-3269}\inst{\ref{aff12},\ref{aff18}}
\and P.~Fosalba\orcid{0000-0002-1510-5214}\inst{\ref{aff73},\ref{aff74}}
\and M.~Frailis\orcid{0000-0002-7400-2135}\inst{\ref{aff10}}
\and E.~Franceschi\orcid{0000-0002-0585-6591}\inst{\ref{aff12}}
\and M.~Fumana\orcid{0000-0001-6787-5950}\inst{\ref{aff49}}
\and S.~Galeotta\orcid{0000-0002-3748-5115}\inst{\ref{aff10}}
\and K.~George\orcid{0000-0002-1734-8455}\inst{\ref{aff70}}
\and B.~Gillis\orcid{0000-0002-4478-1270}\inst{\ref{aff57}}
\and C.~Giocoli\orcid{0000-0002-9590-7961}\inst{\ref{aff12},\ref{aff28}}
\and J.~Gracia-Carpio\inst{\ref{aff69}}
\and A.~Grazian\orcid{0000-0002-5688-0663}\inst{\ref{aff29}}
\and F.~Grupp\inst{\ref{aff69},\ref{aff70}}
\and S.~V.~H.~Haugan\orcid{0000-0001-9648-7260}\inst{\ref{aff75}}
\and W.~Holmes\inst{\ref{aff76}}
\and F.~Hormuth\inst{\ref{aff77}}
\and A.~Hornstrup\orcid{0000-0002-3363-0936}\inst{\ref{aff78},\ref{aff79}}
\and K.~Jahnke\orcid{0000-0003-3804-2137}\inst{\ref{aff80}}
\and M.~Jhabvala\inst{\ref{aff81}}
\and E.~Keih\"anen\orcid{0000-0003-1804-7715}\inst{\ref{aff82}}
\and S.~Kermiche\orcid{0000-0002-0302-5735}\inst{\ref{aff68}}
\and A.~Kiessling\orcid{0000-0002-2590-1273}\inst{\ref{aff76}}
\and M.~Kilbinger\orcid{0000-0001-9513-7138}\inst{\ref{aff6}}
\and B.~Kubik\orcid{0009-0006-5823-4880}\inst{\ref{aff60}}
\and M.~K\"ummel\orcid{0000-0003-2791-2117}\inst{\ref{aff70}}
\and M.~Kunz\orcid{0000-0002-3052-7394}\inst{\ref{aff83}}
\and H.~Kurki-Suonio\orcid{0000-0002-4618-3063}\inst{\ref{aff19},\ref{aff84}}
\and A.~M.~C.~Le~Brun\orcid{0000-0002-0936-4594}\inst{\ref{aff16}}
\and S.~Ligori\orcid{0000-0003-4172-4606}\inst{\ref{aff45}}
\and P.~B.~Lilje\orcid{0000-0003-4324-7794}\inst{\ref{aff75}}
\and V.~Lindholm\orcid{0000-0003-2317-5471}\inst{\ref{aff19},\ref{aff84}}
\and I.~Lloro\orcid{0000-0001-5966-1434}\inst{\ref{aff85}}
\and G.~Mainetti\orcid{0000-0003-2384-2377}\inst{\ref{aff86}}
\and D.~Maino\inst{\ref{aff87},\ref{aff49},\ref{aff88}}
\and E.~Maiorano\orcid{0000-0003-2593-4355}\inst{\ref{aff12}}
\and O.~Mansutti\orcid{0000-0001-5758-4658}\inst{\ref{aff10}}
\and O.~Marggraf\orcid{0000-0001-7242-3852}\inst{\ref{aff9}}
\and M.~Martinelli\orcid{0000-0002-6943-7732}\inst{\ref{aff52},\ref{aff89}}
\and N.~Martinet\orcid{0000-0003-2786-7790}\inst{\ref{aff90}}
\and F.~Marulli\orcid{0000-0002-8850-0303}\inst{\ref{aff24},\ref{aff12},\ref{aff28}}
\and R.~J.~Massey\orcid{0000-0002-6085-3780}\inst{\ref{aff91}}
\and S.~Maurogordato\inst{\ref{aff92}}
\and E.~Medinaceli\orcid{0000-0002-4040-7783}\inst{\ref{aff12}}
\and S.~Mei\orcid{0000-0002-2849-559X}\inst{\ref{aff3},\ref{aff5}}
\and M.~Melchior\inst{\ref{aff93}}
\and Y.~Mellier\inst{\ref{aff94},\ref{aff17}}
\and M.~Meneghetti\orcid{0000-0003-1225-7084}\inst{\ref{aff12},\ref{aff28}}
\and E.~Merlin\orcid{0000-0001-6870-8900}\inst{\ref{aff52}}
\and G.~Meylan\inst{\ref{aff95}}
\and A.~Mora\orcid{0000-0002-1922-8529}\inst{\ref{aff96}}
\and M.~Moresco\orcid{0000-0002-7616-7136}\inst{\ref{aff24},\ref{aff12}}
\and E.~Munari\orcid{0000-0002-1751-5946}\inst{\ref{aff10},\ref{aff11}}
\and R.~Nakajima\orcid{0009-0009-1213-7040}\inst{\ref{aff9}}
\and C.~Neissner\orcid{0000-0001-8524-4968}\inst{\ref{aff97},\ref{aff51}}
\and S.-M.~Niemi\orcid{0009-0005-0247-0086}\inst{\ref{aff46}}
\and C.~Padilla\orcid{0000-0001-7951-0166}\inst{\ref{aff97}}
\and S.~Paltani\orcid{0000-0002-8108-9179}\inst{\ref{aff66}}
\and F.~Pasian\orcid{0000-0002-4869-3227}\inst{\ref{aff10}}
\and K.~Pedersen\inst{\ref{aff98}}
\and V.~Pettorino\inst{\ref{aff46}}
\and G.~Polenta\orcid{0000-0003-4067-9196}\inst{\ref{aff72}}
\and M.~Poncet\inst{\ref{aff99}}
\and L.~A.~Popa\inst{\ref{aff100}}
\and L.~Pozzetti\orcid{0000-0001-7085-0412}\inst{\ref{aff12}}
\and F.~Raison\orcid{0000-0002-7819-6918}\inst{\ref{aff69}}
\and R.~Rebolo\orcid{0000-0003-3767-7085}\inst{\ref{aff56},\ref{aff101},\ref{aff102}}
\and A.~Renzi\orcid{0000-0001-9856-1970}\inst{\ref{aff103},\ref{aff67}}
\and J.~Rhodes\orcid{0000-0002-4485-8549}\inst{\ref{aff76}}
\and G.~Riccio\inst{\ref{aff42}}
\and E.~Romelli\orcid{0000-0003-3069-9222}\inst{\ref{aff10}}
\and M.~Roncarelli\orcid{0000-0001-9587-7822}\inst{\ref{aff12}}
\and E.~Rossetti\orcid{0000-0003-0238-4047}\inst{\ref{aff38}}
\and R.~Saglia\orcid{0000-0003-0378-7032}\inst{\ref{aff70},\ref{aff69}}
\and Z.~Sakr\orcid{0000-0002-4823-3757}\inst{\ref{aff25},\ref{aff104},\ref{aff105}}
\and A.~G.~S\'anchez\orcid{0000-0003-1198-831X}\inst{\ref{aff69}}
\and D.~Sapone\orcid{0000-0001-7089-4503}\inst{\ref{aff106}}
\and B.~Sartoris\orcid{0000-0003-1337-5269}\inst{\ref{aff70},\ref{aff10}}
\and P.~Schneider\orcid{0000-0001-8561-2679}\inst{\ref{aff9}}
\and T.~Schrabback\orcid{0000-0002-6987-7834}\inst{\ref{aff107}}
\and A.~Secroun\orcid{0000-0003-0505-3710}\inst{\ref{aff68}}
\and E.~Sefusatti\orcid{0000-0003-0473-1567}\inst{\ref{aff10},\ref{aff11},\ref{aff14}}
\and G.~Seidel\orcid{0000-0003-2907-353X}\inst{\ref{aff80}}
\and M.~Seiffert\orcid{0000-0002-7536-9393}\inst{\ref{aff76}}
\and S.~Serrano\orcid{0000-0002-0211-2861}\inst{\ref{aff73},\ref{aff108},\ref{aff74}}
\and P.~Simon\inst{\ref{aff9}}
\and C.~Sirignano\orcid{0000-0002-0995-7146}\inst{\ref{aff103},\ref{aff67}}
\and G.~Sirri\orcid{0000-0003-2626-2853}\inst{\ref{aff28}}
\and L.~Stanco\orcid{0000-0002-9706-5104}\inst{\ref{aff67}}
\and J.~Steinwagner\orcid{0000-0001-7443-1047}\inst{\ref{aff69}}
\and P.~Tallada-Cresp\'{i}\orcid{0000-0002-1336-8328}\inst{\ref{aff50},\ref{aff51}}
\and D.~Tavagnacco\orcid{0000-0001-7475-9894}\inst{\ref{aff10}}
\and A.~N.~Taylor\inst{\ref{aff57}}
\and I.~Tereno\orcid{0000-0002-4537-6218}\inst{\ref{aff64},\ref{aff109}}
\and N.~Tessore\orcid{0000-0002-9696-7931}\inst{\ref{aff110}}
\and S.~Toft\orcid{0000-0003-3631-7176}\inst{\ref{aff111},\ref{aff112}}
\and R.~Toledo-Moreo\orcid{0000-0002-2997-4859}\inst{\ref{aff113}}
\and F.~Torradeflot\orcid{0000-0003-1160-1517}\inst{\ref{aff51},\ref{aff50}}
\and I.~Tutusaus\orcid{0000-0002-3199-0399}\inst{\ref{aff104}}
\and L.~Valenziano\orcid{0000-0002-1170-0104}\inst{\ref{aff12},\ref{aff18}}
\and J.~Valiviita\orcid{0000-0001-6225-3693}\inst{\ref{aff19},\ref{aff84}}
\and T.~Vassallo\orcid{0000-0001-6512-6358}\inst{\ref{aff70},\ref{aff10}}
\and G.~Verdoes~Kleijn\orcid{0000-0001-5803-2580}\inst{\ref{aff114}}
\and A.~Veropalumbo\orcid{0000-0003-2387-1194}\inst{\ref{aff36},\ref{aff40},\ref{aff39}}
\and Y.~Wang\orcid{0000-0002-4749-2984}\inst{\ref{aff115}}
\and J.~Weller\orcid{0000-0002-8282-2010}\inst{\ref{aff70},\ref{aff69}}
\and G.~Zamorani\orcid{0000-0002-2318-301X}\inst{\ref{aff12}}
\and F.~M.~Zerbi\inst{\ref{aff36}}
\and E.~Zucca\orcid{0000-0002-5845-8132}\inst{\ref{aff12}}
\and V.~Allevato\orcid{0000-0001-7232-5152}\inst{\ref{aff42}}
\and M.~Ballardini\orcid{0000-0003-4481-3559}\inst{\ref{aff30},\ref{aff116},\ref{aff12}}
\and M.~Bolzonella\orcid{0000-0003-3278-4607}\inst{\ref{aff12}}
\and E.~Bozzo\orcid{0000-0002-8201-1525}\inst{\ref{aff66}}
\and C.~Burigana\orcid{0000-0002-3005-5796}\inst{\ref{aff117},\ref{aff18}}
\and R.~Cabanac\orcid{0000-0001-6679-2600}\inst{\ref{aff104}}
\and A.~Cappi\inst{\ref{aff12},\ref{aff92}}
\and D.~Di~Ferdinando\inst{\ref{aff28}}
\and J.~A.~Escartin~Vigo\inst{\ref{aff69}}
\and L.~Gabarra\orcid{0000-0002-8486-8856}\inst{\ref{aff118}}
\and J.~Mart\'{i}n-Fleitas\orcid{0000-0002-8594-569X}\inst{\ref{aff119}}
\and S.~Matthew\orcid{0000-0001-8448-1697}\inst{\ref{aff57}}
\and N.~Mauri\orcid{0000-0001-8196-1548}\inst{\ref{aff55},\ref{aff28}}
\and R.~B.~Metcalf\orcid{0000-0003-3167-2574}\inst{\ref{aff24},\ref{aff12}}
\and A.~Pezzotta\orcid{0000-0003-0726-2268}\inst{\ref{aff120},\ref{aff69}}
\and M.~P\"ontinen\orcid{0000-0001-5442-2530}\inst{\ref{aff19}}
\and C.~Porciani\orcid{0000-0002-7797-2508}\inst{\ref{aff9}}
\and I.~Risso\orcid{0000-0003-2525-7761}\inst{\ref{aff121}}
\and V.~Scottez\orcid{0009-0008-3864-940X}\inst{\ref{aff94},\ref{aff122}}
\and M.~Sereno\orcid{0000-0003-0302-0325}\inst{\ref{aff12},\ref{aff28}}
\and M.~Tenti\orcid{0000-0002-4254-5901}\inst{\ref{aff28}}
\and M.~Viel\orcid{0000-0002-2642-5707}\inst{\ref{aff11},\ref{aff10},\ref{aff37},\ref{aff14},\ref{aff15}}
\and M.~Wiesmann\orcid{0009-0000-8199-5860}\inst{\ref{aff75}}
\and Y.~Akrami\orcid{0000-0002-2407-7956}\inst{\ref{aff123},\ref{aff124}}
\and S.~Alvi\orcid{0000-0001-5779-8568}\inst{\ref{aff30}}
\and I.~T.~Andika\orcid{0000-0001-6102-9526}\inst{\ref{aff125},\ref{aff126}}
\and S.~Anselmi\orcid{0000-0002-3579-9583}\inst{\ref{aff67},\ref{aff103},\ref{aff127}}
\and M.~Archidiacono\orcid{0000-0003-4952-9012}\inst{\ref{aff87},\ref{aff88}}
\and F.~Atrio-Barandela\orcid{0000-0002-2130-2513}\inst{\ref{aff128}}
\and C.~Benoist\inst{\ref{aff92}}
\and P.~Bergamini\orcid{0000-0003-1383-9414}\inst{\ref{aff87},\ref{aff12}}
\and D.~Bertacca\orcid{0000-0002-2490-7139}\inst{\ref{aff103},\ref{aff29},\ref{aff67}}
\and M.~Bethermin\orcid{0000-0002-3915-2015}\inst{\ref{aff129}}
\and A.~Blanchard\orcid{0000-0001-8555-9003}\inst{\ref{aff104}}
\and L.~Blot\orcid{0000-0002-9622-7167}\inst{\ref{aff130},\ref{aff16}}
\and H.~B\"ohringer\orcid{0000-0001-8241-4204}\inst{\ref{aff69},\ref{aff131},\ref{aff132}}
\and M.~L.~Brown\orcid{0000-0002-0370-8077}\inst{\ref{aff58}}
\and S.~Bruton\orcid{0000-0002-6503-5218}\inst{\ref{aff133}}
\and A.~Calabro\orcid{0000-0003-2536-1614}\inst{\ref{aff52}}
\and B.~Camacho~Quevedo\orcid{0000-0002-8789-4232}\inst{\ref{aff11},\ref{aff37},\ref{aff10},\ref{aff74},\ref{aff73}}
\and F.~Caro\inst{\ref{aff52}}
\and C.~S.~Carvalho\inst{\ref{aff109}}
\and T.~Castro\orcid{0000-0002-6292-3228}\inst{\ref{aff10},\ref{aff14},\ref{aff11},\ref{aff15}}
\and F.~Cogato\orcid{0000-0003-4632-6113}\inst{\ref{aff24},\ref{aff12}}
\and S.~Conseil\orcid{0000-0002-3657-4191}\inst{\ref{aff60}}
\and A.~R.~Cooray\orcid{0000-0002-3892-0190}\inst{\ref{aff134}}
\and M.~Costanzi\orcid{0000-0001-8158-1449}\inst{\ref{aff13},\ref{aff10},\ref{aff11}}
\and O.~Cucciati\orcid{0000-0002-9336-7551}\inst{\ref{aff12}}
\and S.~Davini\orcid{0000-0003-3269-1718}\inst{\ref{aff40}}
\and G.~Desprez\orcid{0000-0001-8325-1742}\inst{\ref{aff114}}
\and A.~D\'iaz-S\'anchez\orcid{0000-0003-0748-4768}\inst{\ref{aff135}}
\and J.~J.~Diaz\orcid{0000-0003-2101-1078}\inst{\ref{aff56}}
\and S.~Di~Domizio\orcid{0000-0003-2863-5895}\inst{\ref{aff39},\ref{aff40}}
\and J.~M.~Diego\orcid{0000-0001-9065-3926}\inst{\ref{aff136}}
\and P.~Dimauro\orcid{0000-0001-7399-2854}\inst{\ref{aff137},\ref{aff52}}
\and A.~Enia\orcid{0000-0002-0200-2857}\inst{\ref{aff38},\ref{aff12}}
\and Y.~Fang\inst{\ref{aff70}}
\and A.~G.~Ferrari\orcid{0009-0005-5266-4110}\inst{\ref{aff28}}
\and A.~Fontana\orcid{0000-0003-3820-2823}\inst{\ref{aff52}}
\and A.~Franco\orcid{0000-0002-4761-366X}\inst{\ref{aff138},\ref{aff139},\ref{aff140}}
\and K.~Ganga\orcid{0000-0001-8159-8208}\inst{\ref{aff3}}
\and J.~Garc\'ia-Bellido\orcid{0000-0002-9370-8360}\inst{\ref{aff123}}
\and T.~Gasparetto\orcid{0000-0002-7913-4866}\inst{\ref{aff10}}
\and V.~Gautard\inst{\ref{aff141}}
\and R.~Gavazzi\orcid{0000-0002-5540-6935}\inst{\ref{aff90},\ref{aff17}}
\and E.~Gaztanaga\orcid{0000-0001-9632-0815}\inst{\ref{aff74},\ref{aff73},\ref{aff142}}
\and F.~Giacomini\orcid{0000-0002-3129-2814}\inst{\ref{aff28}}
\and F.~Gianotti\orcid{0000-0003-4666-119X}\inst{\ref{aff12}}
\and G.~Gozaliasl\orcid{0000-0002-0236-919X}\inst{\ref{aff143},\ref{aff19}}
\and M.~Guidi\orcid{0000-0001-9408-1101}\inst{\ref{aff38},\ref{aff12}}
\and C.~M.~Gutierrez\orcid{0000-0001-7854-783X}\inst{\ref{aff144}}
\and A.~Hall\orcid{0000-0002-3139-8651}\inst{\ref{aff57}}
\and S.~Hemmati\orcid{0000-0003-2226-5395}\inst{\ref{aff145}}
\and C.~Hern\'andez-Monteagudo\orcid{0000-0001-5471-9166}\inst{\ref{aff102},\ref{aff56}}
\and H.~Hildebrandt\orcid{0000-0002-9814-3338}\inst{\ref{aff146}}
\and J.~Hjorth\orcid{0000-0002-4571-2306}\inst{\ref{aff98}}
\and J.~J.~E.~Kajava\orcid{0000-0002-3010-8333}\inst{\ref{aff147},\ref{aff148}}
\and Y.~Kang\orcid{0009-0000-8588-7250}\inst{\ref{aff66}}
\and V.~Kansal\orcid{0000-0002-4008-6078}\inst{\ref{aff149},\ref{aff150}}
\and D.~Karagiannis\orcid{0000-0002-4927-0816}\inst{\ref{aff30},\ref{aff151}}
\and K.~Kiiveri\inst{\ref{aff82}}
\and C.~C.~Kirkpatrick\inst{\ref{aff82}}
\and S.~Kruk\orcid{0000-0001-8010-8879}\inst{\ref{aff34}}
\and J.~Le~Graet\orcid{0000-0001-6523-7971}\inst{\ref{aff68}}
\and L.~Legrand\orcid{0000-0003-0610-5252}\inst{\ref{aff152},\ref{aff153}}
\and M.~Lembo\orcid{0000-0002-5271-5070}\inst{\ref{aff17}}
\and F.~Lepori\orcid{0009-0000-5061-7138}\inst{\ref{aff154}}
\and G.~Leroy\orcid{0009-0004-2523-4425}\inst{\ref{aff155},\ref{aff91}}
\and J.~Lesgourgues\orcid{0000-0001-7627-353X}\inst{\ref{aff156}}
\and L.~Leuzzi\orcid{0009-0006-4479-7017}\inst{\ref{aff12}}
\and T.~I.~Liaudat\orcid{0000-0002-9104-314X}\inst{\ref{aff7}}
\and S.~J.~Liu\orcid{0000-0001-7680-2139}\inst{\ref{aff71}}
\and A.~Loureiro\orcid{0000-0002-4371-0876}\inst{\ref{aff157},\ref{aff33}}
\and J.~Macias-Perez\orcid{0000-0002-5385-2763}\inst{\ref{aff158}}
\and G.~Maggio\orcid{0000-0003-4020-4836}\inst{\ref{aff10}}
\and M.~Magliocchetti\orcid{0000-0001-9158-4838}\inst{\ref{aff71}}
\and G.~A.~Mamon\orcid{0000-0001-8956-5953}\inst{\ref{aff17},\ref{aff94}}
\and F.~Mannucci\orcid{0000-0002-4803-2381}\inst{\ref{aff159}}
\and R.~Maoli\orcid{0000-0002-6065-3025}\inst{\ref{aff160},\ref{aff52}}
\and C.~J.~A.~P.~Martins\orcid{0000-0002-4886-9261}\inst{\ref{aff161},\ref{aff162}}
\and L.~Maurin\orcid{0000-0002-8406-0857}\inst{\ref{aff32}}
\and M.~Miluzio\inst{\ref{aff34},\ref{aff163}}
\and P.~Monaco\orcid{0000-0003-2083-7564}\inst{\ref{aff13},\ref{aff10},\ref{aff14},\ref{aff11}}
\and A.~Montoro\orcid{0000-0003-4730-8590}\inst{\ref{aff74},\ref{aff73}}
\and C.~Moretti\orcid{0000-0003-3314-8936}\inst{\ref{aff37},\ref{aff15},\ref{aff10},\ref{aff11},\ref{aff14}}
\and G.~Morgante\inst{\ref{aff12}}
\and C.~Murray\inst{\ref{aff3}}
\and K.~Naidoo\orcid{0000-0002-9182-1802}\inst{\ref{aff142}}
\and A.~Navarro-Alsina\orcid{0000-0002-3173-2592}\inst{\ref{aff9}}
\and S.~Nesseris\orcid{0000-0002-0567-0324}\inst{\ref{aff123}}
\and F.~Passalacqua\orcid{0000-0002-8606-4093}\inst{\ref{aff103},\ref{aff67}}
\and K.~Paterson\orcid{0000-0001-8340-3486}\inst{\ref{aff80}}
\and A.~Pisani\orcid{0000-0002-6146-4437}\inst{\ref{aff68}}
\and D.~Potter\orcid{0000-0002-0757-5195}\inst{\ref{aff154}}
\and S.~Quai\orcid{0000-0002-0449-8163}\inst{\ref{aff24},\ref{aff12}}
\and P.-F.~Rocci\inst{\ref{aff32}}
\and G.~Rodighiero\orcid{0000-0002-9415-2296}\inst{\ref{aff103},\ref{aff29}}
\and S.~Sacquegna\orcid{0000-0002-8433-6630}\inst{\ref{aff139},\ref{aff138},\ref{aff140}}
\and M.~Sahl\'en\orcid{0000-0003-0973-4804}\inst{\ref{aff164}}
\and D.~B.~Sanders\orcid{0000-0002-1233-9998}\inst{\ref{aff54}}
\and E.~Sarpa\orcid{0000-0002-1256-655X}\inst{\ref{aff37},\ref{aff15},\ref{aff14}}
\and A.~Schneider\orcid{0000-0001-7055-8104}\inst{\ref{aff154}}
\and M.~Schultheis\inst{\ref{aff92}}
\and D.~Sciotti\orcid{0009-0008-4519-2620}\inst{\ref{aff52},\ref{aff89}}
\and E.~Sellentin\inst{\ref{aff165},\ref{aff48}}
\and L.~C.~Smith\orcid{0000-0002-3259-2771}\inst{\ref{aff166}}
\and K.~Tanidis\orcid{0000-0001-9843-5130}\inst{\ref{aff118}}
\and C.~Tao\orcid{0000-0001-7961-8177}\inst{\ref{aff68}}
\and G.~Testera\inst{\ref{aff40}}
\and R.~Teyssier\orcid{0000-0001-7689-0933}\inst{\ref{aff167}}
\and S.~Tosi\orcid{0000-0002-7275-9193}\inst{\ref{aff39},\ref{aff40},\ref{aff36}}
\and A.~Troja\orcid{0000-0003-0239-4595}\inst{\ref{aff103},\ref{aff67}}
\and M.~Tucci\inst{\ref{aff66}}
\and C.~Valieri\inst{\ref{aff28}}
\and A.~Venhola\orcid{0000-0001-6071-4564}\inst{\ref{aff168}}
\and D.~Vergani\orcid{0000-0003-0898-2216}\inst{\ref{aff12}}
\and G.~Verza\orcid{0000-0002-1886-8348}\inst{\ref{aff169}}
\and P.~Vielzeuf\orcid{0000-0003-2035-9339}\inst{\ref{aff68}}
\and N.~A.~Walton\orcid{0000-0003-3983-8778}\inst{\ref{aff166}}}
										   
%%%% please do not edit the affiliation list -- contact ECEB Bureau for changes
\institute{Universit\'e Paris-Saclay, CEA, D\'epartement de Physique des Particules, 91191, Gif-sur-Yvette, France\label{aff1}
\and
Department of Physics and Astronomy, University of California, Davis, CA 95616, USA\label{aff2}
\and
Universit\'e Paris Cit\'e, CNRS, Astroparticule et Cosmologie, 75013 Paris, France\label{aff3}
\and
Observatorio Astron\'omico Nacional, IGN, Calle Alfonso XII 3, E-28014 Madrid, Spain\label{aff4}
\and
CNRS-UCB International Research Laboratory, Centre Pierre Bin\'etruy, IRL2007, CPB-IN2P3, Berkeley, USA\label{aff5}
\and
Universit\'e Paris-Saclay, Universit\'e Paris Cit\'e, CEA, CNRS, AIM, 91191, Gif-sur-Yvette, France\label{aff6}
\and
IRFU, CEA, Universit\'e Paris-Saclay 91191 Gif-sur-Yvette Cedex, France\label{aff7}
\and
Universit\'e Paris Diderot, AIM, Sorbonne Paris Cit\'e, CEA, CNRS 91191 Gif-sur-Yvette Cedex, France\label{aff8}
\and
Universit\"at Bonn, Argelander-Institut f\"ur Astronomie, Auf dem H\"ugel 71, 53121 Bonn, Germany\label{aff9}
\and
INAF-Osservatorio Astronomico di Trieste, Via G. B. Tiepolo 11, 34143 Trieste, Italy\label{aff10}
\and
IFPU, Institute for Fundamental Physics of the Universe, via Beirut 2, 34151 Trieste, Italy\label{aff11}
\and
INAF-Osservatorio di Astrofisica e Scienza dello Spazio di Bologna, Via Piero Gobetti 93/3, 40129 Bologna, Italy\label{aff12}
\and
Dipartimento di Fisica - Sezione di Astronomia, Universit\`a di Trieste, Via Tiepolo 11, 34131 Trieste, Italy\label{aff13}
\and
INFN, Sezione di Trieste, Via Valerio 2, 34127 Trieste TS, Italy\label{aff14}
\and
ICSC - Centro Nazionale di Ricerca in High Performance Computing, Big Data e Quantum Computing, Via Magnanelli 2, Bologna, Italy\label{aff15}
\and
Laboratoire d'etude de l'Univers et des phenomenes eXtremes, Observatoire de Paris, Universit\'e PSL, Sorbonne Universit\'e, CNRS, 92190 Meudon, France\label{aff16}
\and
Institut d'Astrophysique de Paris, UMR 7095, CNRS, and Sorbonne Universit\'e, 98 bis boulevard Arago, 75014 Paris, France\label{aff17}
\and
INFN-Bologna, Via Irnerio 46, 40126 Bologna, Italy\label{aff18}
\and
Department of Physics, P.O. Box 64, University of Helsinki, 00014 Helsinki, Finland\label{aff19}
\and
Department of Physics \& Astronomy, University of Sussex, Brighton BN1 9QH, UK\label{aff20}
\and
School of Physics and Astronomy, University of Nottingham, University Park, Nottingham NG7 2RD, UK\label{aff21}
\and
Department of Astronomy, University of Florida, Bryant Space Science Center, Gainesville, FL 32611, USA\label{aff22}
\and
University Observatory, LMU Faculty of Physics, Scheinerstrasse 1, 81679 Munich, Germany\label{aff23}
\and
Dipartimento di Fisica e Astronomia "Augusto Righi" - Alma Mater Studiorum Universit\`a di Bologna, via Piero Gobetti 93/2, 40129 Bologna, Italy\label{aff24}
\and
Institut f\"ur Theoretische Physik, University of Heidelberg, Philosophenweg 16, 69120 Heidelberg, Germany\label{aff25}
\and
Zentrum f\"ur Astronomie, Universit\"at Heidelberg, Philosophenweg 12, 69120 Heidelberg, Germany\label{aff26}
\and
School of Physics, HH Wills Physics Laboratory, University of Bristol, Tyndall Avenue, Bristol, BS8 1TL, UK\label{aff27}
\and
INFN-Sezione di Bologna, Viale Berti Pichat 6/2, 40127 Bologna, Italy\label{aff28}
\and
INAF-Osservatorio Astronomico di Padova, Via dell'Osservatorio 5, 35122 Padova, Italy\label{aff29}
\and
Dipartimento di Fisica e Scienze della Terra, Universit\`a degli Studi di Ferrara, Via Giuseppe Saragat 1, 44122 Ferrara, Italy\label{aff30}
\and
Univ. Lille, CNRS, Centrale Lille, UMR 9189 CRIStAL, 59000 Lille, France\label{aff31}
\and
Universit\'e Paris-Saclay, CNRS, Institut d'astrophysique spatiale, 91405, Orsay, France\label{aff32}
\and
Astrophysics Group, Blackett Laboratory, Imperial College London, London SW7 2AZ, UK\label{aff33}
\and
ESAC/ESA, Camino Bajo del Castillo, s/n., Urb. Villafranca del Castillo, 28692 Villanueva de la Ca\~nada, Madrid, Spain\label{aff34}
\and
School of Mathematics and Physics, University of Surrey, Guildford, Surrey, GU2 7XH, UK\label{aff35}
\and
INAF-Osservatorio Astronomico di Brera, Via Brera 28, 20122 Milano, Italy\label{aff36}
\and
SISSA, International School for Advanced Studies, Via Bonomea 265, 34136 Trieste TS, Italy\label{aff37}
\and
Dipartimento di Fisica e Astronomia, Universit\`a di Bologna, Via Gobetti 93/2, 40129 Bologna, Italy\label{aff38}
\and
Dipartimento di Fisica, Universit\`a di Genova, Via Dodecaneso 33, 16146, Genova, Italy\label{aff39}
\and
INFN-Sezione di Genova, Via Dodecaneso 33, 16146, Genova, Italy\label{aff40}
\and
Department of Physics "E. Pancini", University Federico II, Via Cinthia 6, 80126, Napoli, Italy\label{aff41}
\and
INAF-Osservatorio Astronomico di Capodimonte, Via Moiariello 16, 80131 Napoli, Italy\label{aff42}
\and
Dipartimento di Fisica, Universit\`a degli Studi di Torino, Via P. Giuria 1, 10125 Torino, Italy\label{aff43}
\and
INFN-Sezione di Torino, Via P. Giuria 1, 10125 Torino, Italy\label{aff44}
\and
INAF-Osservatorio Astrofisico di Torino, Via Osservatorio 20, 10025 Pino Torinese (TO), Italy\label{aff45}
\and
European Space Agency/ESTEC, Keplerlaan 1, 2201 AZ Noordwijk, The Netherlands\label{aff46}
\and
Institute Lorentz, Leiden University, Niels Bohrweg 2, 2333 CA Leiden, The Netherlands\label{aff47}
\and
Leiden Observatory, Leiden University, Einsteinweg 55, 2333 CC Leiden, The Netherlands\label{aff48}
\and
INAF-IASF Milano, Via Alfonso Corti 12, 20133 Milano, Italy\label{aff49}
\and
Centro de Investigaciones Energ\'eticas, Medioambientales y Tecnol\'ogicas (CIEMAT), Avenida Complutense 40, 28040 Madrid, Spain\label{aff50}
\and
Port d'Informaci\'{o} Cient\'{i}fica, Campus UAB, C. Albareda s/n, 08193 Bellaterra (Barcelona), Spain\label{aff51}
\and
INAF-Osservatorio Astronomico di Roma, Via Frascati 33, 00078 Monteporzio Catone, Italy\label{aff52}
\and
INFN section of Naples, Via Cinthia 6, 80126, Napoli, Italy\label{aff53}
\and
Institute for Astronomy, University of Hawaii, 2680 Woodlawn Drive, Honolulu, HI 96822, USA\label{aff54}
\and
Dipartimento di Fisica e Astronomia "Augusto Righi" - Alma Mater Studiorum Universit\`a di Bologna, Viale Berti Pichat 6/2, 40127 Bologna, Italy\label{aff55}
\and
Instituto de Astrof\'{\i}sica de Canarias, V\'{\i}a L\'actea, 38205 La Laguna, Tenerife, Spain\label{aff56}
\and
Institute for Astronomy, University of Edinburgh, Royal Observatory, Blackford Hill, Edinburgh EH9 3HJ, UK\label{aff57}
\and
Jodrell Bank Centre for Astrophysics, Department of Physics and Astronomy, University of Manchester, Oxford Road, Manchester M13 9PL, UK\label{aff58}
\and
European Space Agency/ESRIN, Largo Galileo Galilei 1, 00044 Frascati, Roma, Italy\label{aff59}
\and
Universit\'e Claude Bernard Lyon 1, CNRS/IN2P3, IP2I Lyon, UMR 5822, Villeurbanne, F-69100, France\label{aff60}
\and
Institut de Ci\`{e}ncies del Cosmos (ICCUB), Universitat de Barcelona (IEEC-UB), Mart\'{i} i Franqu\`{e}s 1, 08028 Barcelona, Spain\label{aff61}
\and
Instituci\'o Catalana de Recerca i Estudis Avan\c{c}ats (ICREA), Passeig de Llu\'{\i}s Companys 23, 08010 Barcelona, Spain\label{aff62}
\and
UCB Lyon 1, CNRS/IN2P3, IUF, IP2I Lyon, 4 rue Enrico Fermi, 69622 Villeurbanne, France\label{aff63}
\and
Departamento de F\'isica, Faculdade de Ci\^encias, Universidade de Lisboa, Edif\'icio C8, Campo Grande, PT1749-016 Lisboa, Portugal\label{aff64}
\and
Instituto de Astrof\'isica e Ci\^encias do Espa\c{c}o, Faculdade de Ci\^encias, Universidade de Lisboa, Campo Grande, 1749-016 Lisboa, Portugal\label{aff65}
\and
Department of Astronomy, University of Geneva, ch. d'Ecogia 16, 1290 Versoix, Switzerland\label{aff66}
\and
INFN-Padova, Via Marzolo 8, 35131 Padova, Italy\label{aff67}
\and
Aix-Marseille Universit\'e, CNRS/IN2P3, CPPM, Marseille, France\label{aff68}
\and
Max Planck Institute for Extraterrestrial Physics, Giessenbachstr. 1, 85748 Garching, Germany\label{aff69}
\and
Universit\"ats-Sternwarte M\"unchen, Fakult\"at f\"ur Physik, Ludwig-Maximilians-Universit\"at M\"unchen, Scheinerstrasse 1, 81679 M\"unchen, Germany\label{aff70}
\and
INAF-Istituto di Astrofisica e Planetologia Spaziali, via del Fosso del Cavaliere, 100, 00100 Roma, Italy\label{aff71}
\and
Space Science Data Center, Italian Space Agency, via del Politecnico snc, 00133 Roma, Italy\label{aff72}
\and
Institut d'Estudis Espacials de Catalunya (IEEC),  Edifici RDIT, Campus UPC, 08860 Castelldefels, Barcelona, Spain\label{aff73}
\and
Institute of Space Sciences (ICE, CSIC), Campus UAB, Carrer de Can Magrans, s/n, 08193 Barcelona, Spain\label{aff74}
\and
Institute of Theoretical Astrophysics, University of Oslo, P.O. Box 1029 Blindern, 0315 Oslo, Norway\label{aff75}
\and
Jet Propulsion Laboratory, California Institute of Technology, 4800 Oak Grove Drive, Pasadena, CA, 91109, USA\label{aff76}
\and
Felix Hormuth Engineering, Goethestr. 17, 69181 Leimen, Germany\label{aff77}
\and
Technical University of Denmark, Elektrovej 327, 2800 Kgs. Lyngby, Denmark\label{aff78}
\and
Cosmic Dawn Center (DAWN), Denmark\label{aff79}
\and
Max-Planck-Institut f\"ur Astronomie, K\"onigstuhl 17, 69117 Heidelberg, Germany\label{aff80}
\and
NASA Goddard Space Flight Center, Greenbelt, MD 20771, USA\label{aff81}
\and
Department of Physics and Helsinki Institute of Physics, Gustaf H\"allstr\"omin katu 2, University of Helsinki, 00014 Helsinki, Finland\label{aff82}
\and
Universit\'e de Gen\`eve, D\'epartement de Physique Th\'eorique and Centre for Astroparticle Physics, 24 quai Ernest-Ansermet, CH-1211 Gen\`eve 4, Switzerland\label{aff83}
\and
Helsinki Institute of Physics, Gustaf H{\"a}llstr{\"o}min katu 2, University of Helsinki, 00014 Helsinki, Finland\label{aff84}
\and
SKA Observatory, Jodrell Bank, Lower Withington, Macclesfield, Cheshire SK11 9FT, UK\label{aff85}
\and
Centre de Calcul de l'IN2P3/CNRS, 21 avenue Pierre de Coubertin 69627 Villeurbanne Cedex, France\label{aff86}
\and
Dipartimento di Fisica "Aldo Pontremoli", Universit\`a degli Studi di Milano, Via Celoria 16, 20133 Milano, Italy\label{aff87}
\and
INFN-Sezione di Milano, Via Celoria 16, 20133 Milano, Italy\label{aff88}
\and
INFN-Sezione di Roma, Piazzale Aldo Moro, 2 - c/o Dipartimento di Fisica, Edificio G. Marconi, 00185 Roma, Italy\label{aff89}
\and
Aix-Marseille Universit\'e, CNRS, CNES, LAM, Marseille, France\label{aff90}
\and
Department of Physics, Institute for Computational Cosmology, Durham University, South Road, Durham, DH1 3LE, UK\label{aff91}
\and
Universit\'e C\^{o}te d'Azur, Observatoire de la C\^{o}te d'Azur, CNRS, Laboratoire Lagrange, Bd de l'Observatoire, CS 34229, 06304 Nice cedex 4, France\label{aff92}
\and
University of Applied Sciences and Arts of Northwestern Switzerland, School of Engineering, 5210 Windisch, Switzerland\label{aff93}
\and
Institut d'Astrophysique de Paris, 98bis Boulevard Arago, 75014, Paris, France\label{aff94}
\and
Institute of Physics, Laboratory of Astrophysics, Ecole Polytechnique F\'ed\'erale de Lausanne (EPFL), Observatoire de Sauverny, 1290 Versoix, Switzerland\label{aff95}
\and
Telespazio UK S.L. for European Space Agency (ESA), Camino bajo del Castillo, s/n, Urbanizacion Villafranca del Castillo, Villanueva de la Ca\~nada, 28692 Madrid, Spain\label{aff96}
\and
Institut de F\'{i}sica d'Altes Energies (IFAE), The Barcelona Institute of Science and Technology, Campus UAB, 08193 Bellaterra (Barcelona), Spain\label{aff97}
\and
DARK, Niels Bohr Institute, University of Copenhagen, Jagtvej 155, 2200 Copenhagen, Denmark\label{aff98}
\and
Centre National d'Etudes Spatiales -- Centre spatial de Toulouse, 18 avenue Edouard Belin, 31401 Toulouse Cedex 9, France\label{aff99}
\and
Institute of Space Science, Str. Atomistilor, nr. 409 M\u{a}gurele, Ilfov, 077125, Romania\label{aff100}
\and
Consejo Superior de Investigaciones Cientificas, Calle Serrano 117, 28006 Madrid, Spain\label{aff101}
\and
Universidad de La Laguna, Departamento de Astrof\'{\i}sica, 38206 La Laguna, Tenerife, Spain\label{aff102}
\and
Dipartimento di Fisica e Astronomia "G. Galilei", Universit\`a di Padova, Via Marzolo 8, 35131 Padova, Italy\label{aff103}
\and
Institut de Recherche en Astrophysique et Plan\'etologie (IRAP), Universit\'e de Toulouse, CNRS, UPS, CNES, 14 Av. Edouard Belin, 31400 Toulouse, France\label{aff104}
\and
Universit\'e St Joseph; Faculty of Sciences, Beirut, Lebanon\label{aff105}
\and
Departamento de F\'isica, FCFM, Universidad de Chile, Blanco Encalada 2008, Santiago, Chile\label{aff106}
\and
Universit\"at Innsbruck, Institut f\"ur Astro- und Teilchenphysik, Technikerstr. 25/8, 6020 Innsbruck, Austria\label{aff107}
\and
Satlantis, University Science Park, Sede Bld 48940, Leioa-Bilbao, Spain\label{aff108}
\and
Instituto de Astrof\'isica e Ci\^encias do Espa\c{c}o, Faculdade de Ci\^encias, Universidade de Lisboa, Tapada da Ajuda, 1349-018 Lisboa, Portugal\label{aff109}
\and
Department of Physics and Astronomy, University College London, Gower Street, London WC1E 6BT, UK\label{aff110}
\and
Cosmic Dawn Center (DAWN)\label{aff111}
\and
Niels Bohr Institute, University of Copenhagen, Jagtvej 128, 2200 Copenhagen, Denmark\label{aff112}
\and
Universidad Polit\'ecnica de Cartagena, Departamento de Electr\'onica y Tecnolog\'ia de Computadoras,  Plaza del Hospital 1, 30202 Cartagena, Spain\label{aff113}
\and
Kapteyn Astronomical Institute, University of Groningen, PO Box 800, 9700 AV Groningen, The Netherlands\label{aff114}
\and
Infrared Processing and Analysis Center, California Institute of Technology, Pasadena, CA 91125, USA\label{aff115}
\and
Istituto Nazionale di Fisica Nucleare, Sezione di Ferrara, Via Giuseppe Saragat 1, 44122 Ferrara, Italy\label{aff116}
\and
INAF, Istituto di Radioastronomia, Via Piero Gobetti 101, 40129 Bologna, Italy\label{aff117}
\and
Department of Physics, Oxford University, Keble Road, Oxford OX1 3RH, UK\label{aff118}
\and
Aurora Technology for European Space Agency (ESA), Camino bajo del Castillo, s/n, Urbanizacion Villafranca del Castillo, Villanueva de la Ca\~nada, 28692 Madrid, Spain\label{aff119}
\and
INAF - Osservatorio Astronomico di Brera, via Emilio Bianchi 46, 23807 Merate, Italy\label{aff120}
\and
INAF-Osservatorio Astronomico di Brera, Via Brera 28, 20122 Milano, Italy, and INFN-Sezione di Genova, Via Dodecaneso 33, 16146, Genova, Italy\label{aff121}
\and
ICL, Junia, Universit\'e Catholique de Lille, LITL, 59000 Lille, France\label{aff122}
\and
Instituto de F\'isica Te\'orica UAM-CSIC, Campus de Cantoblanco, 28049 Madrid, Spain\label{aff123}
\and
CERCA/ISO, Department of Physics, Case Western Reserve University, 10900 Euclid Avenue, Cleveland, OH 44106, USA\label{aff124}
\and
Technical University of Munich, TUM School of Natural Sciences, Physics Department, James-Franck-Str.~1, 85748 Garching, Germany\label{aff125}
\and
Max-Planck-Institut f\"ur Astrophysik, Karl-Schwarzschild-Str.~1, 85748 Garching, Germany\label{aff126}
\and
Laboratoire Univers et Th\'eorie, Observatoire de Paris, Universit\'e PSL, Universit\'e Paris Cit\'e, CNRS, 92190 Meudon, France\label{aff127}
\and
Departamento de F{\'\i}sica Fundamental. Universidad de Salamanca. Plaza de la Merced s/n. 37008 Salamanca, Spain\label{aff128}
\and
Universit\'e de Strasbourg, CNRS, Observatoire astronomique de Strasbourg, UMR 7550, 67000 Strasbourg, France\label{aff129}
\and
Center for Data-Driven Discovery, Kavli IPMU (WPI), UTIAS, The University of Tokyo, Kashiwa, Chiba 277-8583, Japan\label{aff130}
\and
Ludwig-Maximilians-University, Schellingstrasse 4, 80799 Munich, Germany\label{aff131}
\and
Max-Planck-Institut f\"ur Physik, Boltzmannstr. 8, 85748 Garching, Germany\label{aff132}
\and
California Institute of Technology, 1200 E California Blvd, Pasadena, CA 91125, USA\label{aff133}
\and
Department of Physics \& Astronomy, University of California Irvine, Irvine CA 92697, USA\label{aff134}
\and
Departamento F\'isica Aplicada, Universidad Polit\'ecnica de Cartagena, Campus Muralla del Mar, 30202 Cartagena, Murcia, Spain\label{aff135}
\and
Instituto de F\'isica de Cantabria, Edificio Juan Jord\'a, Avenida de los Castros, 39005 Santander, Spain\label{aff136}
\and
Observatorio Nacional, Rua General Jose Cristino, 77-Bairro Imperial de Sao Cristovao, Rio de Janeiro, 20921-400, Brazil\label{aff137}
\and
INFN, Sezione di Lecce, Via per Arnesano, CP-193, 73100, Lecce, Italy\label{aff138}
\and
Department of Mathematics and Physics E. De Giorgi, University of Salento, Via per Arnesano, CP-I93, 73100, Lecce, Italy\label{aff139}
\and
INAF-Sezione di Lecce, c/o Dipartimento Matematica e Fisica, Via per Arnesano, 73100, Lecce, Italy\label{aff140}
\and
CEA Saclay, DFR/IRFU, Service d'Astrophysique, Bat. 709, 91191 Gif-sur-Yvette, France\label{aff141}
\and
Institute of Cosmology and Gravitation, University of Portsmouth, Portsmouth PO1 3FX, UK\label{aff142}
\and
Department of Computer Science, Aalto University, PO Box 15400, Espoo, FI-00 076, Finland\label{aff143}
\and
Instituto de Astrof\'\i sica de Canarias, c/ Via Lactea s/n, La Laguna 38200, Spain. Departamento de Astrof\'\i sica de la Universidad de La Laguna, Avda. Francisco Sanchez, La Laguna, 38200, Spain\label{aff144}
\and
Caltech/IPAC, 1200 E. California Blvd., Pasadena, CA 91125, USA\label{aff145}
\and
Ruhr University Bochum, Faculty of Physics and Astronomy, Astronomical Institute (AIRUB), German Centre for Cosmological Lensing (GCCL), 44780 Bochum, Germany\label{aff146}
\and
Department of Physics and Astronomy, Vesilinnantie 5, University of Turku, 20014 Turku, Finland\label{aff147}
\and
Serco for European Space Agency (ESA), Camino bajo del Castillo, s/n, Urbanizacion Villafranca del Castillo, Villanueva de la Ca\~nada, 28692 Madrid, Spain\label{aff148}
\and
ARC Centre of Excellence for Dark Matter Particle Physics, Melbourne, Australia\label{aff149}
\and
Centre for Astrophysics \& Supercomputing, Swinburne University of Technology,  Hawthorn, Victoria 3122, Australia\label{aff150}
\and
Department of Physics and Astronomy, University of the Western Cape, Bellville, Cape Town, 7535, South Africa\label{aff151}
\and
DAMTP, Centre for Mathematical Sciences, Wilberforce Road, Cambridge CB3 0WA, UK\label{aff152}
\and
Kavli Institute for Cosmology Cambridge, Madingley Road, Cambridge, CB3 0HA, UK\label{aff153}
\and
Department of Astrophysics, University of Zurich, Winterthurerstrasse 190, 8057 Zurich, Switzerland\label{aff154}
\and
Department of Physics, Centre for Extragalactic Astronomy, Durham University, South Road, Durham, DH1 3LE, UK\label{aff155}
\and
Institute for Theoretical Particle Physics and Cosmology (TTK), RWTH Aachen University, 52056 Aachen, Germany\label{aff156}
\and
Oskar Klein Centre for Cosmoparticle Physics, Department of Physics, Stockholm University, Stockholm, SE-106 91, Sweden\label{aff157}
\and
Univ. Grenoble Alpes, CNRS, Grenoble INP, LPSC-IN2P3, 53, Avenue des Martyrs, 38000, Grenoble, France\label{aff158}
\and
INAF-Osservatorio Astrofisico di Arcetri, Largo E. Fermi 5, 50125, Firenze, Italy\label{aff159}
\and
Dipartimento di Fisica, Sapienza Universit\`a di Roma, Piazzale Aldo Moro 2, 00185 Roma, Italy\label{aff160}
\and
Centro de Astrof\'{\i}sica da Universidade do Porto, Rua das Estrelas, 4150-762 Porto, Portugal\label{aff161}
\and
Instituto de Astrof\'isica e Ci\^encias do Espa\c{c}o, Universidade do Porto, CAUP, Rua das Estrelas, PT4150-762 Porto, Portugal\label{aff162}
\and
HE Space for European Space Agency (ESA), Camino bajo del Castillo, s/n, Urbanizacion Villafranca del Castillo, Villanueva de la Ca\~nada, 28692 Madrid, Spain\label{aff163}
\and
Theoretical astrophysics, Department of Physics and Astronomy, Uppsala University, Box 516, 751 37 Uppsala, Sweden\label{aff164}
\and
Mathematical Institute, University of Leiden, Einsteinweg 55, 2333 CA Leiden, The Netherlands\label{aff165}
\and
Institute of Astronomy, University of Cambridge, Madingley Road, Cambridge CB3 0HA, UK\label{aff166}
\and
Department of Astrophysical Sciences, Peyton Hall, Princeton University, Princeton, NJ 08544, USA\label{aff167}
\and
Space physics and astronomy research unit, University of Oulu, Pentti Kaiteran katu 1, FI-90014 Oulu, Finland\label{aff168}
\and
Center for Computational Astrophysics, Flatiron Institute, 162 5th Avenue, 10010, New York, NY, USA\label{aff169}}    

% \date{...}

% 
% Put your abstract here
%
   \abstract{We present our methodology for identifying known clusters as counterparts to objects in the Euclid Catalogue of Galaxy Clusters (ECGC). \Euclid is expected to detect a large number of optically selected galaxy clusters over the approximately \num{14000} square degrees of its extragalactic sky survey. Extending out well beyond redshift unity, the catalogue will contain many new high-redshift clusters, while at lower redshifts a fraction of the clusters will have been observed in other surveys. Identifying these known clusters as counterparts to the \Euclid-detected clusters is an important step in the validation and construction of the ECGC to augment information with external observables. We present a set of catalogues and meta-catalogues of known clusters that we have assembled for this step, and we illustrate their application and our methodology using the Dark Energy Survey Year 1 \texttt{RedMaPPer} cluster catalogue in lieu of the future ECGC. In the process of this work, we have constructed and delivered an updated \texttt{EC-RedMaPPer} catalogue with multi-wavelength counterparts.
}
%
% Provide up to five key words:
%
%
% Add short versions of title and author list for page headings
%
   \titlerunning{Validation method for the detections in the ECGC using external data}
   \authorrunning{Euclid Collaboration: J.-B. Melin et al.}
   
   \maketitle
   
   \nolinenumbers
   
%
%-------------------------------------------------------------------
%
%
%   Start the main text of your paper here
%

\section{\label{sec:Intro}Introduction}

The \gls{esa} \Euclid mission, launched in July 2023, began nominal survey operations in February 2024~\citep{EuclidSkyOverview}. Over the course of six years, \Euclid will survey approximately \num{14000}\,deg$^2$ of the extragalactic sky in four photometric bands and with a grism spectrograph from its station at the Earth-Sun Lagrange point two (L2). Cosmological analyses will employ multiple observational probes: galaxy clustering, weak gravitational lensing, photometric \mbox{3$\times$2\,pt} analysis, higher-order statistics, clusters of galaxies, strong gravitational lensing, cross-correlation with \gls{cmb} observables, high-redshift quasars, and cosmic chronometers~\citep{Blanchard-EP7,Scaramella-EP1,EuclidSkyVIS,EuclidSkyNISP}.

One of the signature products of the survey will be the \gls{ecgc}, one of the largest catalogues of optically selected galaxy clusters ever constructed. Clusters serve as valuable laboratories for astrophysical studies, and catalogues of clusters have proven to be powerful cosmological probes \citep[e.g.][]{Vikhlinin2009, Allen2011, Kravtsov2012, Mantz2015, PlanckSZCosmo2016, Schellenberger2017, Bocquet2019, DESCosmo2020, Lesci2022, Ghirardini2024, Bocquet2024}. These studies have used cluster catalogues assembled at millimetre wavelengths, in X-rays, and in the optical \citep[e.g.][]{Bleem2015, PlanckSZCat2016, DESCosmo2020, Hilton2021, Bulbul2024}. 

\Euclid detects clusters in the photometric survey using two detection algorithms ~\citep{Adam-EP3}: \AMICO~\citep{Bellagamba2018,Maturi2019} and \PZWav~\citep{Gonzalez2014}, which have been adapted by the \gls{ec} and implemented in the Science Ground Segment pipeline. Well-characterised sub-catalogues of this large data set will be used for targeted purposes, such as the \Euclid cluster cosmology analysis. 

A critical element in the construction of the \gls{ecgc} is the identification of counterparts of \Euclid detections: the association with and comparison to known clusters from other optical surveys and from other wavebands (e.g. millimetre and X-rays). This step is needed for several reasons:
\begin{itemize}
\item[--] to confirm that detections by \Euclid algorithms are bona fide clusters;
\item[--] to check for newly discovered clusters; 
\item[--] to prepare analyses of scaling relations between \Euclid and external cluster catalogues (i.e. clusters detected in other surveys);
\item[--] to enable multi-wavelength studies; and
\item[--] to characterise the \Euclid cluster selection function.
\end{itemize}
We refer to this association and comparison step as the `external validation' of the detections in the \Euclid cluster catalogue. This type of external validation has successfully been undertaken in the past for other surveys~\citep[e.g.][]{PlanckSZCat2014,Sadibekova2014,PlanckSZCat2016}. We note that the external catalogues may have a fraction (typically under 15\%) of detections that are spurious. However, these are unlikely to spatially match \Euclid clusters if they were obtained at different wavelengths. We stress that full characterisation of the completeness and purity of the \gls{ecgc} is beyond the scope of this article. It will be performed in a separate work based on simulations.

A simple cross-correlation within a fixed angular aperture is not sufficient for reliable matching between \Euclid detections and known clusters. We expect some \Euclid clusters to be affected by projections along the line of sight or to belong to merging systems which leads to miscentring and fragmentation problems. A massive cluster detected in X-ray or at millimetre wavelengths might therefore not necessarily be associated with a single \Euclid detection. Apart from the main cluster component, \Euclid may separately identify merging sub-structures, in addition to foreground and background systems. To address this complexity, we developed a method for validating the \gls{ecgc} detections with external catalogues based on physical cross-correlations: matching based on physical properties of the associated clusters (redshifts and physical sizes). This matching is complemented by visual inspection in cases with ambiguous outcomes (e.g. multiple merging systems or masking issues). 

To prepare the external validation, we compiled a set of catalogues and meta-catalogues of external clusters from various wavebands (millimetre, optical, and X-ray) and grouped them into master tables. We herein present these compilations and our validation procedure. We define meta-catalogue and master table in \cref{sec:cat-metacat-mastertab}, where we detail the catalogues, meta-catalogues, and master tables used in this work. In \cref{sec:metho}, we describe our procedure to validate the \gls{ecgc}, and we illustrate it using the \gls{des} Y1 \gls{rm} catalogue~\citep{Rykoff2016,DESCosmo2020} in lieu of the future \gls{ecgc}. We discuss our results in \cref{sec:discuss}, and then summarise and conclude in \cref{sec:summ}.

For cosmological quantities calculated in this paper, we used a flat $\Lambda$CDM model with $H_0=70\,\si{\kmsMpc}$ and $\Omega_{\rm m}=0.3$. We did not adjust the cosmologies used in the various catalogues and meta-catalogues to the same cosmological model.
Inspection of the source papers tells us the impact of different cosmologies should be minor on our external validation work because the different cosmological parameters are close to the values adopted in this work.

\section{\label{sec:cat-metacat-mastertab}Catalogues, meta-catalogues, and master tables}

We define a `meta-catalogue' as a combination of different source catalogues with cross-identification of clusters and homogenisation of key quantities, notably redshift and a mass proxy. A meta-catalogue thus does not contain duplicate entries. The homogenised quantities facilitate matching with other catalogues through the use of scaling relations between cluster observables. The mass proxy can be, for example, X-ray luminosity (\cref{sec:xsz}), weak lensing mass (\cref{sec:optical}), or velocity dispersion (\cref{sec:mccd}).

A `master table' groups catalogues and meta-catalogues at a higher level, with a single entry per cluster that links to the  associated entries in the grouped catalogues and meta-catalogues. For example, the \mtwoc Galaxy Cluster database\footnote{\url{https://www.galaxyclusterdb.eu/m2c/}} contains a master table that unifies the ComPRASS catalogue~\citep{Tarrio2019}, the Meta-Catalogue of X-ray Clusters~II~\citep{Sadibekova2024}, and the Meta-Catalogue of SZ Clusters~(Tarr{\'i}o et al. in prep.).

We constructed a number of catalogues and meta-catalogues for our validation work which we describe in this section. The catalogues and meta-catalogues in X-rays and at millimetre wavelengths are presented in \cref{sec:xsz}, and those in the optical domain are described in \cref{sec:optical}. For the optical work, we constructed a global Abell catalogue with updated redshifts (\cref{sec:abell}), and we created the \gls{mccd}, a new meta-catalogue based on cluster velocity dispersions (\cref{sec:mccd}). We integrated the global Abell catalogue, the \gls{mccd} and the LC$^2$~\citep{Sereno2015} into an optical master table, presented in  
\cref{sec:opticalmaster}.

In addition to the catalogues and meta-catalogues described in \cref{sec:xsz,sec:optical}, we plan to validate the \Euclid catalogue with complementary catalogues in the X-ray and optical domains; we list these in \cref{sec:complementary}. These catalogues are not included in our matching procedure applied here to the \gls{des} Y1 \gls{rm} catalogue in \cref{sec:metho}, but we will use them in future work on the validation of the \gls{ecgc}.

\subsection{\label{sec:xsz}Sunyaev--Zeldovich and X-ray catalogues and meta-catalogues}

The catalogues and meta-catalogues based on the \gls{icm} are very useful for catalogue validation because they locate the gravitational potential wells of cluster halos. The \gls{icm} gas is detected in X-rays via thermal bremsstrahlung, the photon emission produced by the deceleration of hot electrons deflected by nuclei, and at millimetre wavelengths via the thermal \gls{sz} effect, the inverse Compton scattering of \gls{cmb} photons off of the same electrons.

In this work, we use the following X-ray and \gls{sz} catalogues and meta-catalogues, and master table.
\begin{itemize}
\item[--] The \gls{mcxc} was recently published by~\cite{Sadibekova2024}. This X-ray meta-catalogue is a compilation of clusters from publicly available serendipitous catalogues and catalogues based on the \gls{rosat}
All-Sky Survey.
\item[--] The \gls{mcsz} is based on all publicly available blind \gls{sz} catalogues~(Tarr{\'i}o et al. in prep.) from \Planck, the \gls{spt}, and the \gls{act}.
\item[--] ComPRASS~\citep{Tarrio2019} is a catalogue obtained from the joint extraction of \Planck and \gls{rass} data.
\item[--] The \gls{mcxc}, \gls{mcsz}, and the ComPRASS catalogues are included in the \mtwoc Galaxy Cluster database, where they are compiled into a master table with a single entry for each cluster.
\item[--] The \eROSITA cluster catalogue was released in early 2024 in the form of the general catalogue~\citep{Bulbul2024}, its optical properties~\citep{Kluge2024}, and a cosmological sample~\citep{Ghirardini2024}. We compiled the information from these three publications into a single file for our validation work.
\end{itemize}

These X-ray and \gls{sz} catalogues and meta-catalogues sample well-known clusters at masses above a few $10^{13} M_\odot$ from $z=0$ to $z=0.5$, and above $10^{14} M_\odot$ from $z=0.5$ to $z\sim2$. They will be important for evaluating the efficiency of the \Euclid detection algorithms at high mass. They will also guide understanding of the relationship between \Euclid detections (galaxy over-densities) and gravitational potential wells of massive clusters, as well as the study of cluster substructures detected by \Euclid. It should be noted, however, that these catalogues and meta-catalogues are not complete above $10^{14} M_\odot$ at $z>1$, so some rich \Euclid\/ cluster detections in the range $1<z<2$ may not have X-ray or SZ counterparts in the current datasets. 

\subsection{\label{sec:optical}Optical catalogues and meta-catalogues}

We considered the following catalogues, meta-catalogues, and master table constructed from optical data:
\begin{itemize}
\item[--] The \gls{lc2} is a meta-catalogue of clusters with weak lensing data~\citep{Sereno2015}. We used the latest version (v3.9)\footnote{\url{http://pico.oabo.inaf.it/~sereno/CoMaLit/LC2/}}, which contains 806 single entries.
\item[--] We formatted the Abell cluster catalogue into a single file for our work (\cref{sec:abell}). The catalogue contains \num{5250} entries.
\item[--] We built the \gls{mccd}, a meta-catalogue of clusters with velocity dispersion measurements (\cref{sec:mccd}). It contains \num{1083} clusters.
\item[--] We assembled a master table for the \gls{lc2} and \gls{mccd} meta-catalogues and the Abell catalogue (\cref{sec:opticalmaster}). It contains \num{6367} single entries.
\end{itemize}

The distribution on the sky of the three optical catalogues and meta-catalogues display different levels of homogeneity, with \gls{lc2} being the most inhomogeneous, as can be seen in the left column of \cref{fig:sky_distribution_and_redshift_optical}.

\subsubsection{\label{sec:abell}A global Abell cluster catalogue}

The first Abell catalogue was published by~\cite{Abell1958}. We use the ACO catalogue~\citep{Abell1989} published in four different tables (Tables 3, 4, 5, and 6 of the article), each covering a part of the sky. Table 3 and 4 cover the northern and southern hemisphere respectively, while Table 5 contains supplementary southern clusters and Table 6 the overlap clusters. We combined the four tables into one, keeping a unique entry per cluster with its original coordinates (hence the name `global Abell cluster catalogue'). In the overlap (Table 6), we used the values of the northern catalogue (Table 3) for the count and redshift. 

We then updated the redshifts of the global catalogue when available from NED\footnote{The NASA/IPAC Extragalactic Database (NED)
is operated by the Jet Propulsion Laboratory, California Institute of Technology,
under contract with the National Aeronautics and Space Administration. \url{http://ned.ipac.caltech.edu/}} and/or SIMBAD.\footnote{\url{http://simbad.cds.unistra.fr/simbad/}} When the redshift was available in both NED and SIMBAD, and the two redshifts were consistent ($\Delta z<0.001$), we took the NED redshift. If the two redshifts were not consistent, we adopted the spectroscopic redshift, unless it was based on a single spectrum or it came from a follow-up of another cluster. In the latter two situations, we gave priority to NED over SIMBAD. 

This resulted in a global Abell catalogue with \num{5250} entries, of which \num{3215} have a redshift. Future work will aim to complete missing redshifts with data from the \gls{sdss}\footnote{\url{https://www.sdss.org/}} and the \gls{desi}.\footnote{\url{https://www.desi.lbl.gov/}}

\subsubsection{\label{sec:mccd}The Meta-Catalogue of Cluster Dispersions (MCCD)}

To aid validation of the \gls{ecgc} and to enhance its scientific productivity, we constructed a meta-catalogue of velocity dispersions of known galaxy clusters (the \gls{mccd}). We adopted the following criteria to balance the competing goals of homogenising velocity dispersions (serving as mass proxy) from a wide variety of sources and the desire to create a sufficiently large set of dispersions that could leverage the large-area covered by the \Euclid survey.
\begin{itemize}
\item[--] Adequate rejection of interlopers when defining the cluster members that are used in the dispersion calculation such as through the Shifting Gapper method
\citep{Fadda1996} or by the Clean method \citep{Mamon2013}. 
\item[--] Clusters must have at least 15 galaxy members with spectroscopic redshifts.
\item[--] Dispersions are calculated following the methodology of \cite{Beers1990} over an aperture of at least $R= 0.5$~Mpc.
\item[--] In addition to the originally quoted values, uncertainties in the dispersions are also calculated using equation~(B.1) in~\cite{SerenoEttori2015}.
\item[--] The relative error in the calculated velocity dispersion must be $\leq 25\%$.
\item[--] When clusters have more than one published dispersion, the value based on the largest number of cluster members with spectroscopic redshifts was adopted. 
\item[--] Clusters were matched to the NED database to obtain alternate cluster IDs when available.
\end{itemize}

Red and blue galaxy populations within clusters usually have different spatial and kinematic distributions \citep{Barsanti2016, Cava2017}, and so these populations contribute to the measured velocity dispersion of a cluster in different ways. While it would be preferable to control for the galaxy populations used in the dispersion calculation when creating the meta-catalogue, doing so would require first compiling all the redshifts (and photometric measurements) of the cluster galaxies used in the various published surveys. Because this information is not routinely available, we considered this procedure to be beyond the scope of the current project. 

The aperture over which the velocity dispersion is calculated is important, but as long as the aperture is larger than the core cluster region, its size is not critical to obtaining a homogeneous set of measurements. Beyond a cluster-centric radius of $\sim 0.5$\,Mpc, the integral dispersion profile remains approximately flat out to at least $R_{200}$\footnote{This is the radius at which the mean interior overdensity equals 200 times the critical density at the redshift of the cluster.}\citep{vandermarel2000}. Detailed studies such as \citet{Girardi1993} and \citet{Girardi1998} show that the dispersion profiles decrease only by about 20\% at such large clustercentre radii.  In fact, most of the relevant publications (see below) contain dispersions which were calculated for aperture sizes of about $R_{200}$.

Nevertheless, we developed a procedure to correct all dispersions to an equivalent aperture of $R_{200}$.  For those clusters whose published dispersions 
were not representative of $R_{200}$, we:  1) used the published dispersion, $\sigmav$, to infer $R_{200}$ from a scaling relation \citep{Biviano2017}; 2) used a mean integral velocity dispersion profile to correct 
the measured $\sigma_v$ to $R_{200}$; and then 3) iterated until convergence. As a model for the line-of-sight velocity dispersion profile we used a 4th-order polynomial fit for $\logten[\sigma_v(<R)/\sigma_v(<R_{200})]$ vs $\logten(R/R_{200})$ based on the WINGS data-set \citep{Cava2017}. Note that clusters whose published dispersions were originally given at $R_{200}$ have their values in the radial aperture column 
of the meta-catalogue set to `200' to denote this -- these values do not mean the cluster velocity dispersions used an aperture radius of $R_{\rm ap} = 200\,{\rm Mpc}$.

To construct the meta-catalogue, we used the following procedure:
\begin{enumerate}
\item begin with an updated version (hereinafter referred to as SCv2) of the compilation published in \citet{SerenoEttori2015}, which has \num{3476} galaxy clusters;
\item include the \Planck cluster results published in \citet{Aguado-Barahona2022} for clusters not already in SCv2, and update those clusters in SCv2 where the Planck results have greater number of cluster members and/or lower uncertainties on the dispersions; 
\item update the entries of clusters already in SCv2 that are also in the list maintained by A.\ Biviano (private communication) of galaxy clusters with dispersions determined from at least 200 spectroscopic members; 
\item include \gls{spt} clusters with dispersions as published in \citet{Ruel2014} and/or \citet{Bayliss2016};
\item include clusters from the CODEX/SPIDERS catalogue, as reported in \citet{Damsted2023};
\item downselect from the meta-catalogue after all the previous updates, following the criteria described above;
\item compare remaining entries by RA and Dec coordinates to consolidate duplicate entries, preferentially keeping dispersions based on the greatest number of spectroscopic members while retaining alternate cluster identification information where available. 
\end{enumerate}

\begin{figure}
    \centering
    \includegraphics[width=1.0\linewidth]{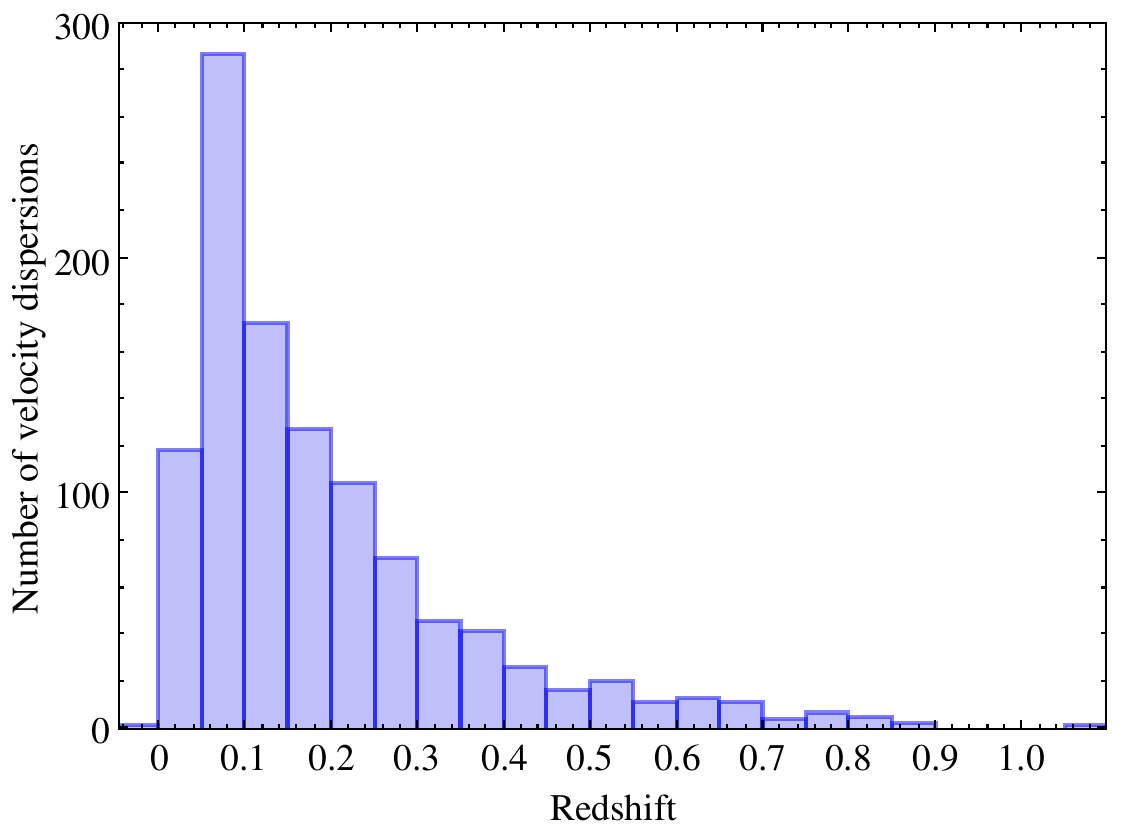}
    \caption{Redshift distribution of 1082 galaxy clusters in the \gls{mccd}.}
    \label{fig:disp-z-hist}
\end{figure}

\begin{figure}
    \centering
    \includegraphics[width=1\linewidth]{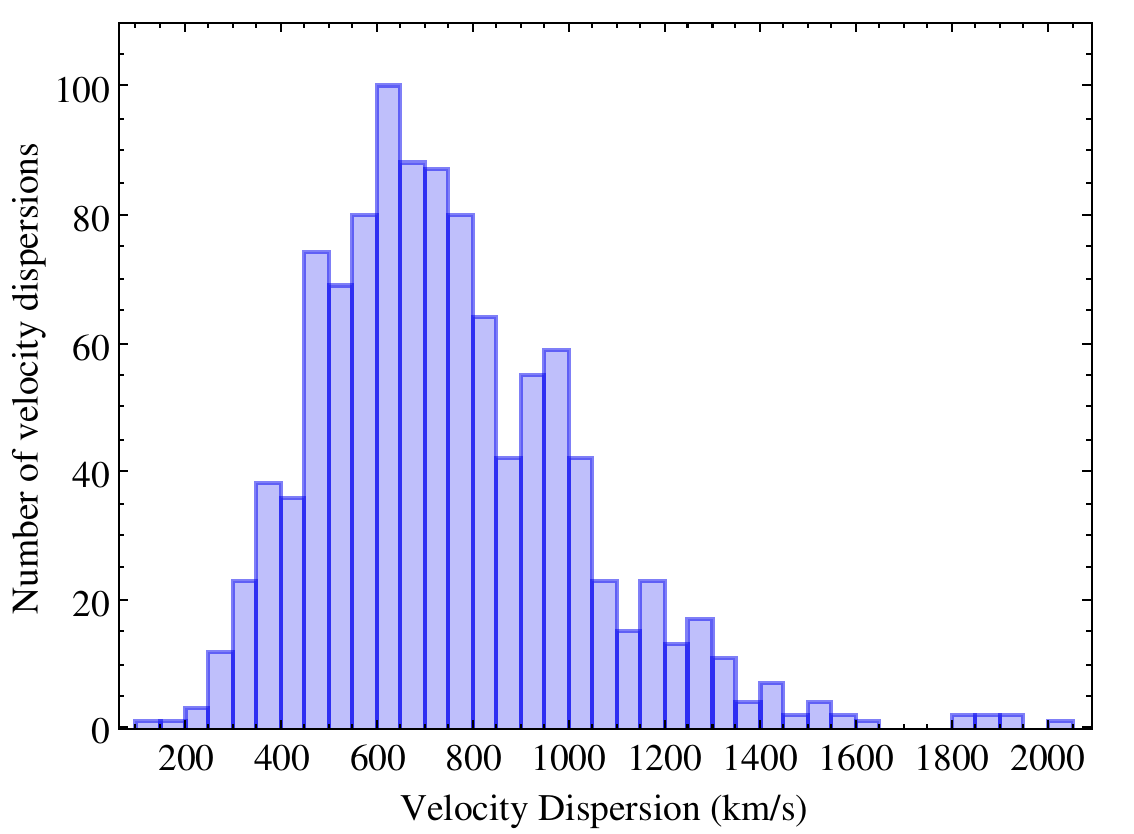}
    \caption{Distribution of velocity dispersions for the 1082 clusters in the \gls{mccd}.}
    \label{fig:disp-disp-hist}
\end{figure}

Following these steps, we compared the meta-catalogue with the Abell and \gls{lc2} cluster catalogues to homogenise entries, coordinates, and redshifts where identical clusters were found. \Cref{fig:disp-z-hist,fig:disp-disp-hist} show the redshift and velocity dispersion distributions of the \gls{mccd}. The current \gls{mccd} contains \num{1082} clusters.\footnote{The catalogue is available at \url{https://zenodo.org/uploads/16970007}. Its contents are listed in Appendix \ref{apdx:MCCD}.}

\subsubsection{\label{sec:opticalmaster}The optical master table}

As with the \gls{mcxc} and \gls{mcsz} meta-catalogues and the ComPRASS catalogue, which were compiled into the \mtwoc master table, we built an optical master table based on the \gls{mccd} and \gls{lc2} meta-catalogues and the Abell catalogue. The construction of the optical master table follows the strategy developed in this work for the validation of the \gls{des} Y1 \gls{rm} catalogue. Details of the matching methods and overall consistency checks are given in \cref{sec:metho}. In particular, we used the methods presented in \cref{sec:2way,sec:nt500} to match the \gls{mccd} and Abell to \gls{lc2}. This was possible because the \gls{lc2} meta-catalogue provides an estimate of the mass for each cluster. We used the third method presented in \cref{sec:nomass} to match \gls{mccd} and Abell because they do not provide estimates for the mass. 

After having performed the three individual matchings, we consolidated them by verifying the consistency across the three. For example, if an \gls{mccd} cluster was associated with an \gls{lc2} cluster, and the \gls{lc2} was associated with an Abell cluster, we checked that the Abell cluster was also associated with the same \gls{mccd} cluster. This overall check allowed us to clarify associations for clusters with sub-components that were not consistently associated across the three matchings. 

\begin{figure}
    \centering
    \includegraphics[width=1.0\linewidth]{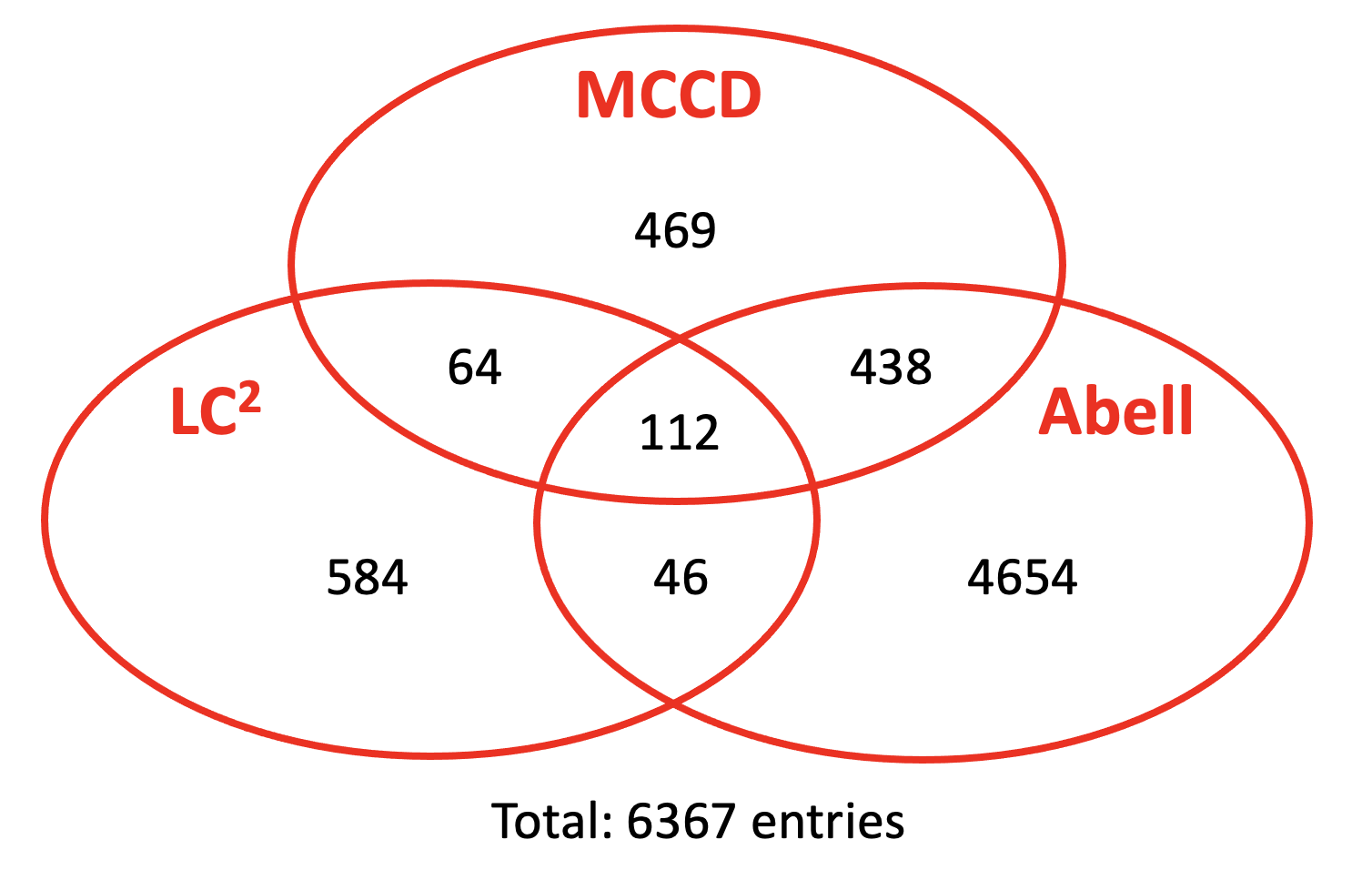}
    \caption{Venn diagram illustrating the structure of the optical master table.}
    \label{fig:optical_master}
\end{figure}

A Venn diagram of the resulting optical master table is presented in \cref{fig:optical_master}. The master table contains \num{4654}, \num{584}, \num{469} unique Abell, \gls{lc2}, \gls{mccd} clusters, respectively. \num{46} clusters are in common between Abell and \gls{lc2}, \num{64} between \gls{mccd} and \gls{lc2}, and \num{438} between \gls{mccd} and Abell; and \num{112} are common to the three catalogues. The optical master table contains \num{6367} single entries.

\subsection{\label{sec:complementary}Complementary catalogues}
Additional, complementary catalogues may also be used to validate the \gls{ecgc}, such as the \XMM serendipitous cluster catalogues XCLASS~\citep{Koulouridis2021}, XXL~\citep{Adami2018}, and  XCS~\citep{Mehrtens2012}. Catalogues based on X-ray+optical or SZ+optical bands (i.e. ICM-selected samples utilising systematic optical confirmation), such as 2XMMi/\gls{sdss}~\citep{Takey2011,Takey2013,Takey2014}, 3XMM/\gls{sdss}~\citep{Takey2016}, MARD-Y3~\citep{Klein2019}, \gls{rass}-MCMF~\citep{Klein2023}, PSZ-MCMF~\citep{Hernandez-Lang2023}, and \gls{spt}-SZ MCMF~\citep{Klein2024}, may also be useful because they are deeper than single-band catalogues. Finally, optical catalogues based on richness will aid validation of the lowest mass detections, for example, \gls{rm}~\citep{Rykoff2014}, Wen et al. catalogues~\citep{Wen2009,Wen2012,Wen2015,Wen2018,Wen2021,Wen2022,Wen2024}, WaZP~\citep{Aguena2021}, and AMICO-KiDS~\citep{Maturi2025}. We did not use these catalogues in the work described in this article, but we will consider using them to validate the \gls{ecgc} and extend the methodology developed herein (\cref{sec:metho}).

\section{\label{sec:metho}Method}
To prepare our validation procedure for the future \gls{ecgc}, we adopted the \gls{des} Y1 \gls{rm}~\citep[][]{Rykoff2016,DESCosmo2020} catalogue\footnote{\url{https://des.ncsa.illinois.edu/releases/y1a1/key-catalogs/key-redmapper}}$^,$\footnote{During the \Euclid internal review of this article, the \gls{des} Y3 \gls{rm} catalogue became public at \url{https://des.ncsa.illinois.edu/releases/y3a2/Y3key-cluster}} as a surrogate for the \gls{ecgc}, to set up and illustrate our method. We chose the \gls{des} Y1 \gls{rm} catalogue because it is based on an optical selection and covers a large fraction of the sky
-- it contains \num{6729} detections between $z=0.2$ and $z=0.86$ with \gls{rm} richness $\lambda>20$ over about $1650\,{\rm deg}^2$.

Our procedure progresses through three steps: 1) match the target catalogue (i.e. the catalogue to be validated; here, the \gls{rm} surrogate, later the \gls{ecgc}) with catalogues and meta-catalogues (\cref{sec:matching}); 2) consolidate the matching using master tables (\cref{sec:conso}); and 3) perform visual inspection of complex cases (\cref{sec:visu}). 

We do not directly match the target catalogue with the master tables because the target catalogue is expected to contain complex (e.g. merging, line-of-sight projected) systems or substructures. They will more likely be properly associated if we perform the matching with catalogues and meta-catalogues at various wavelengths (optical, \gls{sz}, X-ray). Complex structures have different positions in catalogues at different wavelengths, which often leads to inconsistent matching across catalogues at different wavelengths. Using the catalogues and meta-catalogues, instead of the constructed master tables, helps correctly identify such complex systems.

\Cref{tab:used_cats} provides a summary of meta-catalogues and catalogues used in this article, as well as the parameters used for the matching methods (details given below). We discuss the final result of our validation procedure on the \gls{des} Y1 \gls{rm} catalogue in Sect.~\ref{sec:final}.

\subsection{\label{sec:matching}Step one: Matching to catalogues and meta-catalogues}

We used two complementary matching methods for clusters that have mass estimates, which define a physical cluster size (\cref{sec:2way,sec:nt500}), and one matching method based on purely angular distance for clusters lacking mass estimates (\cref{sec:nomass}). In~\cref{sec:2waynt500} we compare the results of the two physical matches presented in~\cref{sec:2way,sec:nt500}.

\subsubsection{\label{sec:2way}Two-way matching}
In the following, we refer to the target catalogue as `\catT' 
and the meta-catalogue or catalogue with mass estimates (given as $\Mfive$\,\footnote{The mass enclosed in a radius $\Rfive$ within which the average cluster mass is 500 times the critical density of the Universe at the cluster's redshift.}) as `\catM' (in practice, the \gls{mcxc}, \gls{mcsz}, ComPRASS, \gls{lc2}, and the \eROSITA\/ catalogues\footnote{The masses are obtained from the X-ray luminosity-mass scaling relation, the SZ flux-mass scaling relation, the weak lensing masses, and the X-ray count rate-mass scaling relation for the \gls{mcxc}, \gls{mcsz}, ComPRASS, \gls{lc2}, and the \eROSITA\/ catalogues, respectively.}).
The procedure for `\tway' matching is the following.
\begin{enumerate}
\item For each cluster in \catT, compute the projected, two-dimensional angular distances to clusters in \catM; keep the closest cluster from \catM.\footnote{Cases with multiple clusters in projection are studied with the $\ntfive$ matching in \cref{sec:nt500}.}
\item For each cluster in \catM, compute the distances to clusters in \catT; keep the closest from \catT.\footnote{In this and in the previous step, we do not impose any separation limit.}
\item Keep only cluster pairs that are the same in steps 1 and 2 (hence the name of the method); reject other \catT clusters, leaving them as unmatched.
\item Plot the separation, $d$, between the \tway matched pairs divided by the characteristic scale $\tfive$\,\footnote{$\tfive=\Rfive/D_{\rm ang}$, where $D_{\rm ang}$ is the angular diameter distance to the cluster.} as a function of $d$. The plot clearly shows two distinct regions: at low $d$ or low $d/\tfive$, the pairs are a `possible good match' (region A); at high $d$ and high $d/\tfive$, the pairs are a `possible bad match' (region B).
\item Define $d_{\rm cut}$ and $(d/\tfive)_{\rm cut}$ to separate regions A and B.
\item Plot the redshift $z_{\rm T}$ of \catT versus the redshift $z_{\rm M}$ of \catM for pairs in region A. Keep pairs with \mbox{$|\Delta z| / (1+z_{\rm M}) < \epsilon_z$} ($\Delta z = z_{\rm T}-z_{\rm M}$) as confirmed matches, and reject the other pairs as bad matches (see below for our chosen value of $\epsilon_z$) because they are line-of-sight projections.
\item Check the distribution of confirmed matches in the initial $d/\tfive$ versus $d$ plot to see if some pairs lie on the border between regions A and B.
\item Plot the redshifts for pairs in region B and use the same parameter $\epsilon_z$ to look for potential good matches, although they are most probably chance associations in redshift. Check the distribution of these potential good matches in the initial $d$ versus $d/\tfive$ plane. If the pair lies close to the border between regions A and B, perform a visual inspection of the cluster pair (\cref{sec:visu}) to decide if the association is a good match. Reject all other pairs as bad matches.
\end{enumerate}

\begin{figure}
\centering
 \includegraphics[width=\hsize]{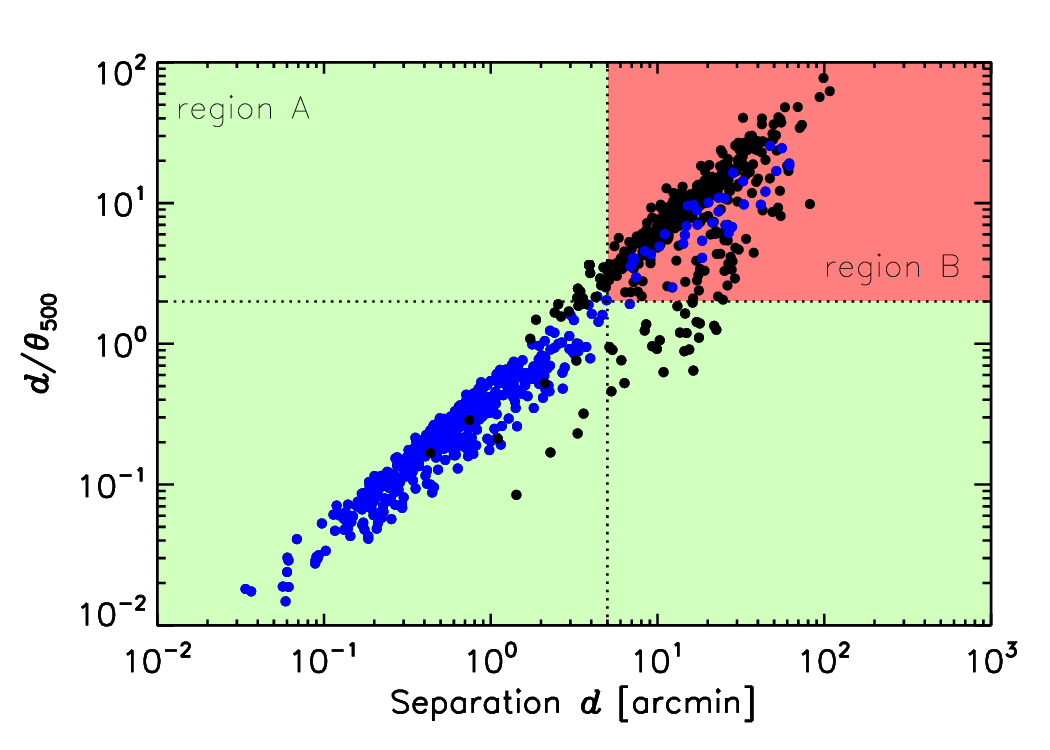}
\caption{Distribution of the \tway matched pairs in the $d$-$d/\tfive$ plane for the matching between the \gls{des} Y1 \gls{rm} and \gls{mcsz} catalogues. Among the 883 pairs, 489 pairs are a priori good matches (region A, green area) and 394 are a priori bad matches (region B, red area). Pairs in blue match in redshift (see \cref{fig:rm_mcsz_z_regA,fig:rm_mcsz_z_regB}, and associated text), while pairs in black do not match in redshift. The blue pairs in region B are most probably chance associations.}  
\label{fig:rm_mcsz_dist}
\end{figure}

We note that the 'two-way' matching procedure first associates on separation (step 3) and then selects a match if the redshift criterion is satisfied (step 6); if the redshift criterion is not satisfied, then no match is made and no further attempt to match to other farther clusters is undertaken. As a consequence, the procedure may lose correct matches that are the 2nd (or nth) closest association in sky separation if the closest association fails to pass the redshift criterion. This is intentional: it allows us to correct the redshift in catalogues in case there is an error in one of the two redshifts, which we can spot and study with a multiple-step process. To avoid missing the 2nd (or nth) closest match, we specifically developed the '$\ntfive$' matching described in Sect.~\ref{sec:nt500}.

For the \gls{des} Y1 \gls{rm} catalogue, we fixed $\epsilon_z=0.03$, which is about five times the $1\sigma$ photometric redshift uncertainty of \gls{rm} clusters~\citep{Rykoff2016}.  In the following, we illustrate the \tway matching with the \gls{des} Y1 \gls{rm} catalogue and the \gls{mcsz} meta-catalogue. We adopted $d_{\rm cut}=5\,{\rm arcmin}$ and $(d/\tfive)_{\rm cut}=2$ to separate regions A and B. \Cref{fig:rm_mcsz_dist} shows $d/\tfive$ vs $d$ for the 883 \tway matches, of which 489 pairs fall in region A and 394 in region B. 

\begin{figure}
\centering
 \includegraphics[width=1.0\hsize]{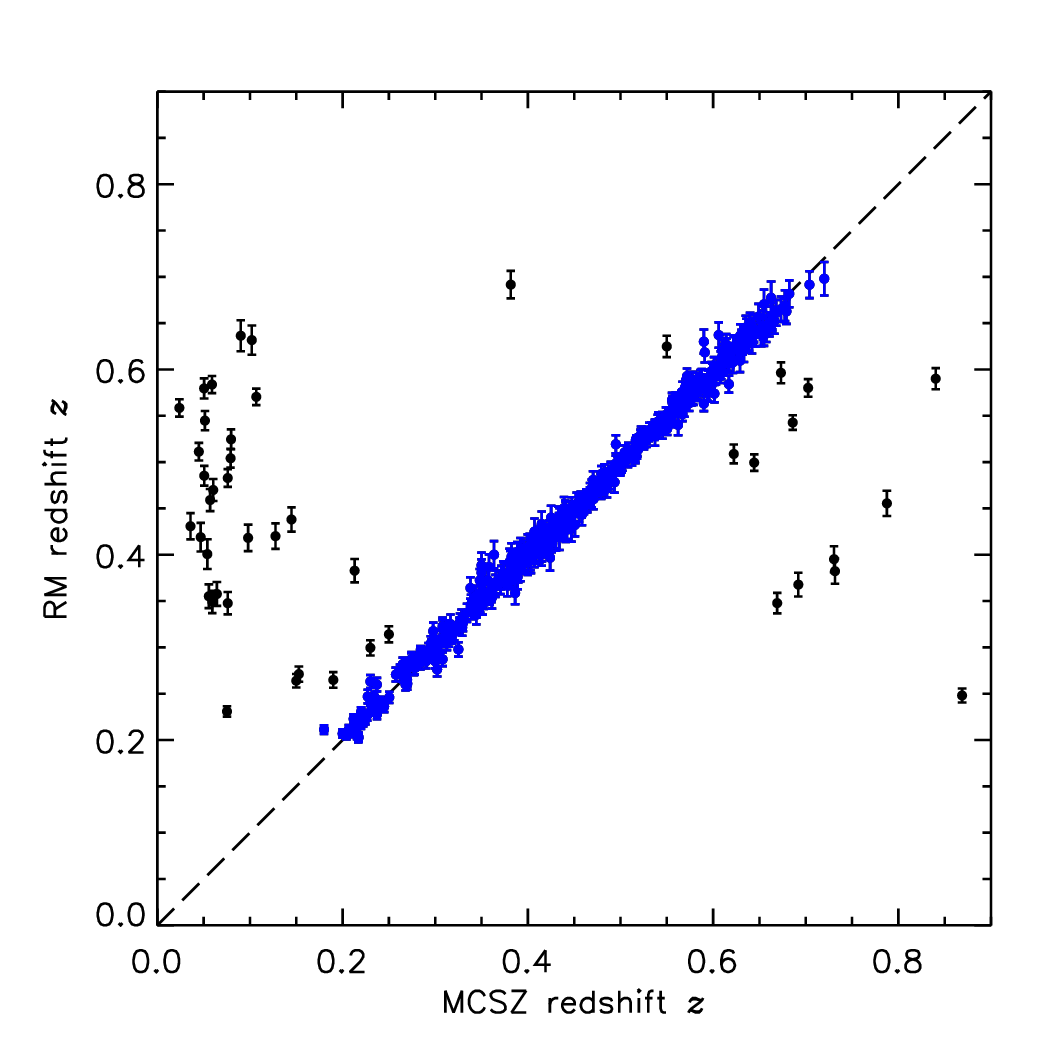}
\caption{\gls{rm} redshift vs \gls{mcsz} redshift for the 489 pairs in region A; 435 pairs (in blue) match in redshift within the criterion \mbox{$|\Delta z| / (1+z_{\rm M}) | < \epsilon_z$} ($\epsilon_z=0.03$), while 54 pairs (in black) do not and are considered projections along the line of sight.}
\label{fig:rm_mcsz_z_regA}
\end{figure}

We then looked at the match in redshift for clusters in region A (see \cref{fig:rm_mcsz_z_regA}). Of these, 435 pairs match in redshift, and we consider these as confirmed associations (good matches); 54 pairs do not match, and we conclude that these are bad matches caused by projection effects. \Cref{fig:rm_mcsz_dist} identifies the 435 pairs of region A that match in redshift as blue dots. Most of them are well within region A, but a few lie at the border with region B.

\begin{figure}
\centering
  \includegraphics[width=1.0\hsize]{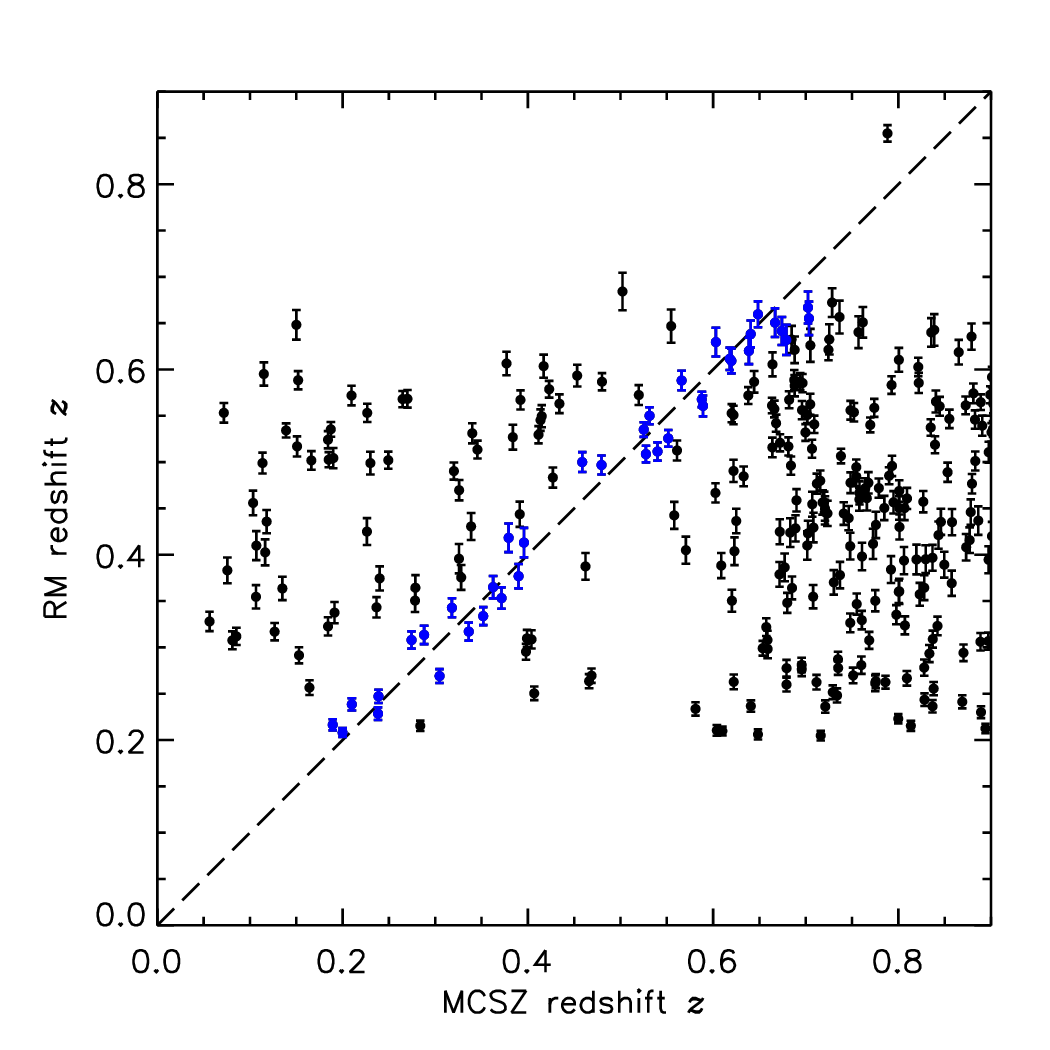}
\caption{\gls{rm} redshift vs \gls{mcsz} redshift for the 394 pairs in region B; 37 pairs (in blue) match in redshift, but they are most probably chance associations given their position in region B and the larger scatter around the one-to-one line (0.025) compared to the scatter of the blue points in \cref{fig:rm_mcsz_z_regA} (0.001).}
\label{fig:rm_mcsz_z_regB}
\end{figure}

We also checked redshifts for the 394 pairs in region B. \Cref{fig:rm_mcsz_z_regB} shows that 37 pairs (in blue) match in redshift, but the general distribution of the pairs (in blue and black) does not exhibit any clustering around the 1:1 relation. These matched pairs are most probably chance associations. In order to confirm this, we examined their position in the $d$-$d/\tfive$ plane (\cref{fig:rm_mcsz_dist}), where they are tagged as blue dots. Most of them are located at large separations ($d>6.8\,{\rm arcmin}$ and $d/\tfive>3$). We therefore classify them as chance associations\footnote{We note, however, that \cite{Kelly2024} found very large miscentring between some \gls{des} Y3 \gls{rm} clusters and their X-ray counterparts in \XMM or Chandra data.} (bad matches). Only two are located within $2<d/\tfive<3$; the situation for these two is less clear, and we study them in detail in \cref{sec:nt500}.

In summary, the \tway matching associated 435 \gls{rm} clusters  with \gls{mcsz} clusters. Two additional matches could be considered in the region $2<d/\tfive<3$ (\cref{sec:nt500}). We applied the same methodology to associate \gls{rm} with \gls{mcxc}, ComPRASS, \gls{lc2}, and \eROSITA. Results on the number of \tway matches are given in \cref{sec:final}, and the adopted values for $d$ and $d/\tfive$ separating regions A and B are given in \cref{tab:used_cats}. We note that the final matching does not depend strongly on the exact choice of $d_{\rm cut}$ and $(d/\tfive)_{\rm cut}$ because we use redshift as an additional criterion in both regions A and B, and we perform a visual inspection for pairs lying at the border of the two regions.

\subsubsection{\label{sec:nt500}$\ntfive$ matching}
The \tway matching procedure is susceptible to projection effects; for example, a given \gls{mcsz} cluster could lie close to a \gls{rm} cluster on the sky, but be at a different redshift, while a second \gls{rm} cluster could be located at a slightly larger distance from the \gls{mcsz} cluster, but closer in redshift. In this case, the \tway matching procedure would fail, incorrectly associating the first \gls{rm} cluster with the \gls{mcsz} cluster. 

We therefore developed a second matching procedure, `$\ntfive$ matching', to compare with the first one. Using the same notation (catalogue T to be validated, known catalogue M with masses and sizes $\tfive$), the procedure of the $\ntfive$ method is the following.
\begin{enumerate}
\item For each cluster in catalogue M, find clusters in catalogue T located within a radius of $n \, \tfive$ (hence the name of the method). We fixed $n=3$. This value corresponds to a radius of $3 \, \tfive$, which is larger than the cluster virial radius and allows us to spot possible substructures or in-falling groups/clusters.
\item Associate the cluster from catalogue M with the cluster in catalogue T (located within $n \, \tfive$) that is closest in redshift.
\item If multiple clusters from catalogue M are associated with the same cluster in catalogue T, keep only the pair with the smallest separation on the sky.  
\item Select pairs for which $|\Delta z| / (1+z_{\rm M}) < \epsilon_z$ as good associations. 
\end{enumerate}

We performed this $\ntfive$ matching between the DES Y1 RM catalogue (catalogue T) and the MCXC-II, MCSZ, ComPRASS, \gls{lc2}, and eROSITA (meta-)catalogues (catalogue M). In this section, we illustrate the method with the \gls{mcsz} and give the results for the other catalogues in \cref{sec:final}.

\begin{figure}
\centering
  \includegraphics[width=1.0\hsize]{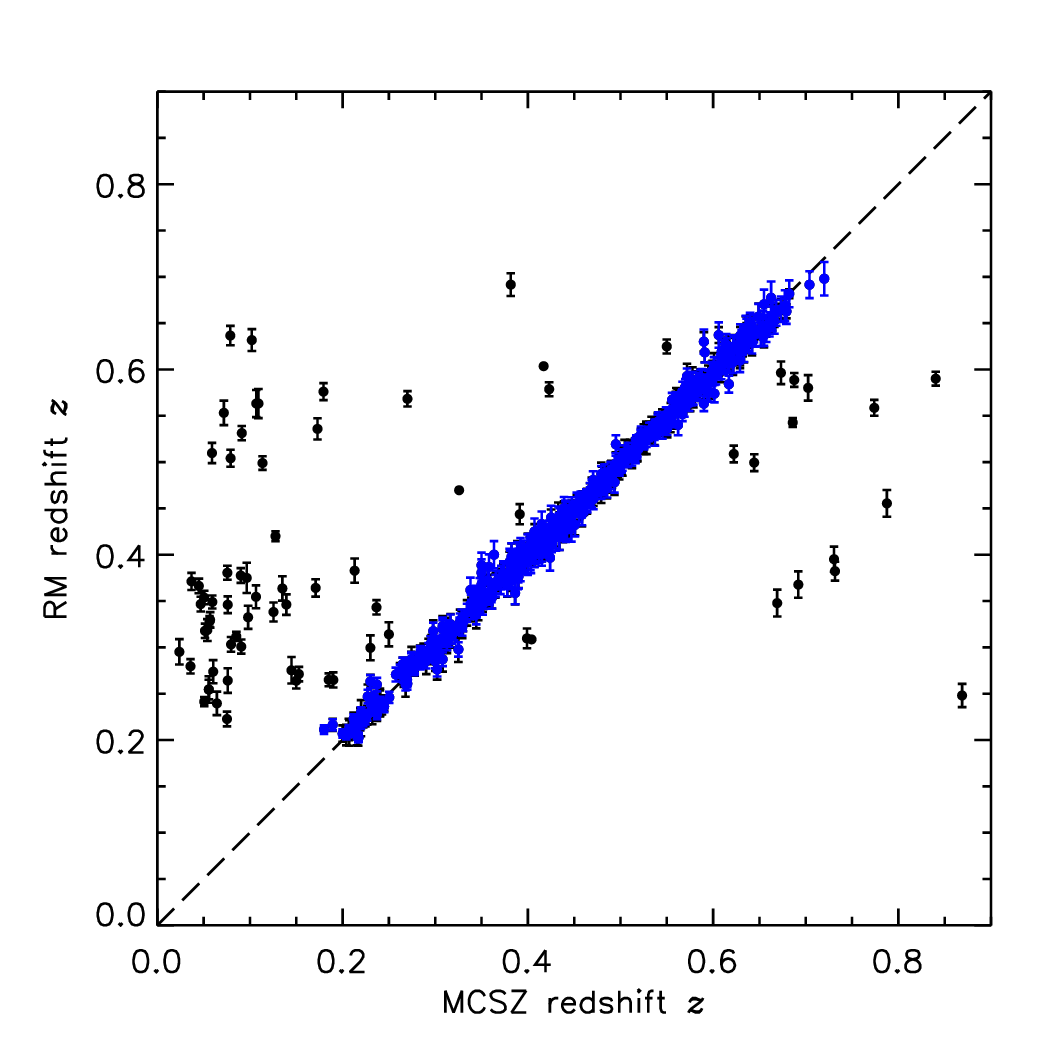}
\caption{\gls{rm} redshift vs \gls{mcsz} redshift for the 517 pairs matched with the $\ntfive$ method. There are 436 pairs (in blue) that match in redshift, and 81 pairs (in black) that are rejected and taken to be projections along the line of sight.}
\label{fig:rm_mcsz_z_nt500}
\end{figure}

After step 2, we found 520 \gls{rm} clusters matching \gls{mcsz} clusters. In step 3, we removed three pairs, leaving 517 matches. In \cref{fig:rm_mcsz_z_nt500}, we show \gls{rm} redshift vs \gls{mcsz} redshift for the 517 pairs. In blue, we show the pairs which verify $|\Delta z| / (1+z_{\rm M}) < \epsilon_z$ with $\epsilon_z=0.03$, giving 436 good matches after step~4; 81 pairs (among the 517) have redshifts that do not match and are thus taken as projections along the line of sight.

\subsubsection{\label{sec:2waynt500}Comparison of the \tway and $\ntfive$ matches}

We compared the results from the \tway and the $\ntfive$ matching. We found 435/436 good associations with the \tway/$\ntfive$ matching, respectively. Of these, 424 are identical between the two methods; two are matched by $\ntfive$, but not by \tway; ten have different associations between \tway and $\ntfive$; and one is matched by \tway, but not by $\ntfive$.

The two pairs that are matched with $\ntfive$ but not with \tway are those falling in region B close to region A at $2<d/\tfive<3$ in \cref{fig:rm_mcsz_dist}. Upon close examination of these two cases, the normalised distance is found to be close to $3\tfive$ and the richness values of the \gls{rm} clusters are small ($\lambda=24.9$ and 35.0), which suggests that these two associations are probable chance associations. We thus reject these matches. We present one of the two cases in Sect.~\ref{sec:visu}.

\begin{figure}
\centering
 \includegraphics[width=\hsize]{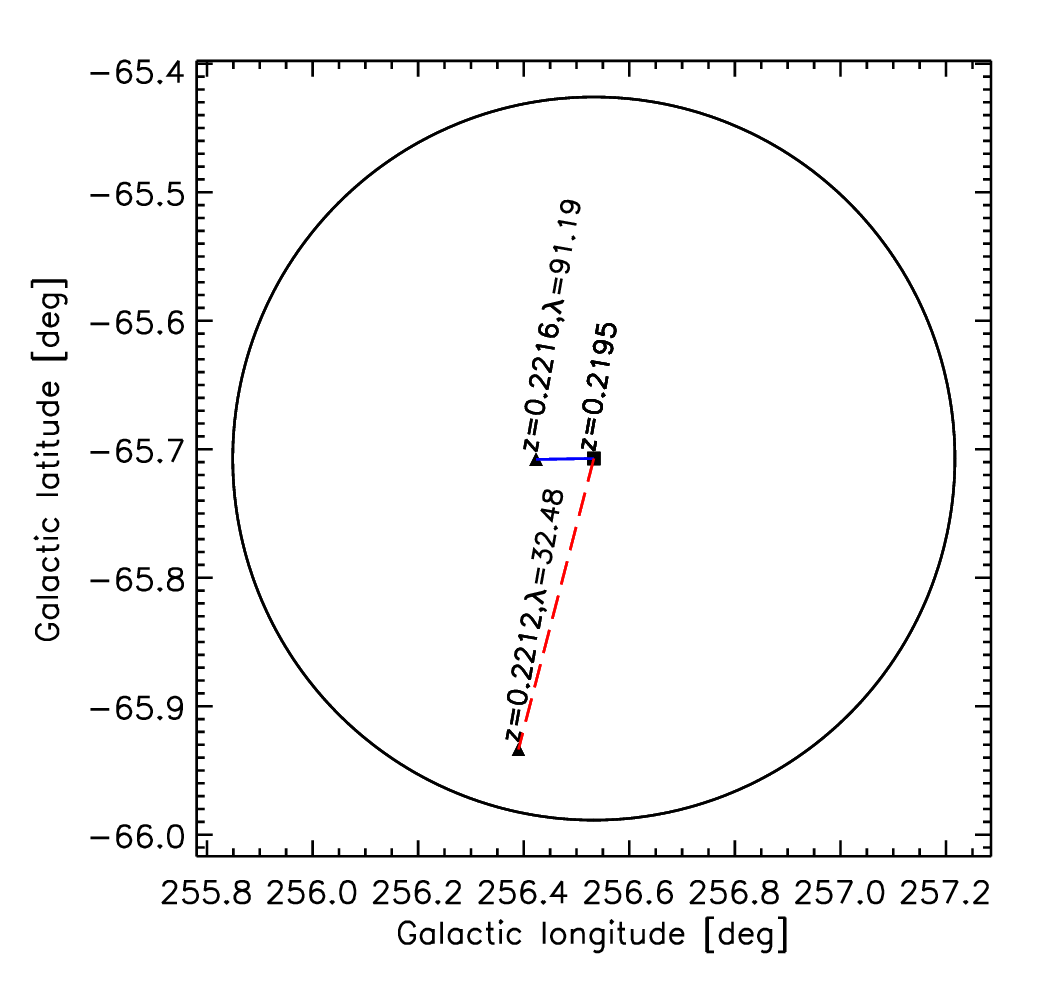}
  \includegraphics[width=\hsize]{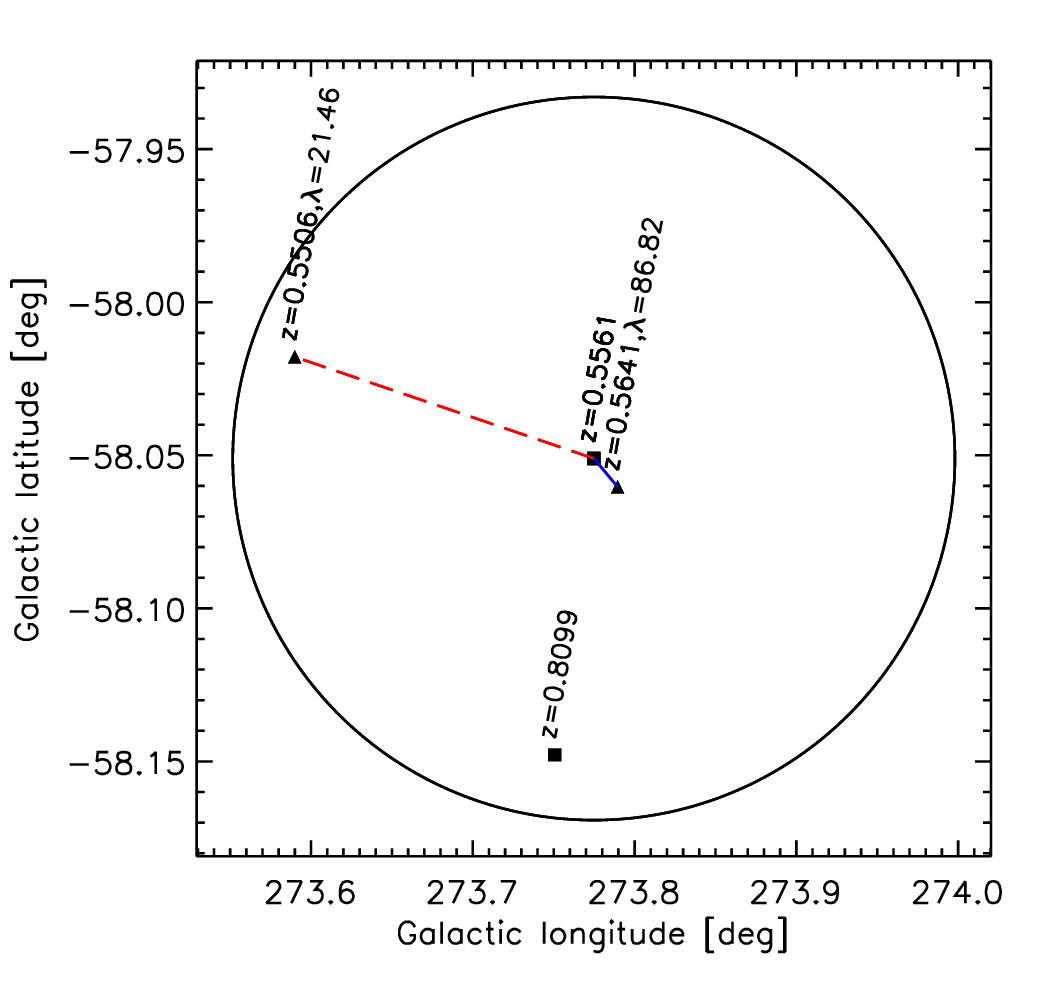}
\caption{Two specific cases (among ten) of clusters matched differently between \tway and $\ntfive$ matching. Symbols are the same as in \cref{fig:rm_mcsz_mis_case1}. In both cases, an \gls{mcsz} cluster (at the centre) is surrounded by two \gls{rm} clusters. The closest \gls{rm} is the richest (\tway match), and we kept it as the correct association. The farthest is less rich ($\ntfive$ match), and we categorised it as a possible substructure or an in-falling cluster.}
\label{fig:rm_mcsz_mis_case23}
\end{figure}

Nine out of the ten clusters matched differently between \tway and $\ntfive$ occur for an \gls{mcsz} cluster associated with two different \gls{rm} clusters. One \gls{rm} cluster is rich and located close to the \gls{mcsz} cluster ($d/\tfive<2$), while the second \gls{rm} cluster is less rich and farther away ($2<d/\tfive<3$) but at the same redshift. Two of these ten cases are shown in \cref{fig:rm_mcsz_mis_case23}. We categorised the \gls{rm} clusters located at $2<d/\tfive<3$ as possible substructures or in-falling clusters, but we kept the \tway matches as the good matches. The last case out of the ten corresponds to an identical situation (a single \gls{mcsz} cluster with two \gls{rm} clusters), but with the farther \gls{rm} cluster being the richest instead of the poorer one. For simplicity, we also kept the \tway match as the good association and categorised the other associations as a possible substructure or an in-falling group/cluster.

\begin{figure}
\centering
 \includegraphics[width=\hsize]{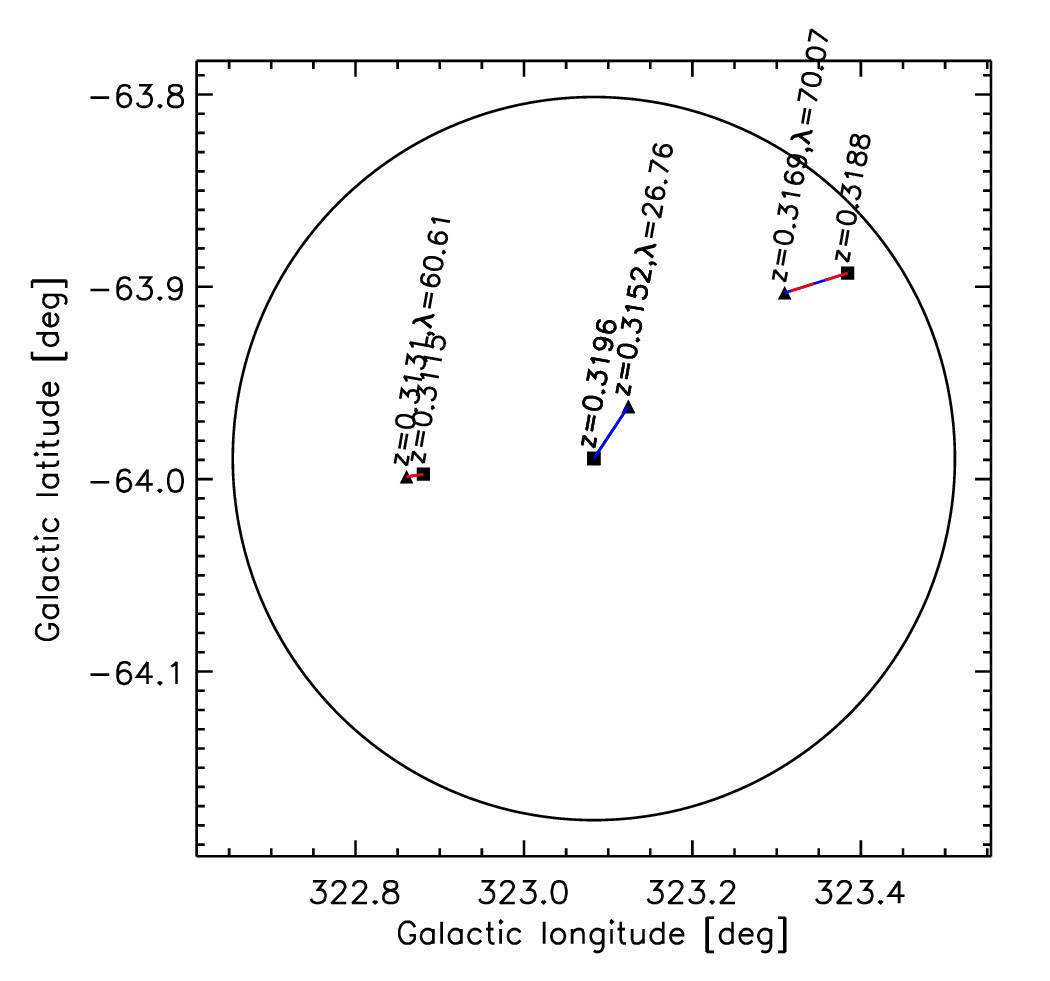}
\caption{Specific case of the \gls{rm} cluster matched with \tway matching but not with $\ntfive$ matching (triangle close to the black square at the centre). The \gls{mcsz} clusters in the field are shown as black squares and the \gls{des} Y1 \gls{rm} clusters as triangles. Redshifts and richnesses of the clusters are labelled next to their symbols. The circle (on the sky) corresponds to a radius of $3\tfive$ with the \gls{mcsz} cluster at the centre. The initial $\ntfive$ association (with the \gls{rm} cluster at $z=0.3169,\lambda=70.07$) was rejected at step 4 of the association process.}
\label{fig:rm_mcsz_mis_case1}
\end{figure}

The situation with the cluster matched with \tway but not with $\ntfive$ is depicted in \cref{fig:rm_mcsz_mis_case1}. The cluster is marked as the black triangle (\gls{rm} cluster) close to the black square (\gls{mcsz} cluster) at the centre. These two clusters are matched with \tway (solid blue connecting line), but not with $\ntfive$. The \gls{mcsz} cluster at the centre ($z=0.3196$) was initially associated with the \gls{rm} cluster ($z=0.3169,\lambda=70.07$) located farther away, but within a radius of $3\tfive$ because of its closer redshift. This initial association corresponds to steps 1 through 2 of the $\ntfive$ association process. In step 3, however, this association was rejected because that \gls{rm} cluster ($z=0.3169,\lambda=70.07$) was associated with another, closer \gls{mcsz} cluster at $z=0.3188$ (straight, dashed, red connecting line). This other association is also a \tway match (solid blue line).

In summary, we kept the associations from the \tway matching as the good matches and used the $\ntfive$ matching to categorise some \gls{rm} detections as possible substructures or in-falling clusters of larger \gls{mcsz} clusters. We thus obtain 435 good matches (via \tway) and ten \gls{rm} clusters (rejecting the two pairs that are matched by $\ntfive$ but not by \tway) as possible substructures or clusters falling into the \gls{mcsz} clusters. The most difficult and problematic cases are thus related to  correlated large-scale structure or to ongoing mergers.

\subsubsection{\label{sec:nomass}Positional matching: Angular separation only}

When the meta-catalogue or catalogue to be matched with the catalogue T (\gls{des} Y1 \gls{rm} and, in the future, the \gls{ecgc}) had no available mass, we performed a simple positional matching in angular distance (angular separation $d<d_\mathrm{cut}$) using the \tway matching procedure (\cref{sec:2way}).
We then checked the redshift consistency between the matched clusters when the redshift was available in the known meta-catalogue or catalogue. The positional matching was performed with the \gls{mccd} meta-catalogue and the Abell catalogue. We used $d_{\rm cut}=2 \, {\rm arcmin}/ 10 \, {\rm arcmin}$ for \gls{mccd}/Abell respectively, and performed visual inspections (\cref{sec:visu}) when the matching was not certain. Results are presented in \cref{sec:final}.

\begin{table*}
\caption{Summary of the meta-catalogues and catalogues used in this article and cross-matched with the \gls{des} Y1 \gls{rm} catalogue.}
\smallskip
\label{tab:used_cats}
\smallskip
\centering
\begin{tabular}{|c|c|c|c|c|c|}
  \hline
  Cluster catalogue & Number of Clusters & Master Table & Observational Band & $d_{\rm cut}$ & $(d/\tfive)_{\rm cut}$ \\
    & & & & {\rm arcmin} &  \\
  \hline
  MCXC-II & \num{2221} & M2C & X-ray & 3 & 1 \\
  MCSZ & \num{5564} & M2C & SZ & 5 & 2 \\
  ComPRASS & \num{2323} & M2C & X-ray and SZ & 5 & 2 \\
  eROSITA & \num{12310} & --- & X-ray & 5 & 3 \\
  MCCD & \num{1083} & optical & optical & 2 & --- \\
  ${\rm LC}^2$ & 806 & optical & optical & 2 & 1 \\
  Abell & \num{5250} & optical & optical & 10 & --- \\
\hline
\end{tabular}
\tablefoot{Columns are from left to right: meta-catalogue or catalogue name; number of clusters; master table to which the meta-catalogue or catalogue belongs; observational band of meta-catalogue or catalogue (X-ray, SZ, optical); $d_{\rm cut}$ value adopted for the \tway matching or for the positional matching when the mass in the meta-catalogue or catalogue is not available; $(d/\tfive)_{\rm cut}$ value used for the \tway matching. We placed '---' in column $(d/\tfive)_{\rm cut}$ for the case of positional matching. In all cases, we used $\epsilon_z=0.03$ for the error on $|\Delta z|/(1+z)$, and we adopted $n=3$ for the $\ntfive$ matching.}
\end{table*}

\subsection{\label{sec:conso}Step 2: Consolidation of the matching using the master tables}

\begin{figure}
\centering
 \includegraphics[width=\hsize]{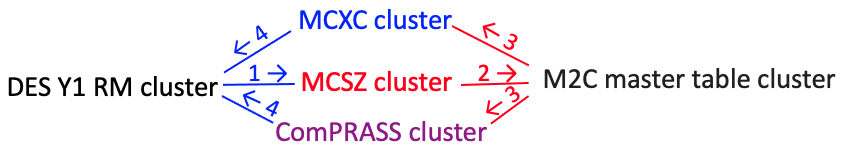}
\caption{Loops used to check consistency between the \gls{des} Y1 \gls{rm} -- \gls{mcsz} matching and the matching performed in the \mtwoc master table. We verified that the loops closed through steps 1, 2, 3, and 4.}
\label{fig:consolidation}
\end{figure}

After matching the catalogues/meta-catalogues with \gls{des} Y1 \gls{rm}, we checked consistency with the matching performed in the \mtwoc (\cref{sec:xsz}) and in the optical (\cref{sec:opticalmaster}) master tables. \Cref{fig:consolidation} illustrates the consolidation procedure for the \gls{des} Y1 \gls{rm} -- \gls{mcsz} matching.
\begin{enumerate}
\item For each \gls{des} Y1 \gls{rm} cluster, check if it is linked to an \gls{mcsz} cluster.
\item Verify that the \gls{mcsz} cluster is in the \mtwoc master table.
\item The \gls{mcsz} cluster could also be associated with an \gls{mcxc} or a ComPRASS cluster in the master table.
\item If so, check that it has been associated in our matching procedure.
\end{enumerate}
When the loops closed (i.e. the four steps above are performed successfully), we considered the \gls{des} Y1 \gls{rm} -- \gls{mcsz} matching to be consolidated.

This consolidation process enabled us to check for numerical bugs in our matching procedure and for inconsistencies between redshifts in our catalogues and redshifts in the master tables due to multiple cluster components along the line of sight (in which case we updated the master tables in adding additional entries corresponding to the individual components). This also allowed us to find and clarify specific cases of confusion (multiple close-by \gls{rm} clusters, \gls{mcsz} and ComPRASS clusters).

\subsection{\label{sec:visu}Step 3: Visual inspection of complex cases}

\begin{figure}
\centering
\includegraphics[width=\hsize]{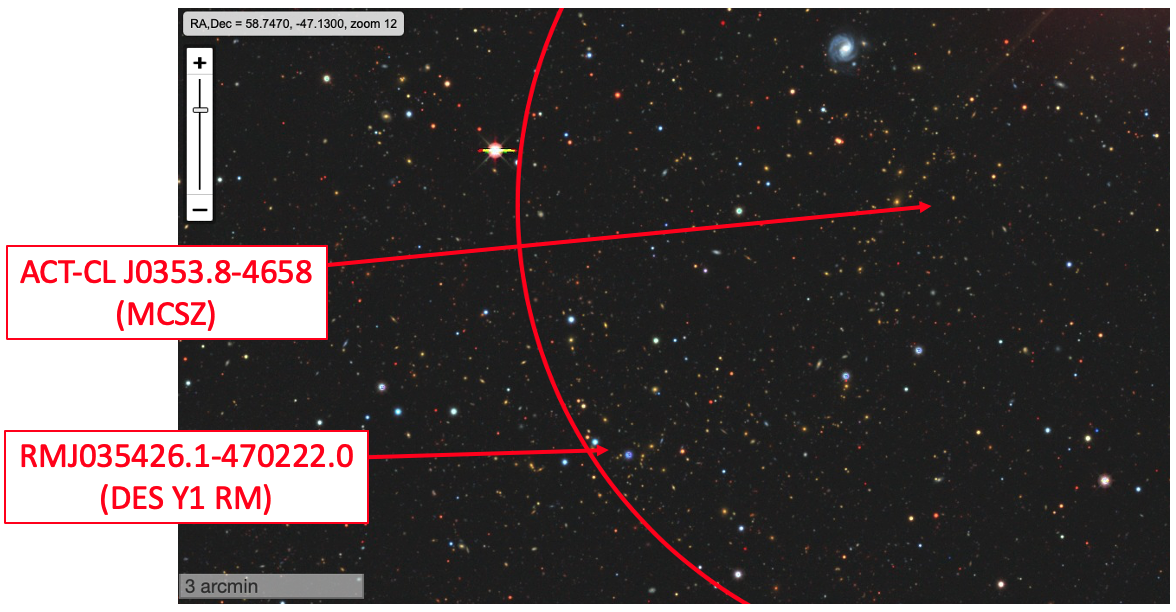}
\caption{Example of visual inspection for the \gls{des} Y1 \gls{rm} -- \gls{mcsz} matching.  \gls{act} cluster ACT-CL J0353.8$-$4658 ($z=0.39$) and \gls{rm} cluster RMJ035426.1$-$470222.0 ($z=0.38$) have consistent redshifts, but they are widely separated on the sky (close to $3\tfive$). We therefore decided not to associate the two. The red circle has a radius of $3\tfive=7.5 \, \rm{arcmin}$ and is centred on the ACT-CL J0353.8$-$4658 position.  The image is extracted from~\url{https://www.legacysurvey.org/viewer}.}
\label{fig:visual}
\end{figure}

We visually inspected the optical images of clusters presenting complex situations. \Cref{fig:visual} shows an example of a visual inspection for the \gls{des} Y1 \gls{rm} -- \gls{mcsz} matching. The \gls{act} cluster ACT-CL J0353.8$-$4658 ($z=0.39$) is located at $7.4 \, {\rm arcmin}$ from RMJ035426.1$-$470222.0 ($z=0.38, \lambda=35.0$). The two redshifts are consistent, but the clusters are widely separated on the sky ($2.97\tfive$). We visually inspected this case to decide if the clusters correspond to a single halo or two different structures. The figure shows two distinct galaxy over-densities. We therefore did not associate the two clusters. We note that the galaxy over-density close to the \gls{act} cluster centre is not detected by \gls{rm}. We verified that this is because it was covered by the \gls{rm} mask.

\subsection{\label{sec:final}Final output of the matching procedure}

\begin{table}
\caption{Number of possible substructures or in-falling groups and clusters for each external catalogue, obtained by comparing the \tway matching with the $\ntfive$ matching.}
\smallskip
\label{tab:res_substruct}
\smallskip
\centering
\begin{tabular}{|c|c|}
  \hline
  Cluster catalogue & Number of substructures  \\
  or meta-catalogue & or in-falling groups/clusters  \\
  \hline
MCXC-II &  2 \\
MCSZ & 10  \\
ComPRASS &  4 \\
eROSITA & 6 \\
${\rm LC}^2$ &  1 \\
\hline
\end{tabular}
\tablefoot{Two examples are given in \cref{fig:rm_mcsz_mis_case23} for the \gls{mcsz}. Results are shown only for \gls{mcxc}, \gls{mcsz}, ComPRASS, \eROSITA, and \gls{lc2} for which both matching procedures were performed.}
\end{table}

The final results of the matching procedure are summarised in \cref{tab:res_substruct,tab:results}. Table~\ref{tab:res_substruct} gives the number of clusters in the \gls{mcxc}, \gls{mcsz}, ComPRASS, \eROSITA and \gls{lc2} for which we  found possible substructures or in-falling groups/clusters in the \gls{des} Y1 \gls{rm} by comparing the \tway and $\ntfive$ matchings. The second column gives the number of clusters belonging to one of our catalogues or meta-catalogues linked to different \gls{des} Y1 \gls{rm} clusters via \tway and $\ntfive$. In these situations, the \tway matching usually provides the main (higher richness) component, while the $\ntfive$ matching identifies possible in-falling groups/clusters (poorer richness).

\Cref{tab:results} gives the overall results. Among the \num{6729} \gls{des} Y1 \gls{rm} clusters, we find \num{1040} counterparts, including 28 from \gls{mcxc}, 435 from \gls{mcsz}, 60 from ComPRASS, 826 from \eROSITA, 48 from \gls{mccd}, 29 from \gls{lc2}, and 32 from Abell (see column one). Unique matches (i.e. matched by a single catalogue) are shown in column two and are dominated by \eROSITA and \gls{mcsz} clusters. The full result of the matching procedure can be downloaded as the \texttt{EC-RedMaPPer} catalogue\footnote{\mbox{\url{https://zenodo.org/records/16962265}}} in the form of the original \gls{des} Y1 \gls{rm} catalogue with additional columns. The description of each additional column is given in \cref{tab:overview}.

We note that we did not include optical catalogues based on richness for this external validation. Doing so would provide additional matches (via positional matching and redshift consistency checks), but would not give any new external mass estimate for the \gls{des} Y1 \gls{rm} clusters.

\subsection{\label{sec:chance}Chance association}

We estimated the rate of chance association by shifting the right ascension of the \gls{rm} detections by +10\,deg and applying the same cross-matching procedure. We obtained 0/7/1/22/2/1/1 matches for \mbox{MCXC-II}/MCSZ/ComPRASS/eROSITA/MCCD/${\rm LC}^2$/Abell, respectively, for the \tway matching. For the $\ntfive$ matching, the numbers are slightly higher with 0/17/4/43/2 matches for \mbox{MCXC-II}/MCSZ/ComPRASS/eROSITA/${\rm LC}^2$, respectively. This represents less than 5\% of the actual number of matches given in the second column of Table~\ref{tab:results}. Chance association is thus small for the catalogues considered in this study, although it is expected to increase for catalogues with higher densities.
In the following section, we discuss the cross-match of quantities from \gls{des} Y1 \gls{rm} and external catalogues/meta-catalogues.

\begin{table*}
\caption{Summary of matching the \gls{des} Y1 \gls{rm} catalogue with external catalogues.}
\smallskip
\label{tab:results}
\smallskip
\centering
\begin{tabular}{|c|c|c|}
  \hline
  Cluster catalogue & Number of matches & Number of unique matches \\
  or meta-catalogue & with DES Y1 RM & with DES Y1 RM \\
  \hline
MCXC-II & 28 & 4 \\
MCSZ & 435 & 175 \\
ComPRASS & 60 & 0 \\
eROSITA & 826 & 572 \\
MCCD & 48 & 4 \\
${\rm LC}^2$ & 29 & 8 \\
Abell & 32 & 4 \\
\hline
In two catalogues & & 188 \\
In three catalogues & & 44 \\
In four catalogues & & 28 \\
In five catalogues & & 9 \\
In six catalogues & & 2 \\
In seven catalogues & & 2 \\
Unmatched & & 5689 \\
\hline
Total & & 6729 \\
\hline
\end{tabular}
\tablefoot{The second column gives the number of matches between the \gls{des} Y1 \gls{rm} catalogue and each individual catalogue considered in this work. The third column gives the number of unique matches (i.e. matched by a single catalogue) for each catalogue or groups of catalogues.}
\end{table*}

\section{\label{sec:discuss}Discussion}

\begin{figure*}
    \centering
    \includegraphics[width=0.45\linewidth]{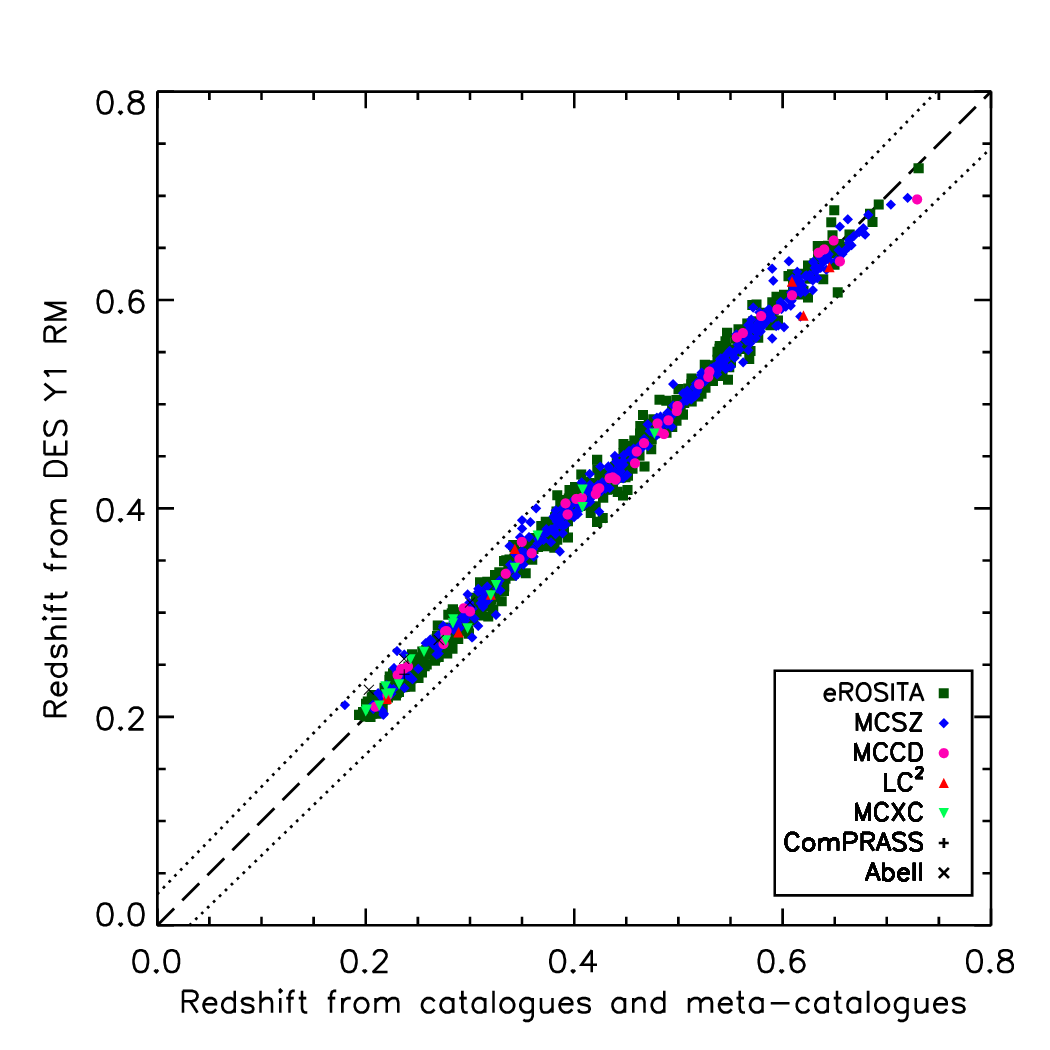} \includegraphics[width=0.45\linewidth]{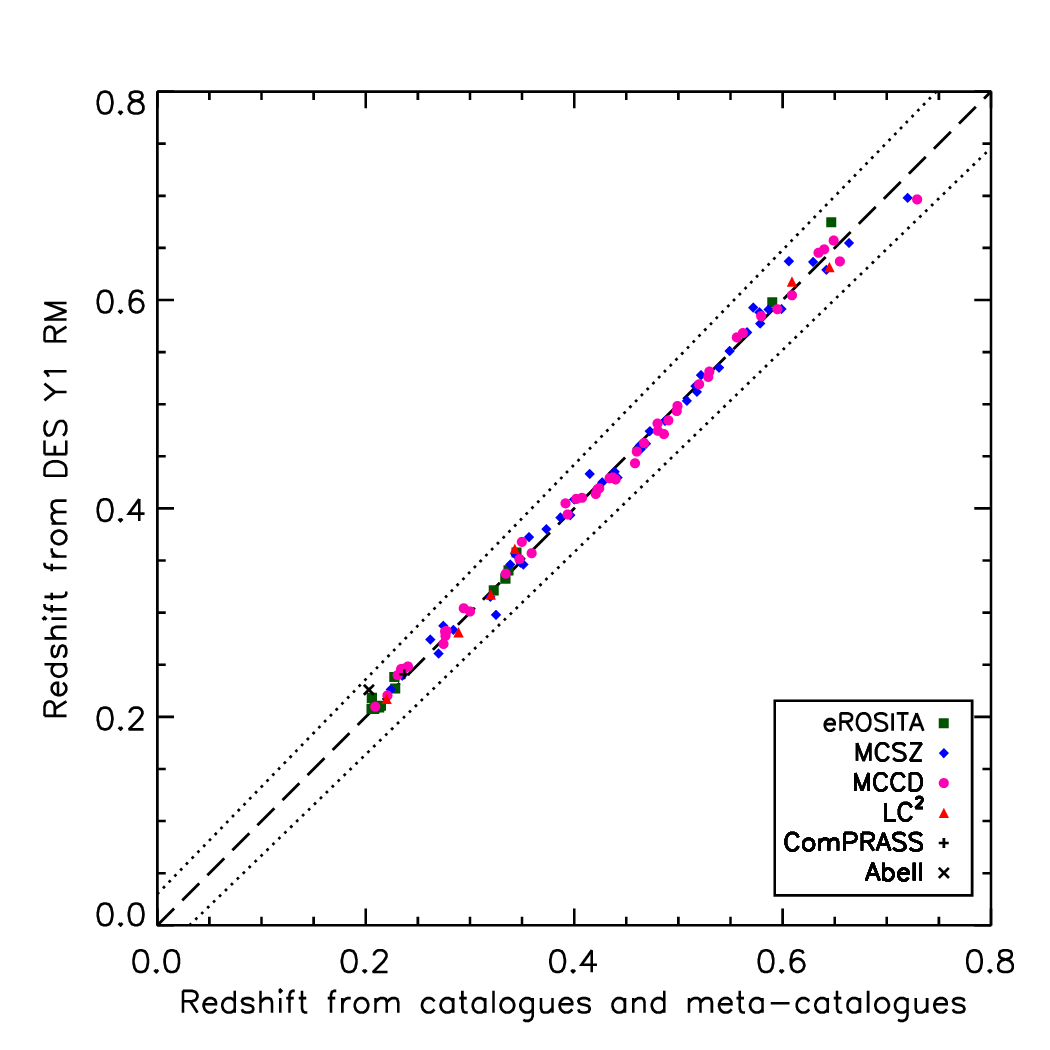}
     \caption{\gls{des} Y1 \gls{rm} photometric redshifts versus redshifts from external catalogues and meta-catalogues. \emph{Left}: all 1040 matches (both photometric and spectroscopic redshifts). \emph{Right}: the 114 matches with spectroscopic redshifts. The dashed line is the unit line, and the dotted lines delineate the maximum error used for cross matching, \mbox{$|\Delta z| / (1+z) = \epsilon_z$} with $\epsilon_z$=0.03.}
    \label{fig:z_comparison}
\end{figure*}

The matching enables us to compare measured properties from the different catalogues and to associate new properties (e.g. mass) to the \gls{des} Y1 \gls{rm} clusters.
\Cref{fig:z_comparison} shows the \gls{des} Y1 \gls{rm} photometric redshifts versus the redshifts from the external catalogues and meta-catalogues. The panel on the left shows all 1040 matches, including both photometric and spectroscopic redshifts. We adopted the following arbitrary priority order for displaying the redshifts: \gls{mccd}, \gls{mcxc}, \gls{mcsz}, \eROSITA, ComPRASS, Abell, and \gls{lc2}. 

The agreement between the \gls{des} Y1 \gls{rm} redshifts and the redshifts from the external catalogues and meta-catalogues is good, as expected given the matching criteria in \cref{sec:metho} -- in particular, \mbox{$|\Delta z| / (1+z_{\rm M}) < \epsilon_z$} with $\epsilon_z$=0.03, delimited by the dotted lines in the figure. The panel on the right shows the 114 matches with spectroscopic redshifts only. The scatter in the relation is largely reduced, as expected given the smaller errors in spectroscopic redshifts compared to photometric redshifts.

\begin{figure*}
    \centering
     \includegraphics[width=0.45\linewidth]{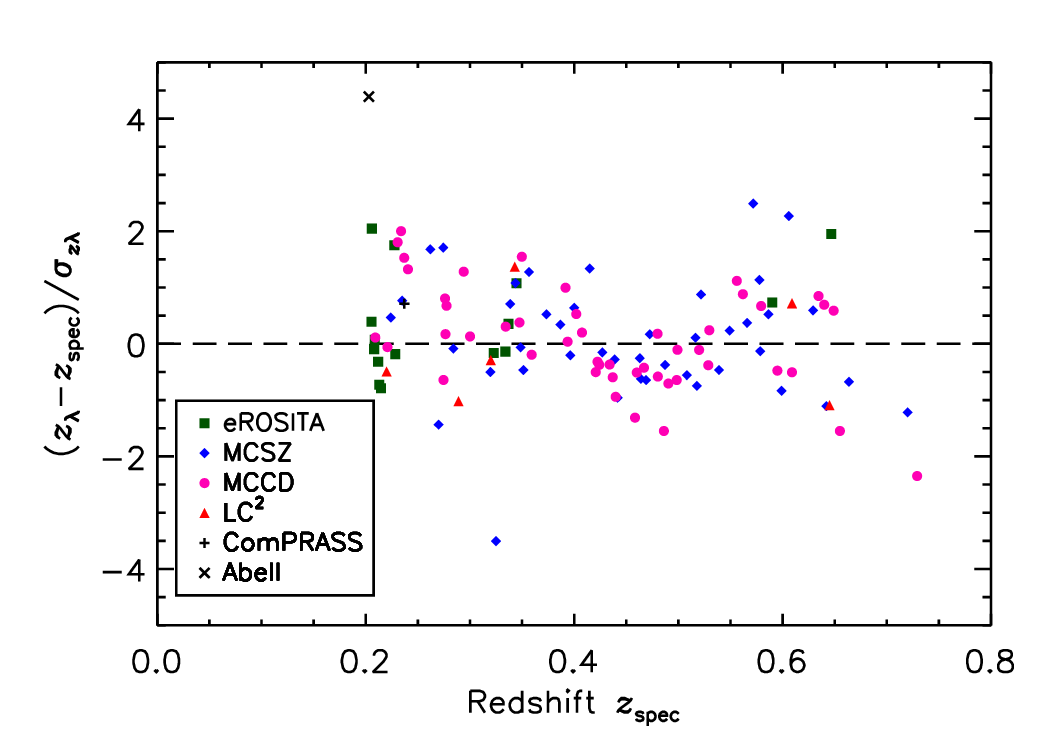} \includegraphics[width=0.45\linewidth]{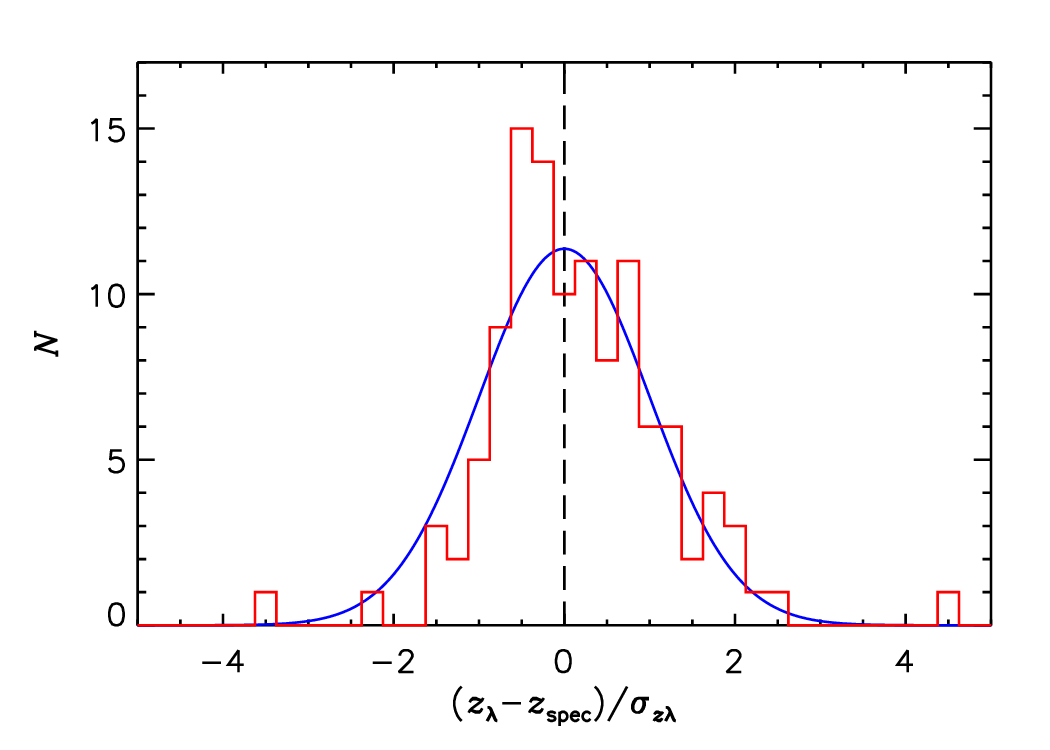}
     \caption{Quantitative assessment of the  \gls{des} Y1 \gls{rm} photometric redshifts and associated errors. \emph{Left}: plot of the normalised scatter in the \gls{des} Y1 \gls{rm} photometric redshifts against matched spectroscopic redshifts as a function of the spectroscopic redshift. \emph{Right}: corresponding histogram in red and normal distribution in blue. The vertical dashed line identifies zero.}
    \label{fig:z_acc}
\end{figure*}

\Cref{fig:z_acc} further quantifies the quality of the photometric redshifts. The panel on the left shows the difference between the \gls{des} Y1 \gls{rm} photometric redshift and the spectroscopic redshift from the right-hand panel of \cref{fig:z_comparison}, divided by the \gls{des} Y1 \gls{rm} photometric redshift error $\sigma_{z\lambda}$, as a function of the spectroscopic redshift. The quantity $\sigma_{z\lambda}$ increases from 0.005 to 0.015 between $z=0$ and $0.4$, decreases from 0.015 to 0.010 between $z=0.4$ and 0.55, and increases again from 0.010 to 0.020 between $z=0.55$ and 0.7. If \gls{des} Y1 \gls{rm} photometric redshifts and associated errors are correctly estimated, the points in the left-hand panel of \cref{fig:z_acc} should scatter around zero with a standard deviation of one, assuming that the errors on spectroscopic redshifts are negligible compared to the errors on photometric redshifts. 

We test this with the histogram in the right-hand panel of \cref{fig:z_acc}. The vertical dashed line indicates zero and the blue curve is a normal distribution with standard deviation of unity. 
The red histogram has a mean value of 0.16 (with error of $1/\sqrt{114}=0.09$) and a standard deviation 1.05 (with error $1/\sqrt{2 \times 114}=0.07$). 
A $\chi^2$ test, however, yields a reduced $\chi_\nu^2=1.95$ with $\nu=41$ degrees of freedom (number of bins in the redshift histogram), statistically rejecting the normal distribution as a good fit. This is largely driven by the peak in the red histogram just below zero, suggesting a small bias (distortion) in the \gls{rm} redshift estimates. The peak is dominated by clusters at $z \sim 0.45$, visible in the left-hand panel of the figure. This result is consistent with the distortion seen in Figure 4 of~\cite{Rykoff2016} and explained by centring failures (clusters with correct photometric redshifts but whose assigned central galaxy is not a cluster member).
This test will be used for the first cluster catalogues produced by  \Euclid in order to check the quality of the cluster photometric redshifts.

\begin{figure}
    \centering
    \includegraphics[width=1.0\linewidth]{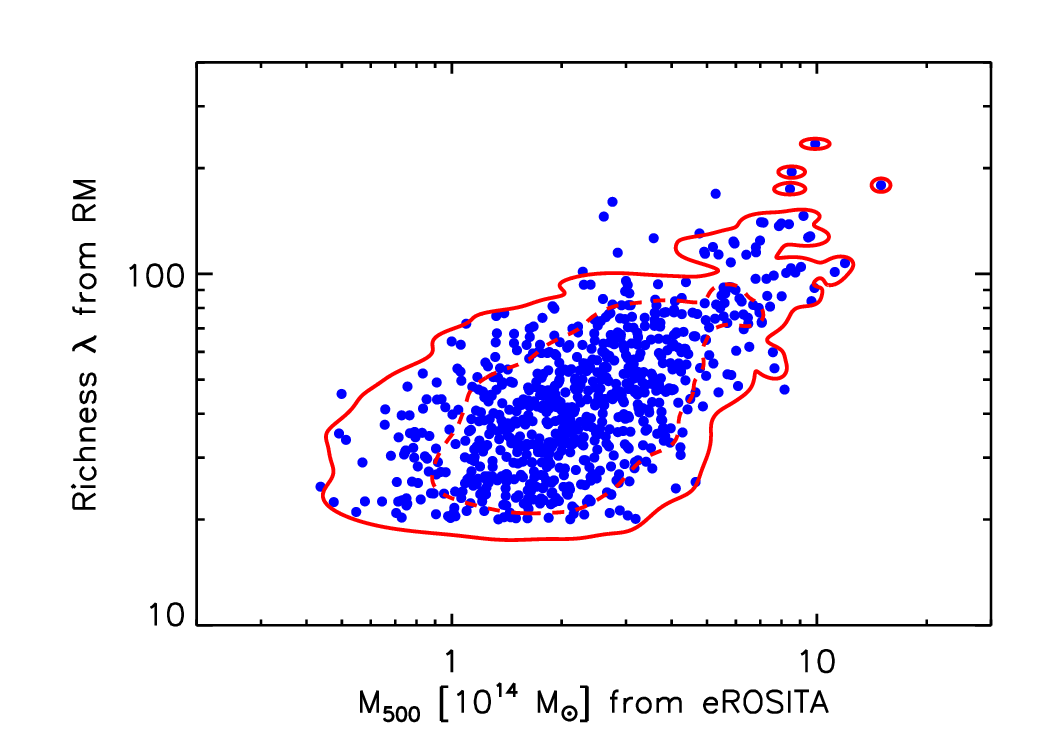}
    \caption{\gls{des} Y1 \gls{rm} richness versus \eROSITA mass for matched clusters. The positive correlation is clear. Red dashed/solid lines are 68\%/95\% confidence limit contours calculated from  a global probability distribution that is constructed by summing the two dimensional Gaussian probability distributions for each cluster with mean equal to  the measured cluster mass and richness and standard deviation corresponding to the given error bars.}
    \label{fig:scatter_plot_richness_erosita}
\end{figure}

We compared the \gls{des} Y1 \gls{rm} richness to mass proxies from the external catalogues and meta-catalogues. One expects a correlation with some scatter -- the richness-mass scaling relation -- if the cluster matches are correct. We present the results for the match between \gls{des} Y1 \gls{rm} and \eROSITA in \cref{fig:scatter_plot_richness_erosita}. One can clearly see the correlation between the two quantities, with more massive eROSITA clusters being richer.

The other relations between the \gls{des} Y1 \gls{rm} richness and mass proxy from external catalogues and meta-catalogues are shown in \cref{fig:scatter_plot_richness_1,fig:scatter_plot_richness_2}. They all manifest clear positive correlation with scatter, and no obvious outliers. We note that the scatter depends on the mass proxy considered, which is expected because the galaxy content of a cluster is more or less closely linked to the other cluster components (gas or dark matter) and observables (X-ray, SZ, velocity dispersion, weak lensing, etc.).

\Cref{fig:sky_distribution_and_redshift_xsz} projects the X-ray and \gls{sz} catalogues and meta-catalogues onto a sky map, and \cref{fig:sky_distribution_and_redshift_optical}
does the same for those in the optical.  In both figures, the left-hand column shows the sky distributions for all the clusters in each catalogue or meta-catalogue in turquoise blue, the clusters in the \gls{des} Y1 \gls{rm} footprint in blue, and the matched clusters in red. The right-hand column plots the redshift histograms for these three samples in each case, with the dashed blue line tracing the full sample, the solid blue line the sub-sample falling within the \gls{des} Y1 \gls{rm} footprint, and the red filled histogram the matched clusters. 

The low number of global matches between the \gls{mcxc} and \gls{des} Y1 \gls{rm} is understood from the difference in the redshift distributions of the catalogues: the \gls{mcxc} lies mostly below $z=0.2$, a region not covered by \gls{des} Y1 \gls{rm}. The situation is similar for the Abell catalogue. In contrast, the \gls{mcsz} covers better the \gls{des} Y1 \gls{rm} redshift range, leading to the higher number of global matches between the two. The high number of matches between eROSITA and DES Y1 RM can be explained by the large number of sources in the \eROSITA catalogue (so the depth of the data).

\gls{des} Y1 \gls{rm} contains nearly all the clusters from \gls{mcxc}, \gls{mcsz}, ComPRASS, \gls{mccd}, and Abell over the redshift range covered by the catalogue ($0.2<z<0.86$). For these catalogues, the small difference between the solid blue line and the red histogram may be explained by the small fraction of bad HEALPix\footnote{\url{http://healpix.sf.net/}} pixels in the \gls{des} Y1 footprint. 
 
 The difference between the solid blue line and the red histogram is larger for \eROSITA and \gls{lc2}.
 This is notable, in particular for \eROSITA, and deserves further investigation.
 
\section{\label{sec:summ}Conclusions}

The method presented herein has proven to be efficient at cross-matching the \gls{des} Y1 \gls{rm} cluster catalogue with external catalogues and meta-catalogues. In particular, the cross-matching procedure allows us to identify complex cases (e.g. merging or multiple systems). Given these results, we plan to use it for validating the \gls{ecgc} detections. A first practical test will be to apply it to the \Euclid Quick Data Release (Q1) data~\citep{Q1-TP001,Q1-SP050}. Possible limitations of the method are time and human resources: the procedure requires detailed manual intervention, and choices involving several persons. The method would thus benefit from comparison to automatic cross-matching procedures, which are also being developed by the \gls{ec}, and by the Rubin\footnote{\url{https://lsstdesc.org/clevar/}} Collaboration. 

The choice of catalogues and meta-catalogues for the external validation is an important aspect of the procedure. In this work, we restricted the cross-matching to large catalogues and meta-catalogues in the X-ray, \gls{sz} and optical, containing the most massive known clusters. In the near future, we aim to include the deeper optical catalogues mentioned in \cref{sec:complementary}, which will allow us to validate lower significance and more distant clusters in the \gls{ecgc}. The deeper optical catalogues will also be useful to validate cluster detections in the \gls{edfall}, where the density of clusters will be higher than in the wide survey. The \XMM\ follow-up of the \gls{edff} will be an additional key element in characterising \Euclid detections in the deep field regime.

\begin{acknowledgements}
This research has made use of the NASA/IPAC Extragalactic Database (NED),
which is operated by the Jet Propulsion Laboratory, California Institute of Technology,
under contract with the National Aeronautics and Space Administration. This research has made use of the SIMBAD database, operated at CDS, Strasbourg, France. Some of the results in this paper have been derived using the healpy and HEALPix packages. 
SAS acknowledges the support of NASA ROSES Grant 12-EUCLID11-0004.
ET acknowledges support by STFC through Imperial College Astrophysics Consolidated Grant ST/W000989/1.
\AckEC
\end{acknowledgements}

%
% Here comes the reference list, generated via bibtex from
% the bibfile AandA.bib
%

\bibliography{Euclid_ev_ecc}

\begin{appendix}

\section{\label{sec:sky_distrib} Richness-mass proxy scatter plots, sky distributions, and redshift histograms for catalogues and meta-catalogues}

\begin{figure}
    \centering
    \includegraphics[width=\hsize]{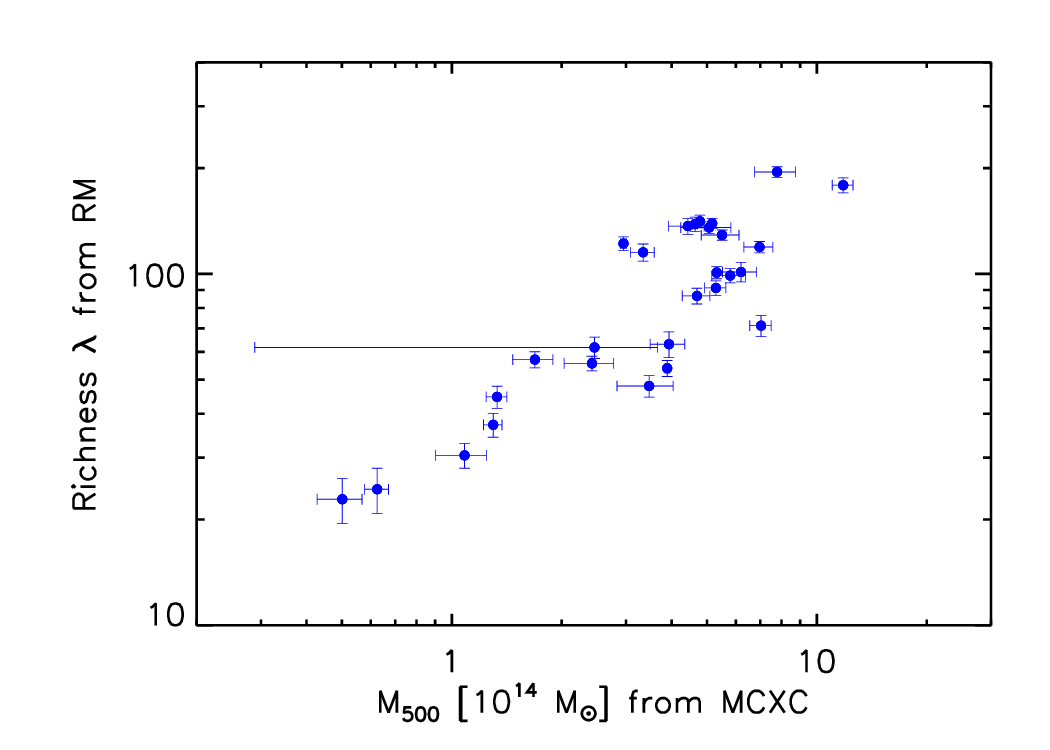}\\
    \includegraphics[width=\hsize]{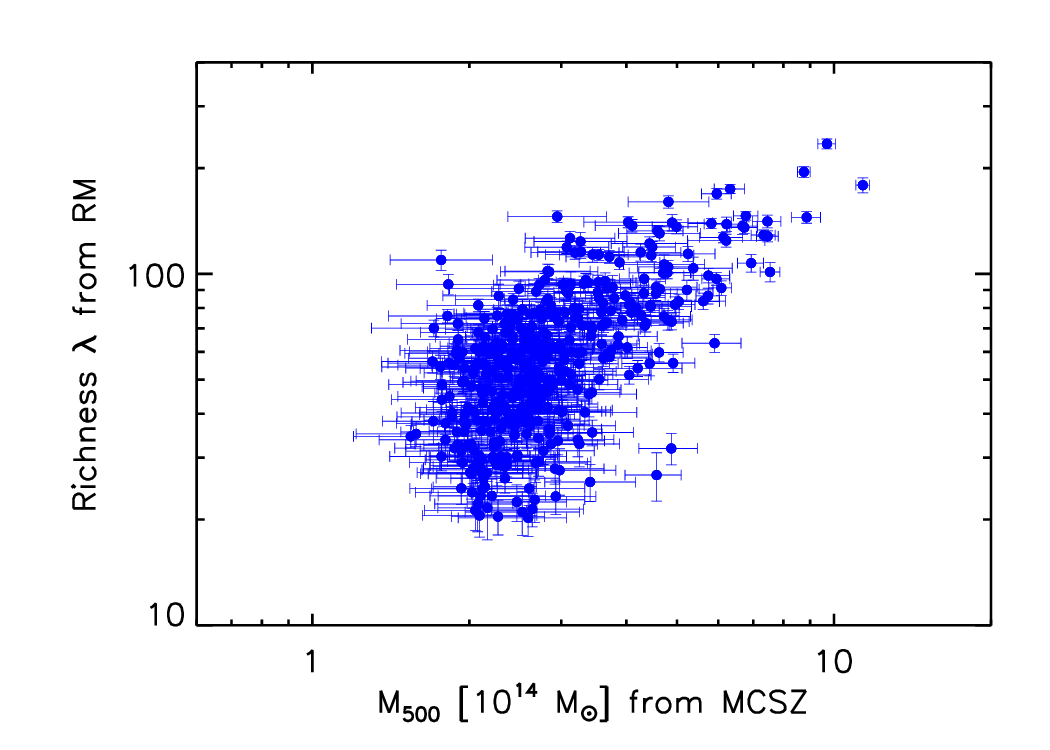}
    \caption{\gls{des} Y1 \gls{rm} richness versus mass proxies for the meta-catalogues. \emph{Top}: \gls{mcxc}; \emph{Bottom}: \gls{mcsz}.}
    \label{fig:scatter_plot_richness_1}
\end{figure}

\begin{figure}
    \centering
    \includegraphics[width=\hsize]{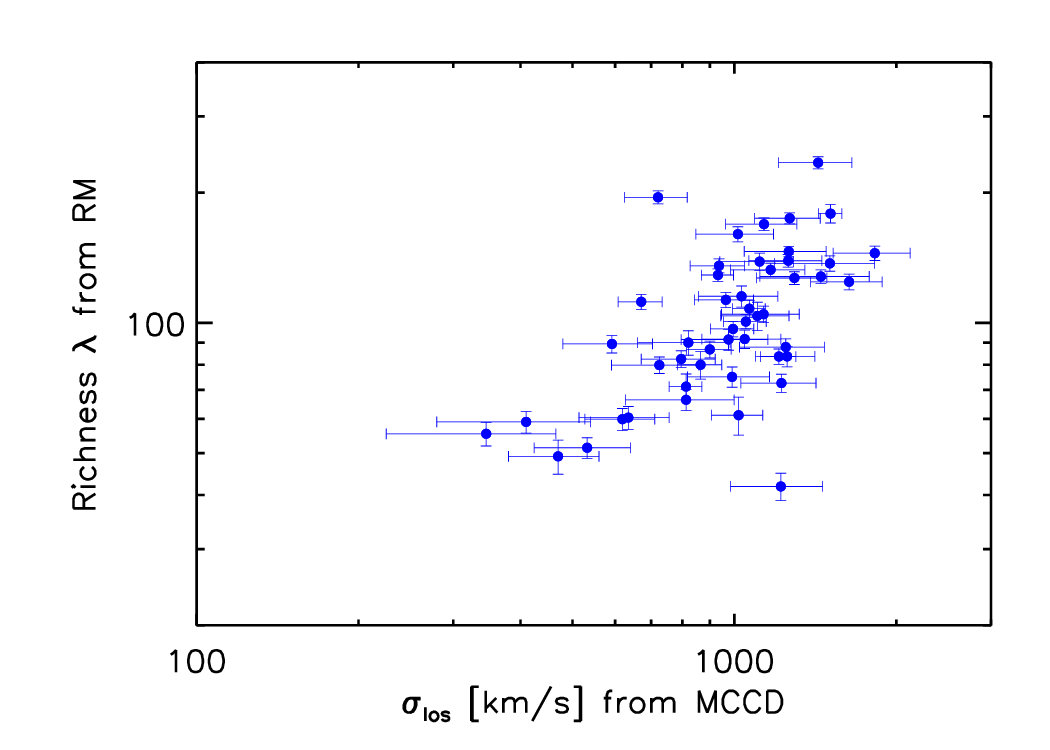}\\
    \includegraphics[width=\hsize]{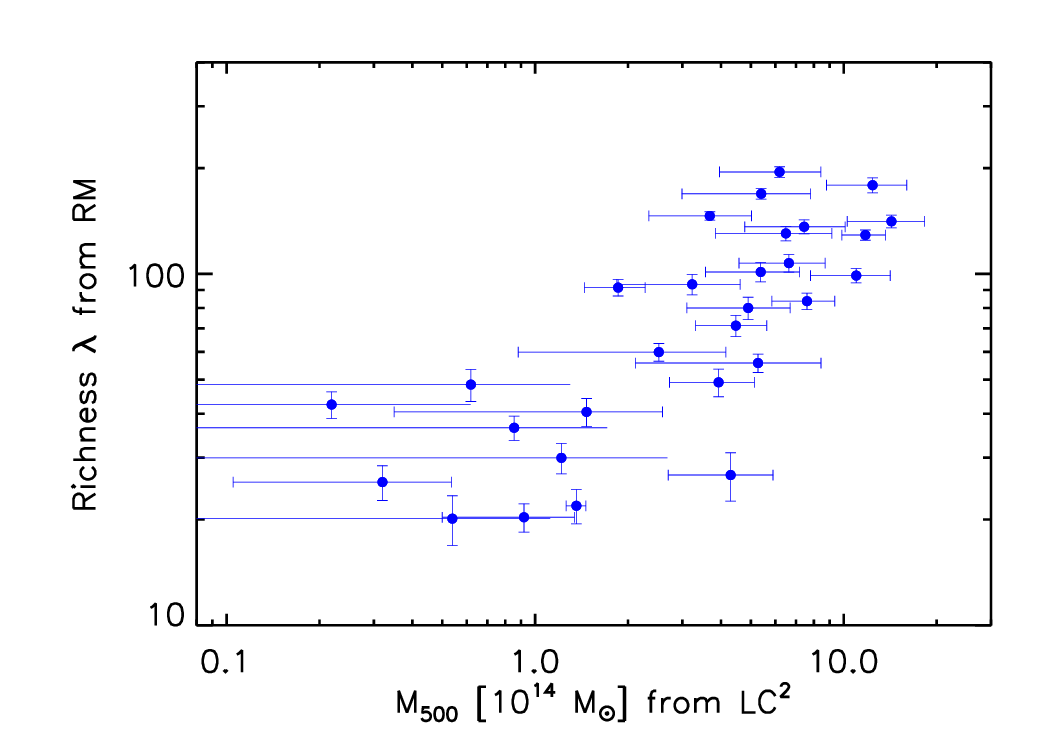}
    \caption{Same as \cref{fig:scatter_plot_richness_1} for \gls{mccd} (\emph{Top}) and \gls{lc2} (\emph{Bottom}). The correlation between richness and the various mass proxies is evident in all cases.}
    \label{fig:scatter_plot_richness_2}
\end{figure}

We present richness-mass proxy scatter plots in \cref{fig:scatter_plot_richness_1,fig:scatter_plot_richness_2} for the \gls{mcxc}, \gls{mcsz}, \gls{mccd}, and \gls{lc2} meta-catalogues. These figures complement \cref{fig:scatter_plot_richness_erosita} in the main text. They demonstrate that \gls{des} Y1 \gls{rm} richness well correlates with the various mass proxies of the meta-catalogues, with differing scatter. The large error bar for the mass of one of the \gls{mcxc} clusters comes from a cluster in the Catalog of X-ray-selected extended galaxy clusters from the \gls{rass}~\citep[RXGCC,][]{Xu2022} with a low-significance measurement.

\begin{figure*}[!ht]
    \centering
    \includegraphics[width=0.45\linewidth]{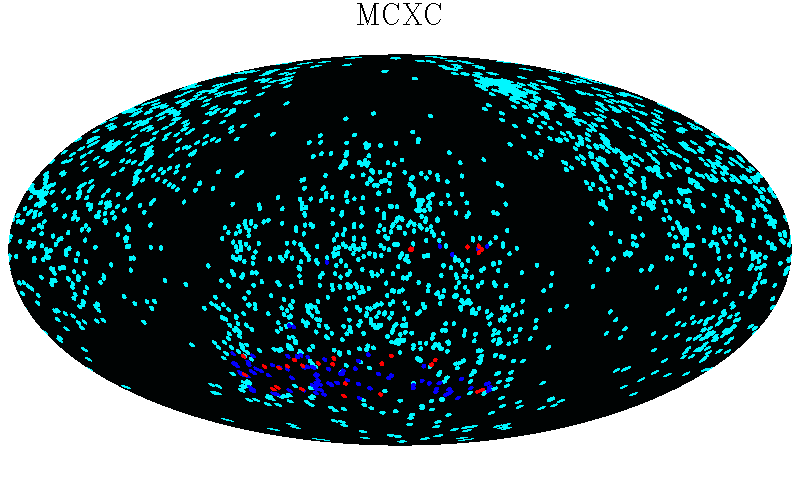} \includegraphics[width=0.45\linewidth]{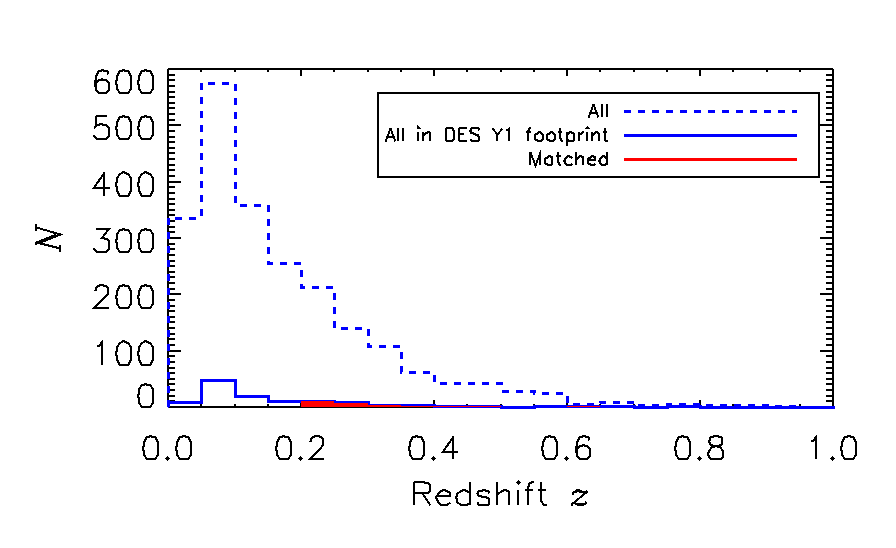}
    \includegraphics[width=0.45\linewidth]{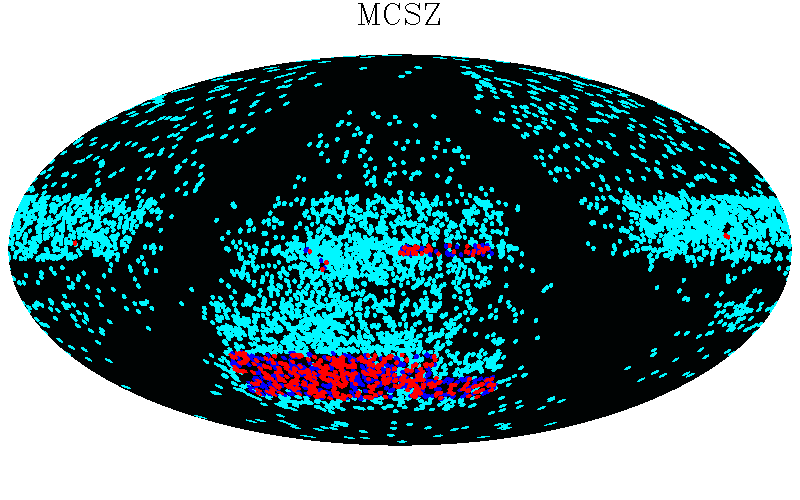} \includegraphics[width=0.45\linewidth]{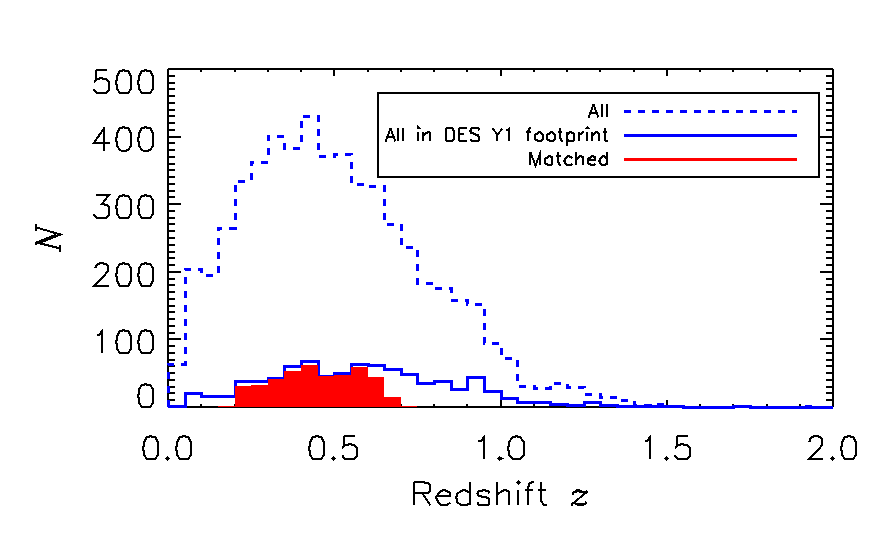}
    \includegraphics[width=0.45\linewidth]{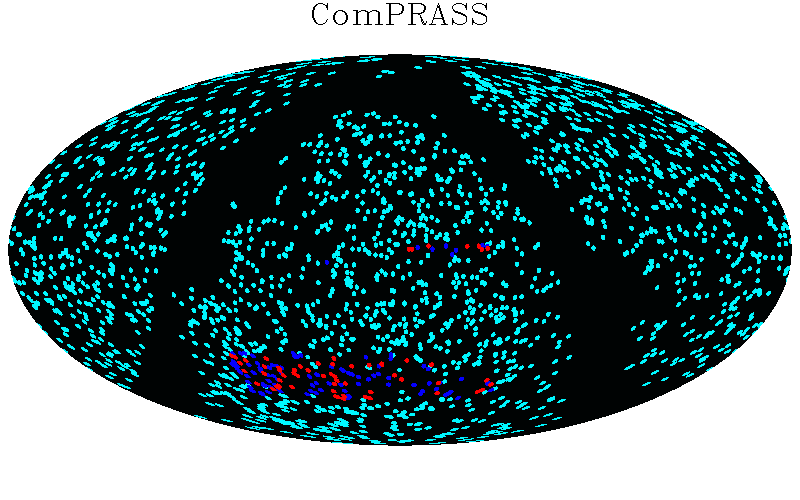} \includegraphics[width=0.45\linewidth]{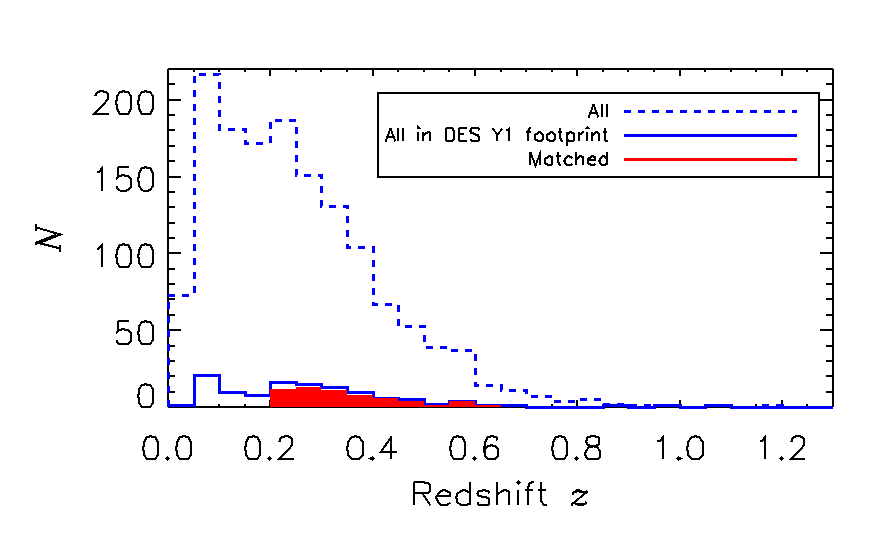}
    \includegraphics[width=0.45\linewidth]{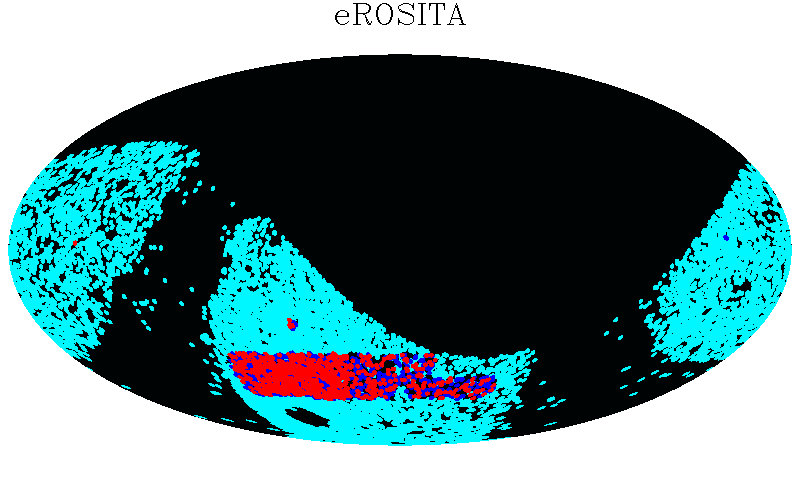} \includegraphics[width=0.45\linewidth]{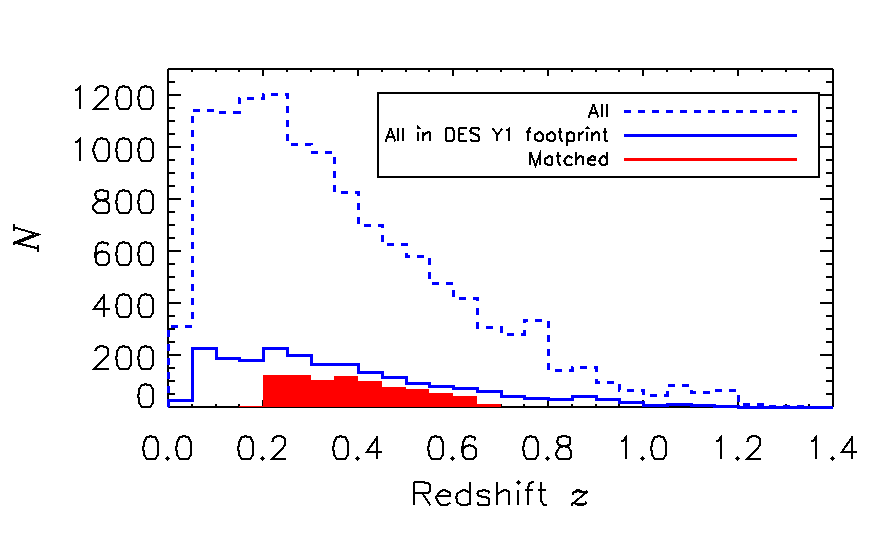}
    \caption{Sky distributions in equatorial coordinates (\emph{left}) and redshift histograms (\emph{right}) of X-ray and SZ catalogues and meta-catalogues. \emph{From top to bottom}: \gls{mcxc}, \gls{mcsz}, ComPRASS, \eROSITA. In the sky distributions, the full catalogue is shown in turquoise blue; those clusters falling within the \gls{des} Y1 \gls{rm} footprint are shown in blue; the clusters matched with \gls{des} Y1 \gls{rm} are shown in red. In the redshift histograms, the dashed blue line traces the full catalogue; the solid blue line, those clusters falling within \gls{des} Y1 \gls{rm} footprint; and the red line, the matched clusters.}
    \label{fig:sky_distribution_and_redshift_xsz}
\end{figure*}

\begin{figure*}[!ht]
    \centering
    \includegraphics[width=0.45\linewidth]{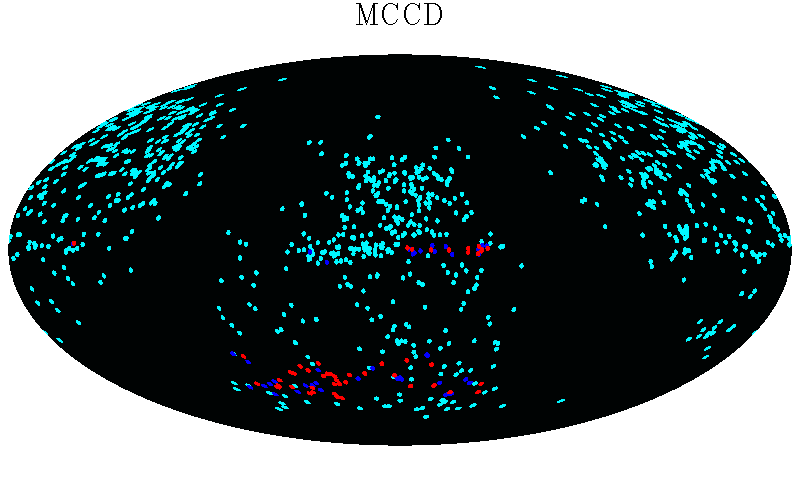} \includegraphics[width=0.45\linewidth]{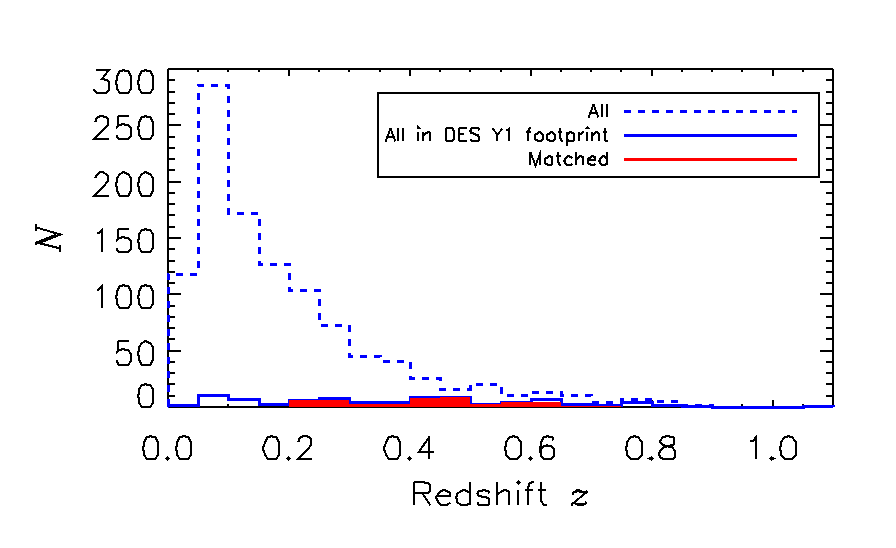}
    \includegraphics[width=0.45\linewidth]{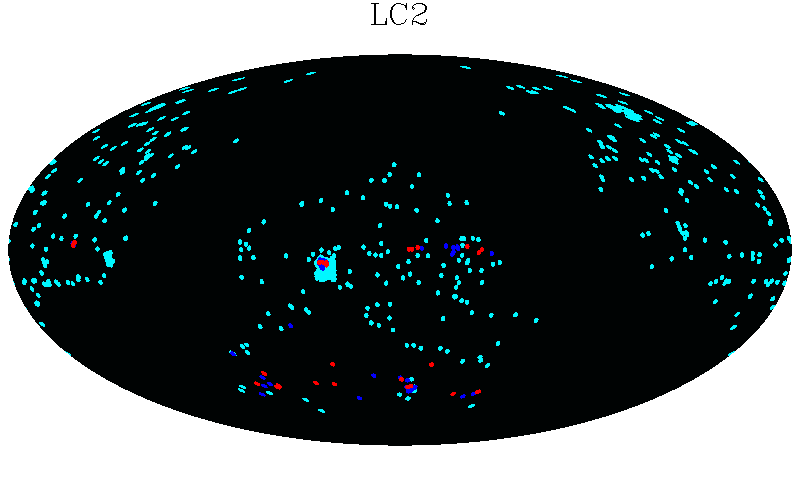} \includegraphics[width=0.45\linewidth]{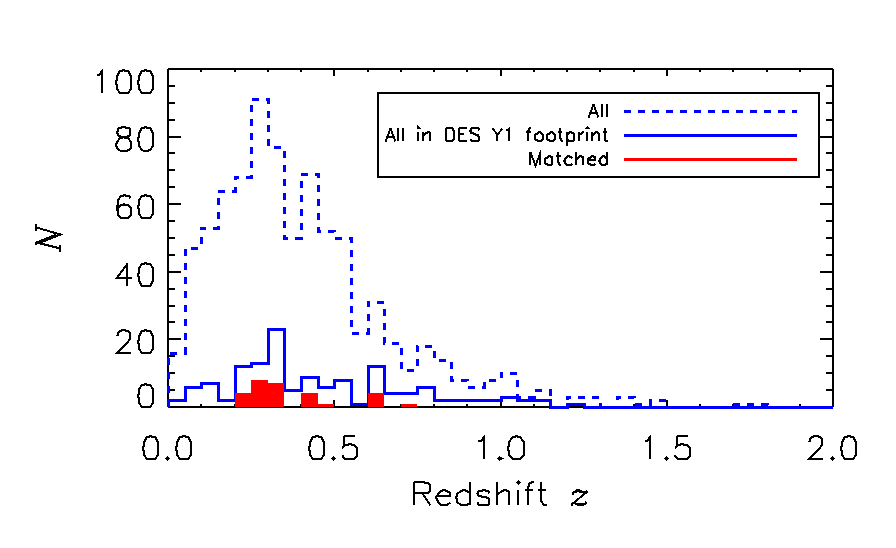}
    \includegraphics[width=0.45\linewidth]{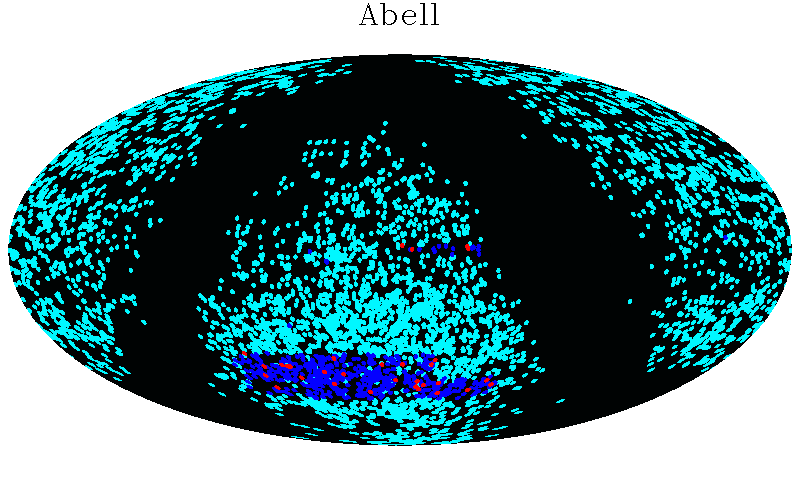} \includegraphics[width=0.45\linewidth]{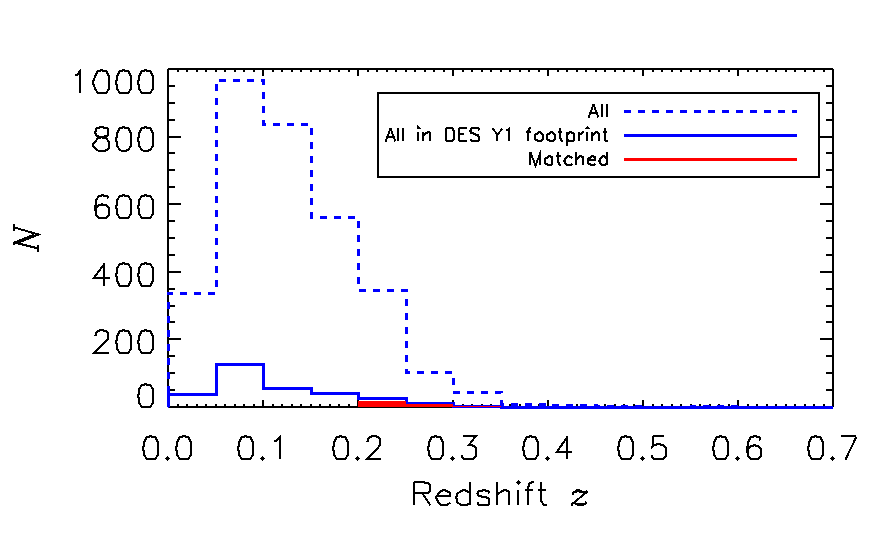}
     \caption{Same as \cref{fig:sky_distribution_and_redshift_xsz} but for the optical meta-catalogues and the \gls{mccd}, \gls{lc2}, Abell catalogues.}
    \label{fig:sky_distribution_and_redshift_optical}
\end{figure*}

\Cref{fig:sky_distribution_and_redshift_xsz,fig:sky_distribution_and_redshift_optical} show, in the left-hand columns, the sky distributions of the clusters in each catalogue and meta-catalogue (turquoise blue), those clusters falling within the \gls{des} Y1 footprint\footnote{We used {\tt redmapper\_y1a1\_public\_v6.4\_zmask.fits}.} (blue), and the clusters matched with \gls{des} Y1 \gls{rm} (red). They also show, in the right-hand columns, the corresponding redshift distributions. \gls{des} Y1 \gls{rm} detects nearly all the clusters from the \gls{mcxc}, the \gls{mcsz}, the ComPRASS catalogue, the \gls{mccd}, and the Abell catalogue over the redshift range $0.2<z<0.86$. It does not detect all eROSITA and ${\rm LC^2}$ clusters in the same redshift range; see main text for a discussion.

\section{\label{sec:fields} Description of the catalogue fields in the validated \gls{des} Y1 \gls{rm} catalogue}

\Cref{tab:overview} summarises the fields added to the \gls{des} Y1 \gls{rm} catalogue after the validation steps. The \gls{mcxc}, \gls{mcsz}, ComPRASS, and \gls{lc2} fields are provided for the \tway and $\ntfive$ matchings, while the \gls{mccd} and Abell fields are provided for the positional matching only because the two catalogues do not give cluster mass estimates.

\begin{table*}
\caption{Summary of the fields added to the \gls{des} Y1 \gls{rm} catalogue after validation.}
\smallskip
\label{tab:overview}
\smallskip
\begin{tabular}{llll}
\hline
     \multicolumn{1}{c}{{\bf Field Name}} &
    \multicolumn{1}{l}{{\bf FORMAT }} &  
    \multicolumn{1}{l}{{\bf UNIT}} &
    \multicolumn{1}{l}{{\bf DESCRIPTION}} \\
\hline
{\tt MCXC\_2WAY\_MATCH}  	& BYTE  &                         	&	Is the RM cluster matched \tway  with a MCXC cluster?  \\
{\tt MCXC\_2WAY\_INDEX}	 	& INTEGER &                         	&	Index of the \tway  match in the MCXC meta-catalogue  \\
{\tt MCXC\_2WAY\_SEP} 		& DOUBLE &  arcmin                &      Separation between the RM cluster and its MCXC \tway  match \\
{\tt MCXC\_2WAY\_NORMSEP}	& DOUBLE &                         	&  	Separation normalised to the $\tfive$ of the MCXC \tway  match \\
{\tt MCXC\_NT500\_MATCH}  	& BYTE  &                         	&	Is the RM cluster matched $\ntfive$ with a MCXC cluster?  \\
{\tt MCXC\_NT500\_INDEX}	 	& INTEGER &                         	&	Index of the $\ntfive$ match in the MCXC meta-catalogue  \\
{\tt MCXC\_NT500\_SEP} 		& DOUBLE &  arcmin                &      Separation between the RM cluster and its MCXC $\ntfive$ match \\
{\tt MCXC\_NT500\_NORMSEP}	& DOUBLE &                         	&  	Separation normalised to the $\tfive$  of the MCXC $\ntfive$ match \\
{\tt MCSZ\_2WAY\_MATCH}  	& BYTE  &                         	&	Is the RM cluster matched \tway  with a MCSZ cluster?  \\
{\tt MCSZ\_2WAY\_INDEX}	 	& INTEGER &                         	&	Index of the \tway  match in the MCSZ meta-catalogue  \\
{\tt MCSZ\_2WAY\_SEP} 		& DOUBLE &  arcmin                &      Separation between the RM cluster and its MCSZ \tway  match \\
{\tt MCSZ\_2WAY\_NORMSEP}	& DOUBLE &                         	&  	Separation normalised to the $\tfive$ of the MCSZ \tway  match \\
{\tt MCSZ\_NT500\_MATCH}  	& BYTE  &                         	&	Is the RM cluster matched $\ntfive$ with a MCSZ cluster?  \\
{\tt MCSZ\_NT500\_INDEX}	 	& INTEGER &                         	&	Index of the $\ntfive$ match in the MCSZ meta-catalogue  \\
{\tt MCSZ\_NT500\_SEP} 		& DOUBLE &  arcmin                &      Separation between the RM cluster and its MCSZ $\ntfive$ match \\
{\tt MCSZ\_NT500\_NORMSEP}	& DOUBLE &                         	&  	Separation normalised to the $\tfive$  of the MCSZ $\ntfive$ match \\
{\tt COMPRASS\_2WAY\_MATCH}  	& BYTE  &                         	&	Is the RM cluster matched \tway  with a ComPRASS cluster?  \\
{\tt COMPRASS\_2WAY\_INDEX}	 	& INTEGER &                         	&	Index of the \tway  match in the ComPRASS catalogue  \\
{\tt COMPRASS\_2WAY\_SEP} 		& DOUBLE &  arcmin                &      Separation between the RM cluster and its ComPRASS \tway  match \\
{\tt COMPRASS\_2WAY\_NORMSEP}	& DOUBLE &                         	&  	Separation normalised to the $\tfive$ of the ComPRASS \tway  match \\
{\tt COMPRASS\_NT500\_MATCH}  	& BYTE  &                         	&	Is the RM cluster matched $\ntfive$ with a ComPRASS cluster?  \\
{\tt COMPRASS\_NT500\_INDEX}	 	& INTEGER &                         	&	Index of the $\ntfive$ match in the ComPRASS catalogue  \\
{\tt COMPRASS\_NT500\_SEP} 		& DOUBLE &  arcmin                &      Separation between the RM cluster and its ComPRASS $\ntfive$ match \\
{\tt COMPRASS\_NT500\_NORMSEP}	& DOUBLE &                         	&  	Separation normalised to the $\tfive$  of the ComPRASS $\ntfive$ match \\
{\tt EROSITA\_2WAY\_MATCH}  	& BYTE  &                         	&	Is the RM cluster matched \tway  with a eROSITA cluster?  \\
{\tt EROSITA\_2WAY\_INDEX}	 	& INTEGER &                         	&	Index of the \tway  match in the eROSITA catalogue  \\
{\tt EROSITA\_2WAY\_SEP} 		& DOUBLE &  arcmin                &      Separation between the RM cluster and its eROSITA \tway  match \\
{\tt EROSITA\_2WAY\_NORMSEP}	& DOUBLE &                         	&  	Separation normalised to the $\tfive$ of the eROSITA \tway  match \\
{\tt EROSITA\_NT500\_MATCH}  	& BYTE  &                         	&	Is the RM cluster matched $\ntfive$ with a eROSITA cluster?  \\
{\tt EROSITA\_NT500\_INDEX}	 	& INTEGER &                         	&	Index of the $\ntfive$ match in the eROSITA catalogue  \\
{\tt EROSITA\_NT500\_SEP} 		& DOUBLE &  arcmin                &      Separation between the RM cluster and its eROSITA $\ntfive$ match \\
{\tt EROSITA\_NT500\_NORMSEP}	& DOUBLE &                         	&  	Separation normalised to the $\tfive$  of the eROSITA $\ntfive$ match \\
{\tt MCCD\_2WAY\_MATCH}  	& BYTE  &                         	&	Is the RM cluster matched \tway  with a MCCD cluster?  \\
{\tt MCCD\_2WAY\_INDEX}	 	& INTEGER &                         	&	Index of the \tway  match in the MCCD meta-catalogue  \\
{\tt MCCD\_2WAY\_SEP} 		& DOUBLE &  arcmin                &      Separation between the RM cluster and its MCCD \tway  match \\
{\tt MCCD\_2WAY\_NORMSEP}	& DOUBLE &                         	&  	Separation normalised to the $\tfive$ of the MCCD \tway  match \\
{\tt LC2\_2WAY\_MATCH}  	& BYTE  &                         	&	Is the RM cluster matched \tway  with a ${\rm LC}^2$ cluster?  \\
{\tt LC2\_2WAY\_INDEX}	 	& INTEGER &                         	&	Index of the \tway  match in the ${\rm LC}^2$ meta-catalogue \\
{\tt LC2\_2WAY\_SEP} 		& DOUBLE &  arcmin                &      Separation between the RM cluster and its ${\rm LC}^2$ \tway  match \\
{\tt LC2\_2WAY\_NORMSEP}	& DOUBLE &                         	&  	Separation normalised to the $\tfive$ of the ${\rm LC}^2$ \tway  match \\
{\tt LC2\_NT500\_MATCH}  	& BYTE  &                         	&	Is the RM cluster matched $\ntfive$ with a ${\rm LC}^2$ cluster?  \\
{\tt LC2\_NT500\_INDEX}	 	& INTEGER &                         	&	Index of the $\ntfive$ match in the ${\rm LC}^2$ meta-catalogue  \\
{\tt LC2\_NT500\_SEP} 		& DOUBLE &  arcmin                &      Separation between the RM cluster and its ${\rm LC}^2$ $\ntfive$ match \\
{\tt LC2\_NT500\_NORMSEP}	& DOUBLE &                         	&  	Separation normalised to the $\tfive$  of the ${\rm LC}^2$ $\ntfive$ match \\
{\tt ABELL\_2WAY\_MATCH}  	& BYTE  &                         	&	Is the RM cluster matched \tway  with a Abell cluster?  \\
{\tt ABELL\_2WAY\_INDEX}	 	& INTEGER &                         	&	Index of the \tway  match in the Abell catalogue  \\
{\tt ABELL\_2WAY\_SEP} 		& DOUBLE &  arcmin                &      Separation between the RM cluster and its Abell \tway  match \\
{\tt ABELL\_2WAY\_NORMSEP}	& DOUBLE &                         	&  	Separation normalised to the $\tfive$ of the Abell \tway  match \\
\hline
\end{tabular}
\end{table*}

\section{\label{apdx:MCCD} Description of the catalogue fields in the MCCD}

We list the contents of the fields in the MCCD in Table~\ref{tab:mccd}

\begin{table*}
\caption{Fields in the MCCD}
\smallskip
\label{tab:mccd}
\smallskip
\begin{tabular}{llll}
\hline
     \multicolumn{1}{c}{{\bf Field Name}} &
    \multicolumn{1}{l}{{\bf FORMAT }} &  
    \multicolumn{1}{l}{{\bf UNIT}} &
    \multicolumn{1}{l}{{\bf DESCRIPTION}} \\
\hline
{\tt MCCD-INDEX}  	        & SHORT  &      & Unique numerical ID \\
{\tt MCCD-NAME-PRE}         & STRING &      & Primary source catalogue name \\
{\tt MCCD-NAME-NUM}  	    & STRING &      & Primary source catalogue ID \\
{\tt MCCD-RA-DECIMAL}  	    & DOUBLE &degrees&  \\
{\tt MCCD-DEC-DECIMAL}  	& DOUBLE &degrees&  \\
{\tt MCCD-SPECZ}  	        & DOUBLE &      &  cluster spectroscopic redshift\\
{\tt MCCD-REFERENCE}  	& STRING &      &  Primary source publication \\
{\tt MCCD-NED-NAME-PRE}  	& STRING &      &  NED cluster name catalogue \\
{\tt MCCD-NED-NAME-NUM}  	& STRING &      &  NED cluster name ID \\
{\tt MCCD-RA-NED}  	        & STRING & hms  &  \\
{\tt MCCD-DEC-NED}  	    & STRING & dms  &  \\
{\tt MCCD-NED-Z}  	        & DOUBLE &      &  NED cluster redshift \\
{\tt MCCD-N-MEMBERS}  	    & SHORT  &      &  Number of spectroscopic member galaxies \\
{\tt MCCD-R-APERTURE}  	    & FLOAT  &Mpc h$^{-1}$ &  Aperture of $\sigma$ in primary source \\
{\tt MCCD-SIGMA-LOS}  	    & DOUBLE & km s$^{-1}$ &  Velocity dispersion in primary source \\
{\tt MCCD-ERR-SIGMA-LOS}  	& FLOAT  & km s$^{-1}$&  Uncertainty in $\sigma$\\
{\tt MCCD-SIGMA-LOS-R200}  	& FLOAT  & km s$^{-1}$&  Velocity dispersion within R200 \\
{\tt MCCD-ERR-SIGMA-LOS-STAND} & DOUBLE &      &  Standardised uncertainty in $\sigma$\\
\hline
\end{tabular}
\label{LastPage}
\end{table*}
\end{appendix}

\end{document}